\documentclass[aps,prb,twocolumn,floatfix,footinbib,superscriptaddress]{revtex4}
\usepackage[english]{babel}
\usepackage[utf8]{inputenc}
\usepackage{graphicx}
\usepackage{fancyhdr}
\usepackage[active]{srcltx}
\usepackage{setspace}
\usepackage{float}
\usepackage{amsmath}
\usepackage{amsfonts}
\usepackage[]{natbib}
\usepackage{color}

\usepackage{xargs}                      % Use more than one optional parameter in a new commands
\usepackage[pdftex,dvipsnames]{xcolor}  % Coloured text etc.
\usepackage[colorinlistoftodos,prependcaption,textsize=tiny]{todonotes}
\newcommandx{\unsure}[2][1=]{\todo[linecolor=red,backgroundcolor=red!25,bordercolor=red,#1]{#2}}
\newcommandx{\change}[2][1=]{\todo[linecolor=blue,backgroundcolor=blue!25,bordercolor=blue,#1]{#2}}
\newcommandx{\info}[2][1=]{\todo[linecolor=OliveGreen,backgroundcolor=OliveGreen!25,bordercolor=OliveGreen,#1]{#2}}
\newcommandx{\improvement}[2][1=]{\todo[linecolor=Plum,backgroundcolor=Plum!25,bordercolor=Plum,#1]{#2}}
\newcommandx{\thiswillnotshow}[2][1=]{\todo[disable,#1]{#2}}

\begin{document}

\title{Spin-orbit coupling effects in zinc-blende InSb and wurtzite InAs nanowires: Realistic calculations with multiband $\vec{k} \cdot \vec{p}$ method}

\author{Tiago Campos}
\affiliation{Instituto de F\'isica de S\~ao Carlos, Universidade de S\~ao Paulo, 13566-590 S\~ao Carlos, S\~ao Paulo, Brazil}
\affiliation{Institute for Theoretical Physics, University of Regensburg, 93040 Regensburg, Germany}

\author{Paulo E. Faria Junior}
\affiliation{Instituto de F\'isica de S\~ao Carlos, Universidade de S\~ao Paulo, 13566-590 S\~ao Carlos, S\~ao Paulo, Brazil}
\affiliation{Institute for Theoretical Physics, University of Regensburg, 93040 Regensburg, Germany}

\author{Martin Gmitra}
\affiliation{Institute for Theoretical Physics, University of Regensburg, 93040 Regensburg, Germany}
\affiliation{Institute of Physics, P. J. Šafárik University in Košice, Park Angelinum 9, 04001 Košice, Slovakia}

\author{Guilherme M. Sipahi}
\affiliation{Instituto de F\'isica de S\~ao Carlos, Universidade de S\~ao Paulo, 13566-590 S\~ao Carlos, S\~ao Paulo, Brazil}
\affiliation{Department of Physics, State University of New York at Buffalo, Buffalo, New York 14260, USA}

\author{Jaroslav Fabian}
\affiliation{Institute for Theoretical Physics, University of Regensburg, 93040 Regensburg, Germany}

%===============================================================================

\begin{abstract}
A systematic numerical investigation of spin-orbit fields in the conduction bands of
III-V semiconductor nanowires is performed. Zinc-blende InSb nanowires are considered
along [001], [011], and [111] directions, while wurtzite InAs nanowires are studied
along [0001] and [10$\overline{1}$0] or [11$\overline{2}$0] directions. Robust multiband
$\vec{k} \cdot \vec{p}\,$ Hamiltonians are solved by using plane wave expansions of
real-space parameters. In all cases the linear and cubic spin-orbit coupling parameters
are extracted for nanowire widths from 30 to 100 nm. Typical spin-orbit energies
are on the $\mu$eV scale, except for InAs wurtzite nanowires grown along [10$\overline{1}$0]
or [11$\overline{2}$0], in which the spin-orbit energy is about meV, largely
independent of the wire diameter. Significant spin-orbit coupling is obtained by applying
a transverse electric field, causing the Rashba effect. For an electric field of about 4 mV/nm
the obtained spin-orbit energies are about 1 meV for both materials in all investigated
growth directions. The most favorable system, in which the spin-orbit effects
are maximal, are InAs WZ nanowires grown along [1010] or [11$\overline{2}$0], since
here spin-orbit energies are giant (meV) already in the absence of electric field.
The least favorable are InAs WZ nanowires grown along [0001], since here even
the  electric field does not increase the spin-orbit energies beyond 0.1 meV.
The presented results should be useful for investigations of optical orientation,
spin transport, weak localization, and superconducting proximity effects in semiconductor
nanowires.
\end{abstract}

\maketitle

%===============================================================================

\section{Introduction}
\label{sec:intro}

The ultimate goal of spintronics is to enhance the functionalities of electronic
 devices by exploring the spin degree of freedom.~\cite{Zutic2004:RMP,Fabian2007} In
 low dimensional semiconductor nanostructures the control of spin allows to
 transfer information between spin and light,~\cite{Ivchenko2008,Chen2014,Zutic2014,FariaJunior2015,FariaJunior2017}
 can realize topological states of matter~\cite{Bernevig2006} and helical states
 in 1D nanowires~\cite{Oreg2010:PRL,Kloeffel2011,Oreg2014,Schmidt2016} that are
 essential in the search for Majorana zero energy states.~\cite{Lutchyn2010:PRL, Oreg2010:PRL}
 In particular, semiconductor nanowires with strong spin-orbit coupling
 (SOC), such as InSb and InAs, in proximity with an s-wave superconductor
 may support such zero energy bound edge state, when time reversal symmetry is
 broken by a  magnetic field.~\cite{Mourik2012:Science,Deng2012:NL,Das2012:NatPhys,albrecht2016exponential,Deng1557,Zhang2018}

In the absence of space inversion symmetry, in addition to orbital splittings
 at high symmetry points and lines, SOC is also manifested by the spin-splitting
 of the energy bands and by the appearance of a spin texture on  energy surfaces.
 The spin-splitting can arise from two main contributions: the bulk inversion
 asymmetry, known as BIA,~\cite{Dresselhaus1955} and the structural inversion
 asymmetry, known as SIA.~\cite{Bychkov1984a} The former is present in materials,
 such as III-V semiconductors, lacking space inversion in the primitive cell.
 The latter appears due to quantum confinement, at interfaces in heterostructures,
 and in the presence of an applied electric field. Tuning the interplay between
 different sources of SOC can lead to persistent spin helices~\cite{Koralek2009,Fu2016,Schliemann2017:RMP},
 spin field-effect transistors~\cite{Schliemann2003}, $g$-factor
 anisotropies~\cite{Erlingsson2010,Sandoval2013,sandoval2016electron} and
 significant changes in the spin relaxation times~\cite{Furthmeier2016,Kammermeier2018}.

Experimentally, reliable determination of the SOC strength in nanowires is a
 challenging task.~\cite{Rainis2014} Distinct setups yield differing values.~\cite{Zawadzki2004,Thorgilsson2012,Kammermeier2018}
 For example, SOC strengths in the same material, extracted from weak antilocalization
 measurements, come out different.~\cite{Nilsson2009,Nowak2013,VanWeperen2015,Scherubl2016}
 These distinct values are due to the electron-electron contribution (Hartree potential)
 to the Rashba SOC term,~\cite{Bernardes2007,Calsaverini2008,Jespersen2018,2018arXiv180109905W} i. e.,
 the fields induced by the gates lead to a charge unbalance in the system. This
 charge unbalance gives rise to the Hartree potential which is strongly dependent
 on the system configuration and has a large contribution to the Rashba SOC term.
 Theoretically, it is common to use reduced models for the semiconductor conduction band.~\cite{Bychkov1984a,Governale2004,Zutic2004:RMP,Fabian2007}
 In these models, SOC enters as an empirical parameter that can assume a wide
 range of values for the same system depending on what is measured.~\cite{Mireles2001,Scheid2008,Thorgilsson2012,Kammermeier2016,Kammermeier2018}

Motivated by the hybrid semiconductor-superconductor proposal~\cite{Sau2010:PRL,Oreg2010:PRL}
 as a platform for the zero-energy Majorana bound states, that uses semiconductor
 nanowires with large SOC, we investigate the role of BIA and SIA SOC terms
 in free-standing zinc-blende (ZB) InSb and wurtzite (WZ) InAs nanowires. In particular,
 we address theoretically how the quantum confinement, given by the nanowire diameter,
 and the orientation of the nanowire (growth direction) modifies the main
 parameter---the SOC energy---that  defines if the system can (or cannot) host
 Majorana zero-energy excitations.

Both multiband tight-binding and $\vec{k} \cdot \vec{p}\,$
 methods~\cite{Niquet2006,Liao2015,Luo2016,Soluyanov2016,Kammhuber2016,Marcelina2017,Winkler2017} have
 been successful in the determination of the electronic and spintronic properties
 of nanowires. Here we use robust multiband $\vec{k} \cdot \vec{p}\,$ models: a
 14-band Kane model~\cite{winkler2003spin,Fabian2007,Pfeffer1996} to treat ZB
 InSb nanowires, and 8-band model~\cite{FariaJunior2016}
 to treat WZ InAs nanowires, under the envelope function approximation and plane wave
 expansion. The BIA SOC for ZB is taken into account with the addition of the
 extra conduction bands (in comparison with the 8-band model) and their explicit
 coupling parameters; in the WZ case we also include often neglected linear-in-momentum
 SOC terms in the 8-band model.~\cite{Chuang1996,Beresford2004,Rinke2008,Fu2008}
 We also apply an electric field, transverse to the nanowires axes, to investigate
 the Rashba effect and extract the field-dependent spin-orbit parameters.

We give the essential spin-orbit splitting parameters and effective masses for
 the lowest conduction subbands, for a set of hexagonal nanowires, from 30 to 100 nm,
 oriented along different directions: [001], [110], and [111] for ZB InSb,
 and [0001] and [1010] or [11$\overline{2}$0] WZ InAs nanowires. In the absence
 of electric field the spin-orbit energies of ZB InSb nanowires are tiny, on the order of
 micro eVs. However, due to the presence of a linear spin-orbit splitting in the bulk,
 WZ InAs nanowires exhibit giant splittings already in the absence of the field. Although
 symmetry suppresses the spin-orbit energy for [0001] nanowires, the splitting is about
 1 meV in [1010] or [11$\overline{2}$0] cases.

Under an applied electric field, ZB InSb nanowires can exhibit spin splittings as on the
 meV scale, in the fields of a few mV/nm. We find that this Rashba effect does not depend
 essentially on the growth direction, nor on the nanowire diameter. However, the case of WZ InAs
 nanowires is curious. The electric field does not significantly increase the spin splitting
 for [0001] directions. For example, the spin-orbit energy reaches only 20 $\mu eV$ for
 fields of 4 mV/nm, hardly enough to be practical as a platform for topological superconductivity.
 This material system is rather unfavorable in this sense. On the other hand, the spin-orbit
 energy of WZ InAs nanowires grown along  [1010] or [11$\overline{2}$0] retain their
 meV spin-orbit energies, not being influenced much by the field. We conjecture that
 this is true even in the presence of gating interfaces, meaning that the bulk effect
 dominates over the interfaces and electric fields which further reduce space inversion
 symmetry. The robustness and large value of the spin-orbit energy in these systems
 makes us suggest them as favorable systems.

Spin-orbit coupling in semiconductor nanowires has recently been investigated.
 Kammermeier {\it et al.}~\cite{Kammermeier2016} devised a theoretical framework
 to calculate the weak antilocalization effects in cylindrical nanowires and have successfully
 reproduced the values of $\alpha_{\textrm{R}} \approx $ 0.1 -- 0.3 meV$\cdot$nm
 for ZB InAs nanowires.~\cite{Hansen2005,Dhara2009,Roulleau2010,PhysRevB.82.235303,Liang2012}
 Winkler {\it et al.}~\cite{Winkler2017}, using multiband
 $\vec{k} \cdot \vec{p}\,$ for WZ InAs oriented along [0001] direction and
 tight-binding model for ZB InSb oriented along [111] direction, found an increase,
 of one order of magnitude or more, in the g-factor of excited conduction
 subbands due to the spin-orbit coupling. Luo {\it et al.}~\cite{Luo2017} found a
 giant Rashba effect of holes in semiconductor nanowires. Using an atomistic
 approach, they found that the hole Rashba coefficient of ZB InAs nanowires
 under an applied electric field of the order of 0.5 mV/nm is about 2 -- 5
 times larger than the electron Rashba coefficient. Consistent with our results,
 they also found a saturation of the electron Rashba coefficient with increasing
 nanowire diameter. Moreover, W{\'o}jcik {\it et al.}~\cite{2018arXiv180109905W}
 using a 2-band $\vec{k} \cdot \vec{p}\,$ model (by folding down the 8-band Kane model),
 in a self-consistent framework were able to accurately reproduce the
 results for ZB InSb nanowires from Ref.~\onlinecite{VanWeperen2015}. Unlike
 in our work, which provides setup-free spin-orbit parameters, these authors
 fixed the nanowire diameter and orientation and studied how SOC
 changes with distinct gate configurations and charge profiles. Nevertheless,
 in their calculations they found for a ZB InSb nanowire with $L \approx 100$ nm
 for a fixed chemical potential of 0.2 eV and an applied electric field of
 4 mV/nm, a Rashba coefficient of about $\alpha_{\textrm{R}} = 2\alpha \approx 50$ meV$\cdot$nm
 which is in good agreement with our results of $\alpha \approx 19$ meV$\cdot$nm.
 Furthermore, for WZ InAs nanowires oriented along [0001] direction with 100 nm
 in diameter, the authors in Ref.~\onlinecite{albrecht2016exponential} reported
 $\alpha_{\textrm{R}} = 8$ meV$\cdot$nm also in agreement with values
 reported in Ref.~\onlinecite{Das2012:NatPhys}, which is about 1.6 times larger
 than our reported value of $2\alpha \approx 5$ meV$\cdot$nm for a 4 mV/nm
 applied field, although in both reports it is not clear the value of the applied
 electric field. Moreover the authors in Ref.~\onlinecite{Zhang2015a}
 experimentally detected that for WZ CdSe nanowires th Dresselhaus SOC
 is absent for nanowires oriented along [0001] direction but present
 for [11$\overline{2}$0] direction, which is consistent with our results.

This paper is organized as follows:  In section \ref{sec:nwmodel} we present
 the geometric schematics of the nanowires we have simulated specifying the
 orientations and coordinate axes. Following, we introduce the respective
 $\vec{k} \cdot \vec{p}\,$ models we used as well as the numeric procedure
 employed to calculate the energies and states of the nanowires. After that,
 in section \ref{sec:modelHam}, we present the model Hamiltonian including
 SOC and its energy dispersion. With the expression from the energy dispersion
 we then apply a fitting procedure to the lowest conduction subband of the
 nanowires. In section \ref{sec:ZB} we discuss the specifics of the effective
 masses and SOC in ZB type structures following, in subsections
 \ref{subsec:ZB001}-\ref{subsec:ZB111}, with a detailed examination of the
 SOC in the distinct nanowire orientations. In section \ref{sec:WZ}
 and subsections \ref{susec:WZ0001} and \ref{susec:WZ1010} we do the same
 but for WZ crystal phase. Next, in section \ref{sec:Discussion} we discuss
 the essential SOC effects in nanowires from the perspective of finding
 a topological quantum phase transition in superconducting nanowires.
 Finally, we conclude in section \ref{sec:conc}.

%===============================================================================

%-------------------------------------
\begin{figure}[ht]
\begin{center}
\includegraphics[width=.3\textwidth]{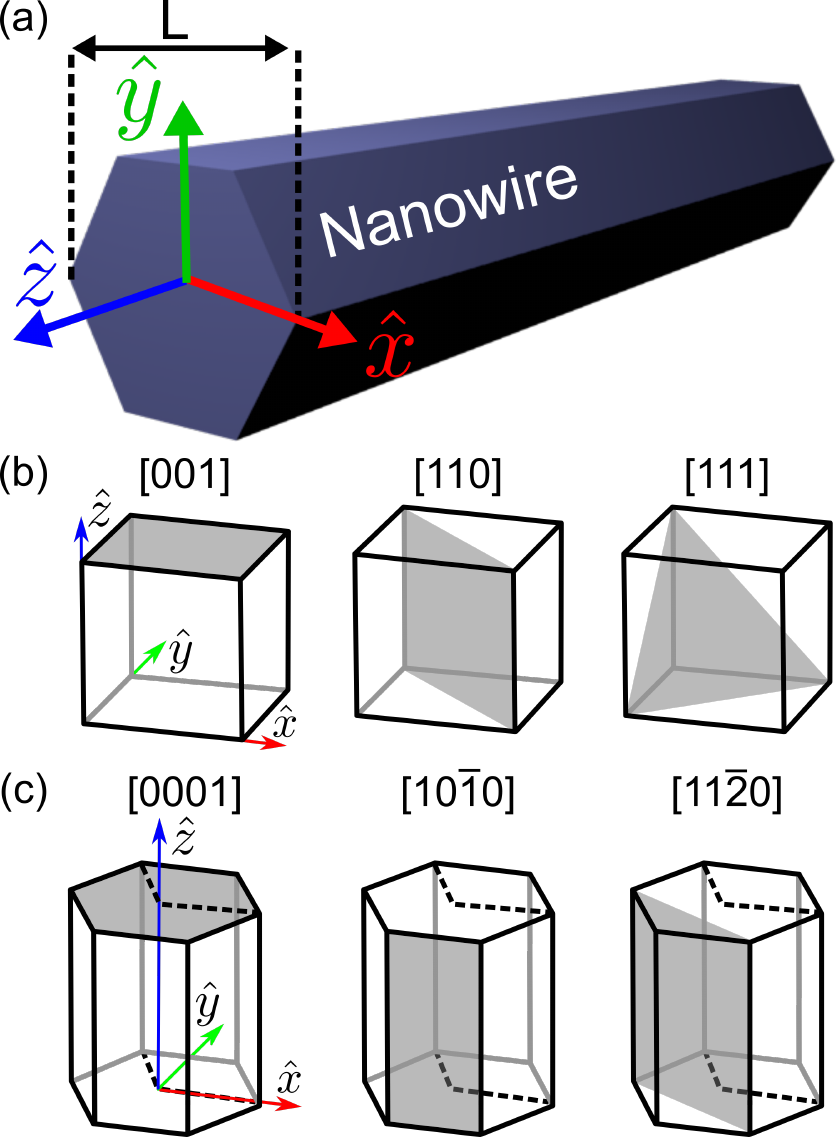}
\caption{(a) Schematics of a hexagonal nanowire and the coordinate axes we use
          in the text. The nanowire growth direction is along $z$, and the
          electric field is applied along the $y$-direction. The wire diameter
          is $L$, which is the distance between the opposite vertices of
          the hexagon.  In (b) and (c) we show the growth planes (shaded regions)
          inside the conventional unit cells for ZB  and WZ crystals, respectively.
          In (c) the dashed lines represent  the primitive unit cell.
          The coordinate axes with respect to the crystal orientations are also
          indicated for [001] ZB and [0001] WZ structures. These are the
          same axes as used in the confinement geometry (a).  }
\label{fig:NWscheme}
\end{center}
\end{figure}
%-----------------------------------------

\section{Nanowire modeling}
\label{sec:nwmodel}

By carefully controlling growth conditions,~\cite{Dick2010,krogstrup2010,Hjort2014}
 semiconductor nanowires using III-V compounds can be synthesized with pure
 ZB or WZ crystal phases,~\cite{Dick2010} but also with a
 mixed phase.~\cite{Panse2011} Furthermore, it is possible to obtain nanowires
 with a variety of cross-sections, such as hexagonal, circular, square and etc,
 grown along different directions. Typically, ZB nanowires grow along [111]
 directions with hexagonal cross-section,~\cite{Wang20081,Caroff2009:Nano,0268-1242-25-2-024005}
 while [001] nanowires have typically square cross-section.~\cite{Krishnamachari2004,Wang20081,0268-1242-25-2-024005,Zhang2015}
 As for [110] oriented ZB nanowires there can be several cross-section
 configurations from trapezoidal to diamond-like shapes and non-regular
 hexagons.~\cite{Wang20081,0268-1242-25-2-024005,Xu2012,Zhang2014,Xin2015,Hallberg2016}
 WZ nanowires can be typically fabricated along [0001] with hexagonal cross-section,
 while WZ nanowires grown along $[11\overline{2}0]$, and $[10\overline{1}0]$
 present square cross-section.~\cite{0268-1242-25-2-024005,Hjort2014,krogstrup2015}
 On the other hand, from a theoretical perspective nanowires are usually
 treated with cylindrical or square cross-section which simplifies the calculations
 without sacrificing the underlying physical features of the system.~\cite{Persson2006,Peeters2008,Kammermeier2016,2018arXiv180102616A}

In this paper we consider ZB InSb and WZ InAs nanowires with {\it hexagonal}
 cross-section,~\cite{noteCS} oriented along different (growth) directions, as shown
 in Fig. \ref{fig:NWscheme}. We calculated ZB nanowires oriented along [001],
 [110], and [111] directions and WZ nanowires oriented along [0001],
 $[11\overline{2}0]$, and $[10\overline{1}0]$ directions. The diameter of the nanowires
 is denoted as $L$, defined as the largest distance between vertices, see
 Fig. \ref{fig:NWscheme}(a). Our cartesian system has its $z$-axis along the
 growth direction, while the quantum confinement is in the $xy$-plane. For
 a ZB nanowire grown along [001], this would mean that $x$ is [100],
 $y$ is [010], and $z$ is [001]. For a WZ nanowire grown along [0001],
 the hexagonal atomic arrangement is compatible with the hexagonal confinement
 of the nanowire; see Fig. \ref{fig:NWscheme} (c). For all other orientations
 the relation between the cartesian coordinates and the crystallographic
 orientations is discussed in the corresponding sections.

Bulk ZB crystals are invariant under $T_{d}$, and WZ crystals
 under $C_{6v}$ symmetry operations.  Because space inversion symmetry is
 broken in these two crystal phases, the energy bands of both ZB and WZ nanowires exhibit
 generic spin-split due to the (bulk-inversion) asymmetry. To simulate
 realistic experimental conditions we also apply an external electric field
 in the cross-sectional plane of the nanowires. Our intention is to give
 {\it benchmark} results, instead of very specific experimental conditions
 with metal gates and electrodes attached to nanowires, as well as heterostructure
 charging effects, which would require self-consistent treatment. We wish to
 rather provide estimates of how large spin-splitting one can expect if a given
 electric field, from whatever environment, acts on the confined electron
 gas in the nanowire.

The electric field introduces additional spin-orbit splitting, which in
 general interferes with BIA SOC. The resulting spin-orbit splitting can be
 said to be due to structure-inversion asymmetry (the Rashba effect), although
 this terminology is not unique, and we simply refer to the spin-orbit splitting
 without any labels, but stating the material, confinement geometry, nanowire
 orientation, and the electric field. Electric field also reduces the mirror
 symmetry, resulting in further orbital splitting of the conduction band
 subbands, as show in Figs. \ref{fig:bsInSb001Y}, \ref{fig:bsInSb110Y}, \ref{fig:bsInSb111X},
 \ref{fig:bsInAs0001Y} and \ref{fig:bsInAs1010Y}.

In order to calculate the electronic structure of ZB InSb and
 WZ InAs semiconductor nanowires and extract the spin-orbit splitting
 of the lowest conduction subbands, we employ the multiband $\vec{k} \cdot \vec{p}\,$
 method combined with the envelope function approximation. Our $\vec{k} \cdot \vec{p}\,$
 Hamiltonians describe realistically the bulk cases as the reference points.

For ZB InSb we use a 14-band extended Kane model,~\cite{Cardona1988,Pfeffer1990,winkler2003spin,Fabian2007}
 which provides the relevant SOC features in the lowest
 conduction band via coupling to higher conduction bands.~\cite{note14kp} The involved
 bands are: the lowest $s$ conduction and the lowest three $p$ conduction
 bands, as well as the three highest $p$ valence bands (heavy and light,
 and spin-orbit split-off bands). Including the spin degree of freedom
 to these bands, we end up with a total of 14 bands. We use the parameters~\cite{deltaminus}
 for InSb from Ref.~\onlinecite{winkler2003spin}. If we took an 8-band
 model~\cite{Luttinger1955,Kane1966}, with only the lowest $s$ conduction
 subband, we would fail to describe properly the bulk spin-orbit splitting
 because this model lacks the couplings that generates the bulk spin-splitting observed
 in ZB III-V semiconductors.

%-------------------------------------
\begin{figure}[ht]
\begin{center}
\includegraphics[width=.45\textwidth]{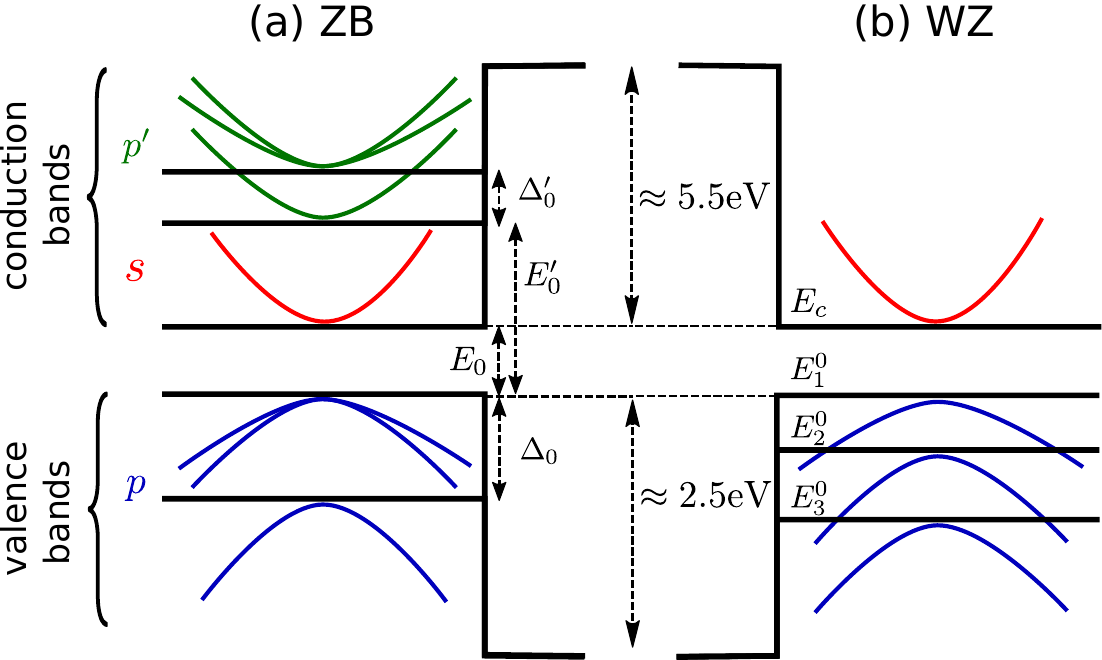}
\caption{ Schematics of the band alignment. (a) ZB 14-band model with band
          alignment to vacuum. The band gap energy is $E_{0}$ for the $s$
          conduction band and $E^{\prime}_{0}$ for the $p^{\prime}$ conduction band.
          In the valence band, $\Delta_{0}$, is the the spin-orbit coupling
          splitting, while in the $p^{\prime}$ conduction band it is described by the
          $\Delta^{\prime}_{0}$ parameter. (b) WZ 8-band model with band alignment
          to vacuum. The $p$ valence bands have an extra splitting (due to the crystal field)
          at $\Gamma$-point, making them only spin (and no longer orbital) degenerate.
          The energy labels for WZ are also indicated: $E_c$ is the conduction band
          minimum, and $E_i^0$, $i$=1, 2, and 3, are the valence band maxima. The
          conduction bands offset to the vacuum is 5.5 eV, and the valence bands offset
          is 2.5 eV following Ref~\onlinecite{Xia1993}. }
\label{fig:bandmismatch}
\end{center}
\end{figure}
%-------------------------------------

For our WZ InAs nanowires we use the 8-band $\vec{k} \cdot \vec{p}\,$
 model with $k$-dependent SOC terms,~\cite{FariaJunior2016} that reproduces
 very well the bulk SOC features in the vicinity of Gamma point. Such $k$-dependent terms are usually
 neglected in conventional 8-band WZ Hamiltonians,~\cite{voon2009k,Miao2012,FariaJunior2014}
 but they are needed to accurately describe bulk spin-orbit effects in
 WZ InAs \cite{FariaJunior2016}.

To model quantum confinement we use the envelope function approximation.~\cite{Bastard1981,Baraff1991,Burt1992,bastard}
 Essentially, this treatment applied to the multiband $\vec{k} \cdot \vec{p}\,$
 Hamiltonians means substituting bulk wave numbers $k_{x(y)}$ by operators
 $-i\,\partial / \partial x(y)$, keeping $k_z$ a parameter, thus transforming
 the bulk multiband Hamiltonian into a set of coupled linear differential
 equations. To solve these coupled differential equations we employ the
 plane wave expansion,~\cite{Rodrigues2000,Mei2007,vukmirovc2008plane,FariaJunior2014,ehrhardt2014multi,Budagosky2017}
 that is, the Fourier transform of the real-space dependent parameters.
 In narrow gap semiconductors, real-space treatment of confined systems
 can lead to spurious solutions and special treatment~\cite{Jiang2014,Ma2014,Luo2016}
 should be applied to eliminate them, while using Fourier transformation
 the spurious solutions are easily identifiable and controllable~\cite{vukmirovc2008plane}.
 The plane wave expansion works by creating, effectively, periodically
 repeated systems of nanowires with vacuum in-between. To treat the vacuum
 we follow the suggested values, in Ref~\onlinecite{Xia1993}, of the band
 offset for the conduction band as $5.5$ eV, and for the valence band
 as $2.5$ eV. In Fig. \ref{fig:bandmismatch} we show a scheme of the used
 band alignment of the semiconductor with the vacuum. In Appendix A
 we discuss the plane wave expansion approach and its numerical
 implementation.

%-------------------------------------
\begin{figure}[ht]
\begin{center}
\includegraphics[width=.45\textwidth]{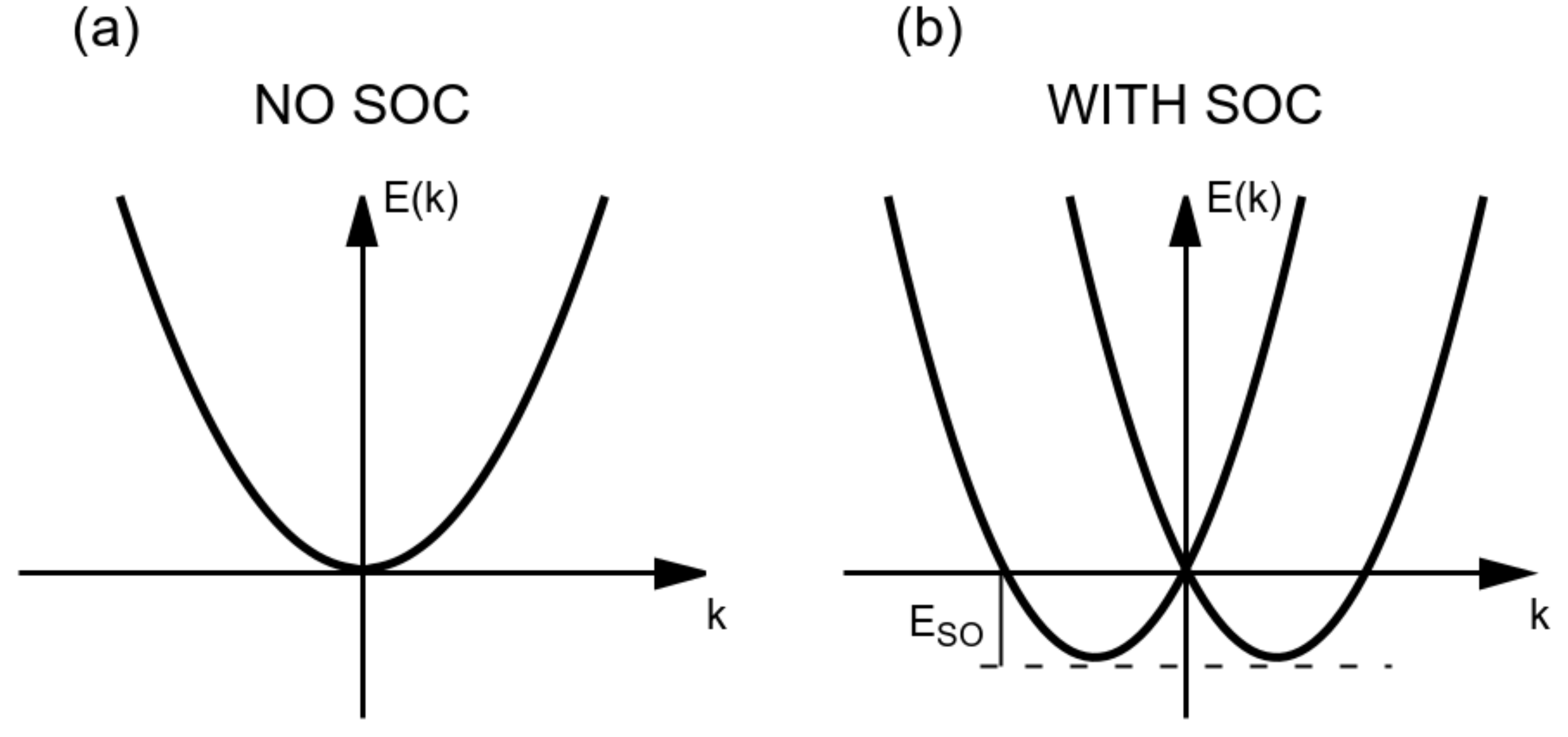}
\caption{Schematical description of  SOC effects at the $\Gamma$ point.
(a) Without SOC the conduction
         bands in the bulk and also in nanowires are parabolic, described by
         the effective mass $m^*$. (b) In the presence of SOC, the spin
         degeneracy of the bands is lifted (due to the lack of inversion
         symmetry, either atomic structure or confinement) and the band
         structure comprises two shifted parabolas, indicating the presence
         of a $k$-linear spin-orbit field. The new minima of the parabolas
         are at energy $E_{SO}$, which is a measure of the strength of SOC.  }
\label{fig:PedagSOC}
\end{center}
\end{figure}
%-------------------------------------

In addition to confinement, we also apply electric field across the nanowires,
 along $x$ and $y$ directions, see Fig. \ref{fig:NWscheme}(a) for the choice of
 coodinate system. For example, if the electric field is along $y$, the voltage drop along
 the nanowire is
 \begin{equation}
  V_{\textrm{ele}} =  e\,E\,y
  \label{eq:PotEle}
 \end{equation}
\\ where $e$ is the modulus of the electron charge and $E$ is the applied
 electric field. The values of $y$ range from $0$ to $L$, thus thicker
 nanowires have larger values of $V_{\textrm{ele}}$ for a fixed value of $E$.
 For sufficiently large values of $V_{\textrm{ele}}$, the confinement profiles
 of conduction and valence energy bands overlap, therefore closing the gap
 of the system. In Appendix B we show an example of such case. Although
 interesting physical phenomena can be found in gapless systems,~\cite{PhysRevB.85.195114,ADFM:ADFM201505357}
 in this study we focus on values of electric fields that do not overlap
 conduction and valence energy band profiles, i. e., we are considering
 gapped systems.

\section{Model Hamiltonians}
\label{sec:modelHam}

We fit our numerical data to effective Hamiltonians in order to extract
 useful parameters such as effective masses and SOCs. In
 general, the conduction bands of our nanowires follow the $2\times2$
 Hamiltonian
 \begin{equation}
  \label{eq:2by2Ham}
  H=H_{0}+H_{\textrm{BIA}}+H_{\textrm{SIA}}.
 \end{equation}
Here, $H_{0}$ is the effective mass Hamiltonian, expressing the parabolic
 dispersion near $\Gamma$-point. The remaining two terms express the bulk-inversion
 (BIA) and structure-inversion (SIA) asymmetry induced SOCs.
 For conduction electrons, which form orbitally non-degenerate bands, these
 spin-orbit Hamiltonians are conventionally written as
 \begin{equation}
  \label{eq:generalSOCHam}
  H^{\prime}= \vec{\Omega}(\vec{k})\cdot\vec{\sigma},
  %H^{\prime}= \frac{\hbar}{2}\vec{\Omega}(\vec{k})\cdot\vec{\sigma},
 \end{equation}
 \\ where $\Omega_{\bf k}$ is the spin-orbit field. Time reversal symmetry requires
 it to be an odd function of momentum,
 $\vec{\Omega}(\vec{k})=-\vec{\Omega}(-\vec{k})$. Otherwise the functional
 form of the spin-orbit field is restricted by the crystal and confinement
 symmetry.

Each structure has its own functional form of the effective mass and spin-orbit
 field, based on the symmetry. In the following sections we discuss the specific
 forms and present effective masses, as well as spin-orbit field parameters up
 to cubic-in momentum terms
 \begin{equation} \label{eq:DE}
  \Delta E = 2\left(\alpha k + \gamma k^3\right).
 \end{equation}
Apart from $\alpha$ and $\gamma$, an important measure of the strength of
 SOC is the spin-orbit energy,
 \begin{equation}
  E_{\textrm{SO}} = \frac{\left(2\,\alpha\right)^2 m^*}{2 \hbar^2},
 \end{equation}
\\ where $m^*$ is the effective mass of the conduction band. The spin-orbit
 energy is indicated in Fig. \ref{fig:PedagSOC}.

\begin{figure}[ht]
\begin{center}
\includegraphics[width=.5\textwidth]{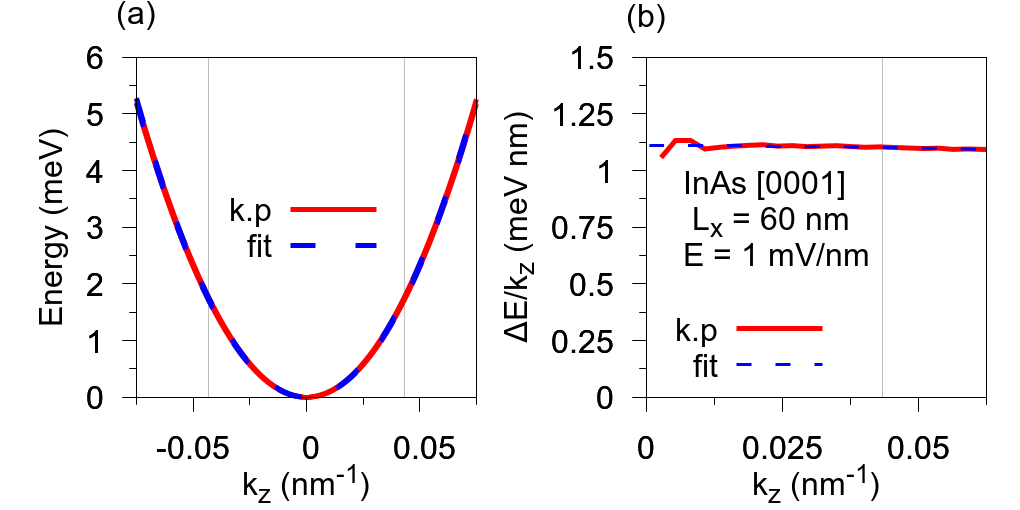}
\caption{Fitting procedure example. (a) Cubic fitting to the $\vec{k} \cdot \vec{p}\,$
         calculated lowest conduction subband structure around $\Gamma$-point
         of InAs WZ nanowire oriented along [0001]. The wire diameter is 60 nm,
         and the applied electric field 1 mV/nm. (b) Spin-orbit splitting
         of the subband in (a) divided by momentum. This line is fit to
         obtain the linear spin-orbit coefficient $\alpha$ as well as the
         cubic spin-orbit coefficient $\gamma$. The thin vertical lines
         correspond to the fitting range which was taken as $\approx 1\%$
         of the Brillouin zone.}
\label{fig:fittEx}
\end{center}
\end{figure}

Our goal is to provide a reliable fitting of these effective models to the
 numerical calculations using the multiband $\vec{k} \cdot \vec{p}\,$ Hamiltonians.
 The fitting procedure is illustrated in Fig. \ref{fig:fittEx}. We fit the
 lowest conduction band, as calculated with the $\vec{k} \cdot \vec{p}\,$
 method, using a cubic fitting, i. e., up to third order in the momentum,
 see Eq. \ref{eq:DE}. The agreement is in general excellent. From that
 fitting we obtain the effective mass of the lowest conduction band. The
 subband's spin-splitting, induced either by the structure itself, or by
 the applied electric field, is then divided by momentum, providing a
 nice way of obtaining the linear spin-orbit splitting parameter $\alpha$
 as the intersection with the vertical axis. A quadratic fitting to this
 curve determines the cubic coefficient $\gamma$. In Appendix C we discuss
 the effects of higher conduction subbands.

%-------------------------------------
%-------------------------------------

%==========================ZINCBLENDE InSb===============

\section{Zinc-blende I\MakeLowercase{n}S\MakeLowercase{b} nanowires}
\label{sec:ZB}

The cubic structure of ZB semiconductors allows to approximate
 the lowest conduction band (also in confined structures) by the parabolic
 dependence near $\Gamma$-point
\begin{equation}
H_{0} = \frac{\hbar^{2}}{2m^*}
\left( k^{2}_{x} + k^{2}_{y} + k^{2}_{z}\right),
\end{equation}
 where $m^*$ is the conduction electron effective mass. For bulk ZB InSb
 $m^{*} \approx 0.015\,m_{0}$~\cite{winkler2003spin}, where $m_{0}$ is the
 free electron mass. However the effective mass value changes with the quantum
 confinement as we will discuss latter in the paper.

The spin-orbit splitting of the conduction bands in ZB InSb nanowires
 can be qualitatively discussed using the expression obtained by
 Dresselhaus for spin-orbit coupling in ZB III-V semiconductors. For bulk ZB III-V
 semiconductors Dresselhaus found that~\cite{Dresselhaus1955}
\begin{equation}
\vec{\Omega}_{\rm BIA}  =  \gamma^{ZB}\left[k_{x}\left(k_{y}^{2}-%
k_{z}^{2}\right),k_{y}\left(k_{z}^{2}-k_{x}^{2}\right),k_{z}
\left(k_{x}^{2}-k_{y}^{2}\right)\right].
\label{eq:Dresselhaus_bulk}
\end{equation}

\begin{figure}[ht]
%\begin{center}
\includegraphics[width=0.9\linewidth]{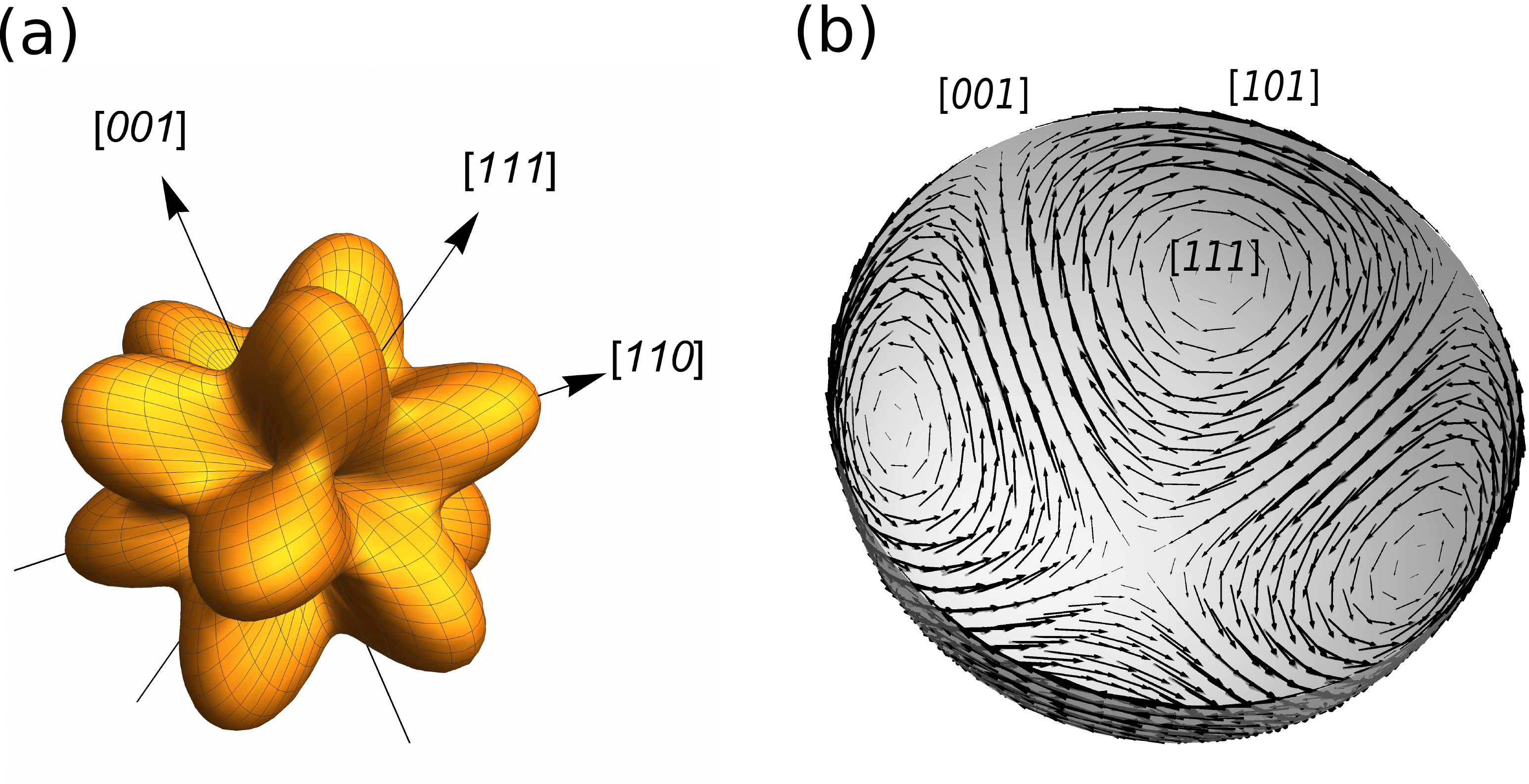}
\caption{Topology of the Dresselhaus spin-orbit coupling field. (a) Spherical
         plot of the magnitude of the Dresselhaus spin-orbit field in the
         momentum space. The crystallographic axes are indicated. The field
         vanishes for [001] and [111] directions (and their equivalents).
         The field has a maximum along [110] directions. (b) Dresselhaus
         vector field over a Fermi sphere. The vortices are along [111],
         and saddle points along [001] directions, indicating the spin-orbit
         fields in the quantum wells grown along these orientations.  }
\label{fig:ZBField}
%\end{center}
\end{figure}

The spin-orbit splitting of the conduction band is increasing as a cubic power
 of the momentum, away from the $\Gamma$-point. There is no linear-in momentum
 splitting in the bulk. A spherical plot of $\vec{\Omega}_{\rm BIA}$ is shown in
 Fig. \ref{fig:ZBField}(a). The field vanishes for [001] and [111] directions, as is clear
 from Eq. \ref{eq:Dresselhaus_bulk}. Maximal spin-orbit splittings are along
 the site diagonals, [110]. A recent DFT calculation~\cite{Gmitra2016} found
 for ZB InSb, that the bulk cubic coefficient is $\gamma^{ZB}_{\rm InSb} \approx 200$ meV$\cdot\,\rm nm ^3$.

%-------------------------------------
%-------------------------------------

We also project the vector field $\vec{\Omega}_{\rm BIA}$ on a
 Fermi sphere, in Fig. \ref{fig:ZBField}(b). The field has saddle points along [001],
 which is the familiar vector pattern for the Dresselhaus field in [001] grown
 quantum wells. Along body diagonals, [111], the field has vortices, resembling the
 Rashba texture. Finally, along [011] the field does not vanish, but has
 a strong component perpendicular to the momentum. A simple
 counting of the winding numbers for the field indexes (points where the
 field vanishes)---6 saddle points of winding number -1 each, and
 8 vortices of winding number +1 each---gives the total winding number of 2, which
 is the Euler characteristic of a sphere, in line with the Poincaré-Hopf
 theorem.~\cite{guillemin1974differential}

In the following subsections we discuss separately the spin physics of
 hexagonal nanowires along the three growth directions, [001], [110], and
 [111]. It is worth mention that our band structure is in agreement
 with previously published results.~\cite{Persson2006,Peeters2008,Liao2015,Luo2016}

%-------------------------------------InSb 001-----------------------------------------------------

\subsection {[001] growth direction}
\label{subsec:ZB001}

\begin{figure}[ht]
%\begin{center}
\includegraphics[width=1\linewidth]{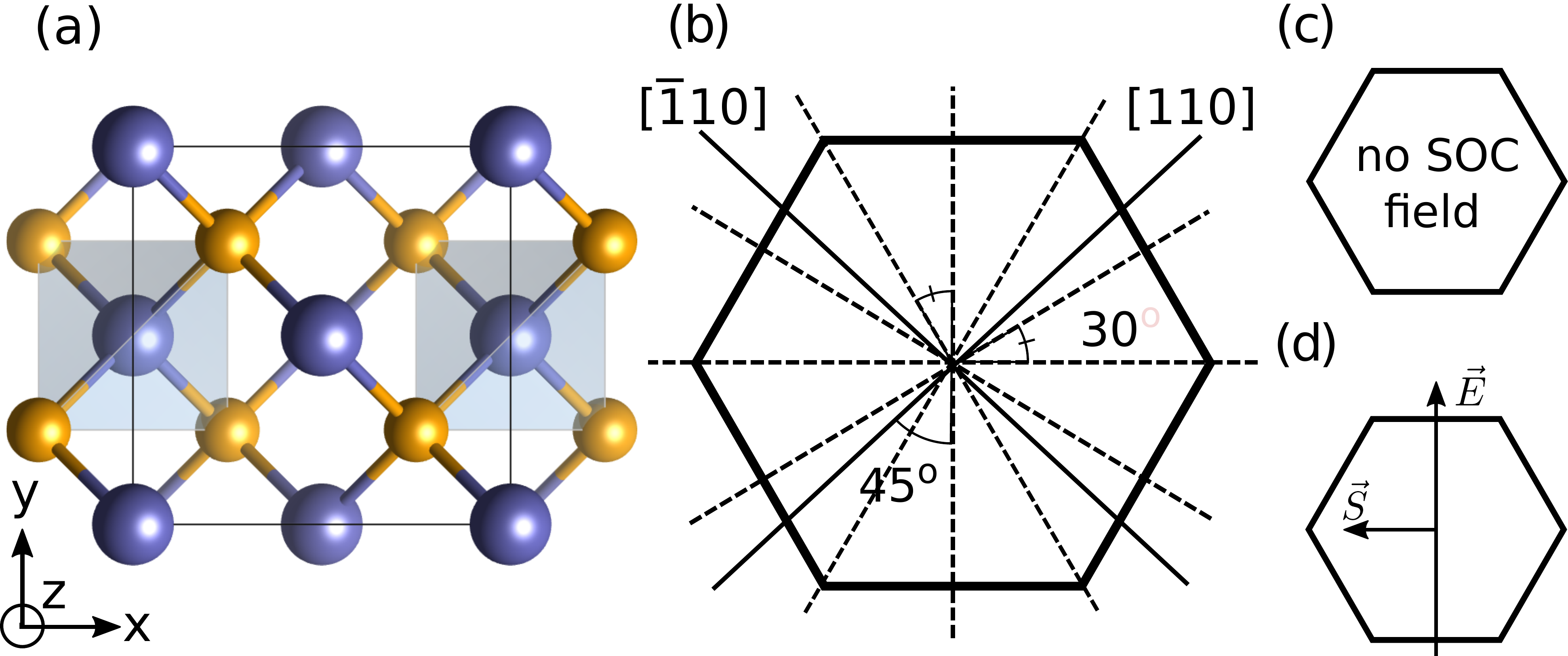}
\caption{Symmetry analysis of [001] oriented ZB nanowire. (a) Atomic arrangement
         along [001] orientation of a ZB structure with indicated $x$ and $y$
         axes. (b) Mirror symmetry planes of the atomic structure (solid)
         and of the hexagonal confinement (dashed). (c) and (d) are the spin
         projections without and with the applied electric field, respectively.
         In (c), the resulting spin projection along the nanowire axis is
         due to the mismatch between the atomic arrangement and the quantum
         confinement, and is absent if the growth direction would lie in a
         mirror symmetry plane (which is the case for square and circular
         nanowires). (a) was produced using the VESTA software.~\cite{Momma:db5098} }
\label{fig:ZB001}
%\end{center}
\end{figure}

The cross-section of the atomic structure of a ZB semiconductor
 along [001] direction is shown in Fig. \ref{fig:ZB001}(a). The principal
 symmetry axes are along [110] and $[1\overline{1}0]$, which are also the
 normal vectors of the mirror symmetry planes. This symmetry is not compatible
 with the chosen hexagonal confinement, resulting in the absence of a mirror
 symmetry plane in the nanowire structure, see  Fig. \ref{fig:ZB001}(b). For
 this growth direction $x = [100]$, $y= [010]$, and $z = [001]$.

%-------------------------------------
%-------------------------------------

In the bulk, the spin-orbit field vanishes for momenta along [001], as is
 clear from the Dresselhaus expression, Eq. \ref{eq:Dresselhaus_bulk}. However,
 the disorientation of the hexagonal confinement in our nanowires leads to
 a finite, linear in momentum spin-orbit field. This field (that is, the
 spin quantization axis) points in the [001] direction, so the spin-orbit
 splitting is proportional to $k_z \sigma_z$. Such a term would not be allowed
 if the growth direction would lie in a mirror symmetry plane (which is the
 case for square and circular nanowires).~\cite{Luo2011} The orientation
 of the spin caused by SOC in ZB nanowires, without applied electric field,
 is shown in Fig. \ref{fig:ZB001}(c). By applying an electric field, the
 simple SIA model tells us that the spin is oriented perpendicular to both
 the direction of the field and the electron's velocity, which in the case
 of electric field along $y$ direction makes the spins oriented along $x$,
 see Fig. \ref{fig:ZB001}(d).

\begin{figure}[ht]
\begin{center}
\includegraphics[width=.4\textwidth]{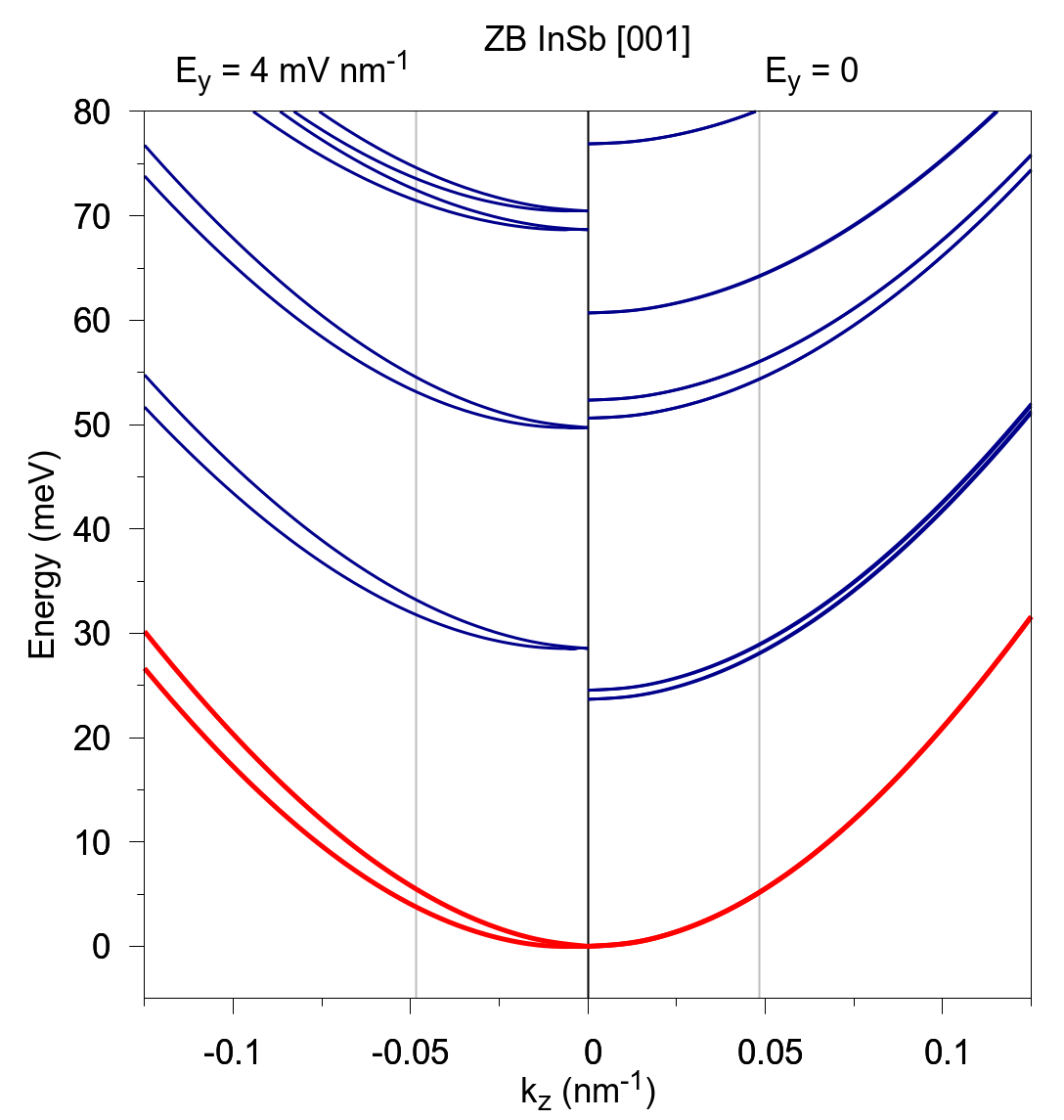}
\caption{Calculated electronic subband dispersion for a $L = 60$ nm ZB InSb
         hexagonal nanowire oriented along [001] direction. The leftmost subband dispersion
         (negative $k_{z}$)  corresponds to an electric field of $E_{y}=4 $mV/nm and
         the rightmost subband dispersion (positive $k_{z}$) to a zero applied electric
         field. The thin vertical lines correspond to the fitting range which was taken as
         $\approx 1\%$ of the Brillouin zone. }
\label{fig:bsInSb001Y}
\end{center}
\end{figure}

Figure \ref{fig:bsInSb001Y} shows the calculated electronic subband structure
 for a ZB InSb hexagonal nanowire along [001], calculated using the 14-band
 $\vec{k} \cdot \vec{p}\,$ method. The conduction subbands are shown in the
 absence and presence of a transverse electric field along the $y$ direction.
 In the absence of the electric field, the mismatch between the crystal structure
 and the confinement profile tells us that the lowest conduction subband should
 be spin-split. This is explained by directly quantizing the Dresselhaus
 field, Eq. \ref{eq:Dresselhaus_bulk}, in the $x$ and $y$ directions.
 Indeed, considering that the expectation values of operators $\hat{k}_x = -i \partial/\partial x$
 and $\hat{k}_y = -i \partial/\partial y$ vanish for the ground state $|0\rangle$,
 we obtain
\begin{equation}
 \vec{\Omega}_{\textrm{001}}  =  \gamma^{ZB}\left[0,0,\kappa^2 k_{z}\right],
\end{equation}
 where $\kappa^2$ is the expectation value of $\hat{k}_x^2 - \hat{k}_y^2$
 in the ground state: $\kappa^2 = \langle 0 |\hat{k}_x^2|0\rangle - \langle 0 |\hat{k}_y^2|0\rangle$.
 Because our nanowire does not have the $x \to y$ symmetry, $\kappa^2$ does not,
 in general, vanish, and the lowest conduction subband exhibits a weak spin-orbit
 field oriented along the growth direction $z$.

%-------------------------------------
%-------------------------------------

To get an estimate for the linear splitting, we approximate
\begin{equation}
\label{eq:approxBIA}
2\alpha = \gamma^{ZB} \kappa^2 \approx f_a \frac{\gamma^{ZB}}{L^2},
\end{equation}
 where $f_a$ is an anisotropy factor quantifying the difference between
 $x$ and $y$ directions. This factor should be on the order of 0.1 (this
 is a guess), which say, for $L = 60$ nm, we would then get $\alpha \approx 3$ $\mu$eV$\cdot$nm
 for InSb. This is indeed a tiny value and it is bellow our numerical
 precision as discussed in Appendix A. Therefore we regard it as zero.

\begin{figure}[ht]
\begin{center}
\includegraphics[width=.5\textwidth]{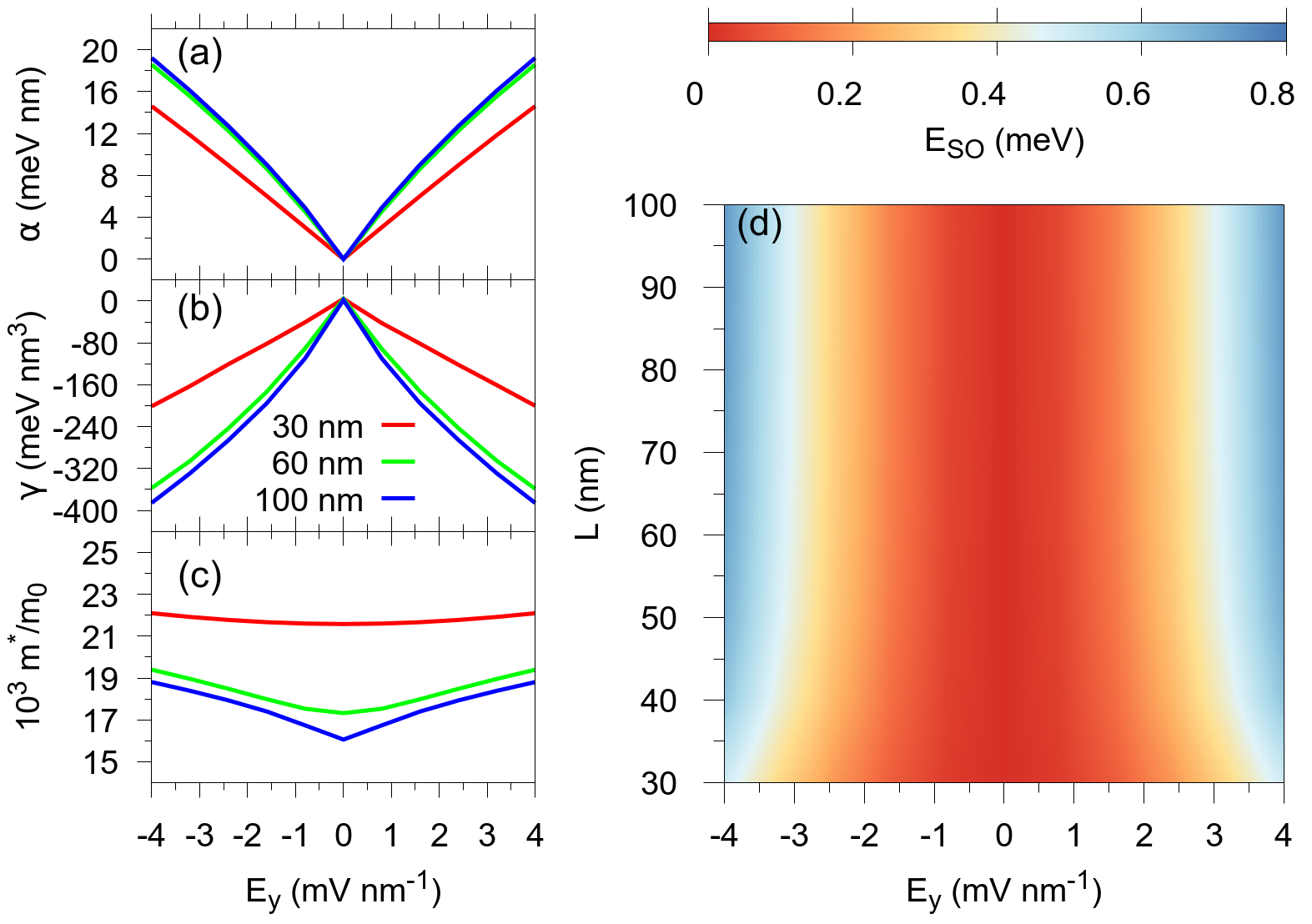}
\caption{Extracted (a) linear, $\alpha$,  and (b) cubic, $\gamma$, spin-orbit
         splitting coefficients, (c) effective masses and (d) spin-orbit
         coupling energy for different diameters $L$ as indicated, for InSb
         ZB nanowires oriented along [001].}
\label{fig:SOC_InSb_001_Y}
\end{center}
\end{figure}

However, the splitting is strongly enhanced in the presence of the electric
 field, whose effect is nicely visible already on the scale of Fig. \ref{fig:bsInSb001Y}.
 The extracted linear and cubic spin-orbit coefficients $\alpha$ and $\gamma$,
 as functions of $E_y$ are plotted in Fig. \ref{fig:SOC_InSb_001_Y}(a)-(b).
 The linear coefficient is typically 10 meV$\cdot$nm for electric fields
 of a few mV/nm. Cubic coefficients are about 400 meV$\cdot$nm$^3$.
 In Fig. \ref{fig:SOC_InSb_001_Y}(c) we see that the confinement influences
 the effective mass of the lowest conduction subbands. The effective mass
 for nanowires with $L \agt 50$ nm is already within 20\% of the bulk electron
 mass. For thinner nanowires (30 nm) the effective mass reaches values 0.02 $m$.

%-------------------------------------
%-------------------------------------

 Finally, in Fig. \ref{fig:SOC_InSb_001_Y}(d) we provide the full map of
 the extracted spin-orbit strength $E_{\rm SO}$ as a function of both the
 electric field $E_y$ and the diameter of the nanowire $L$. For a given
 electric field in the considered range there is not much variation of the
 spin-orbit strength with respect to the nanowire diameter. The electric field
 is the most critical control parameter to tune the spin-orbit splitting.
 The obtained spin-orbit energies for the ZB InSb nanowires
 are about $E_{\rm SO} = 0.8$ meV for fields of 4 meV/nm. The scaling
 with the electric field is quadratic, since $ E_{\rm SO} \sim \alpha^2$, and
 $\alpha$ grows linearly with increasing electric field.

%----------------------------------------InSb 110---------------------------------------------------------------

\subsection {[110] growth direction}
\label{subsec:ZB110}

\begin{figure}[ht]
%\begin{center}
\includegraphics[width=0.5\linewidth]{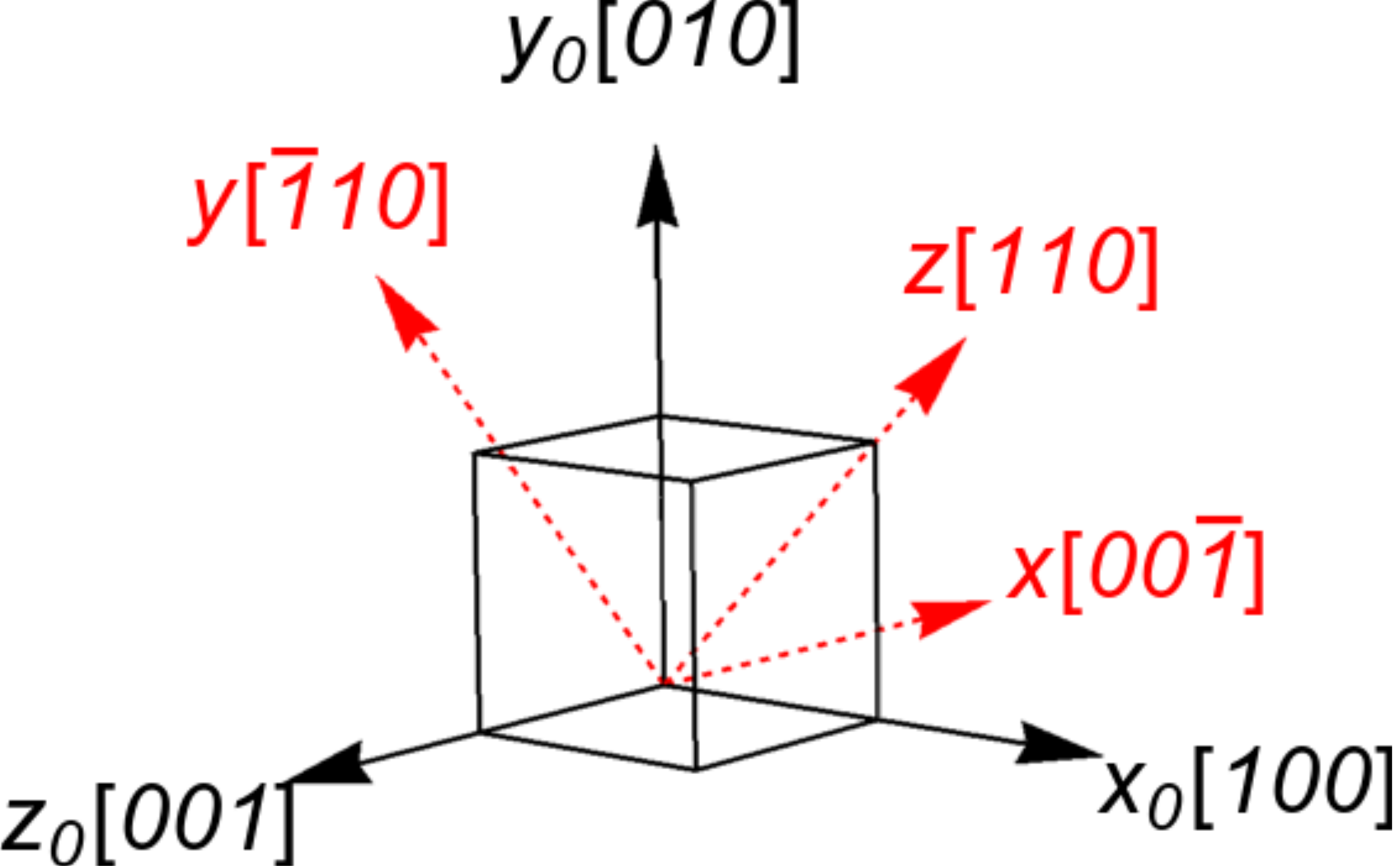}
\caption{ Scheme of the coordinate system with the growth direction along [001],
          and transverse plane spanned by indicated rotated $x$ and
          $y$ axes. }
\label{fig:rotated[011]}
%\end{center}
\end{figure}

\begin{figure}[ht]
%\begin{center}
\includegraphics[width=1\linewidth]{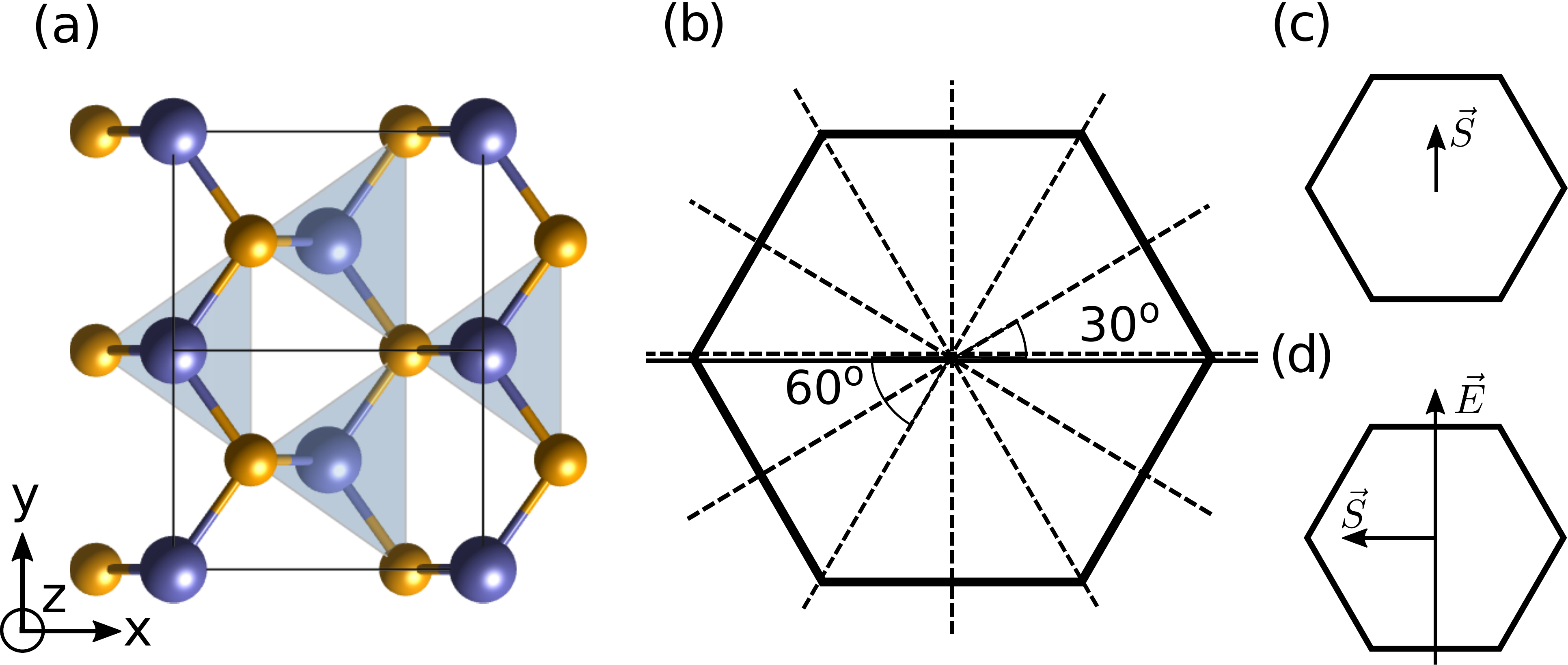}
\caption{Symmetry analysis of [110] oriented ZB nanowire.(a) Atomic arrangement
         along [110] orientation of a ZB structure with indicated $x$ and
         $y$ axes. (b) Mirror symmetry planes of the atomic structure (solid)
         and of the hexagonal confinement (dashed). (c) and (d) are the spin
         projections without and with the applied electric field, respectively.
         (a) was produced using the VESTA software.~\cite{Momma:db5098} }
\label{fig:ZB110}
%\end{center}
\end{figure}

We now rotate the coordinate system such that the nanowire axis is along
 $z =  [110]$. The new cartesian system is shown in Fig. \ref{fig:rotated[011]}:
 axis $x = [00\overline{1}]$ and $y = [\overline{1}10]$. The cross-section
 of the atomic structure of a ZB semiconductor along [110] direction
 is shown in Fig. \ref{fig:ZB110}(a). The hexagonal confinement reduces
 the structural symmetry, retaining only one mirror plane, spanned by $y$
 and $z$ (making the system symmetric as $y \to - y$). The compatibility
 of the atomic structure along [110] and of the confinement is shown in Fig. \ref{fig:ZB110}(b).

%-------------------------------------
%-------------------------------------

The Dresselhaus spin-orbit field, Eq. \ref{eq:Dresselhaus_bulk}, for ZB
 structures with rotated coordinates as shown in Fig. \ref{fig:rotated[011]},
 transforms according to the functional form,~\cite{Kammermeier2016}

\begin{eqnarray}
\label{eq:Dress110}
\vec{\Omega}_{110}^{\textrm{ZB}} & \!=\! & \!\frac{\gamma^{\textrm{ZB}}}{2}\!\left[-4k_{x}k_{y},2k_{x}^{2}-k_{z}^{2}+k_{y}^{2},2k_{x}^{2}+k_{z}^{2}-k_{y}^{2}\right]\!\!\left[\begin{array}{c}
k_{z}\\
k_{z}\\
k_{y}
\end{array}\right].\nonumber \\
 &  & \phantom{}
\end{eqnarray}

The coordinates of momenta $k_x$, $k_y$, and $k_z$, are with respect to
 the rotated axes with unit vector $\hat{k}_z$ pointing along [011], $k_x$
 along [00$\overline{1}$], and $k_y$ along [$\overline{1}$10].

%-------------------------------------
%-------------------------------------

When we quantize the spin-orbit field along $x$ and $y$, we get linear
 spin-orbit splitting for the free motion along $z$ proportional to
 $k_{z} \sigma_{y}$. The orientation of the spin caused by SOC
 in ZB nanowires without applied electric field is along $y$, as shown
 in Fig. \ref{fig:ZB110}(c).  In an electric field along $y$, the spin orientation
 points along $y$, see Fig. \ref{fig:ZB110}(d).

\begin{figure}[ht]
\begin{center}
\includegraphics[width=.4\textwidth]{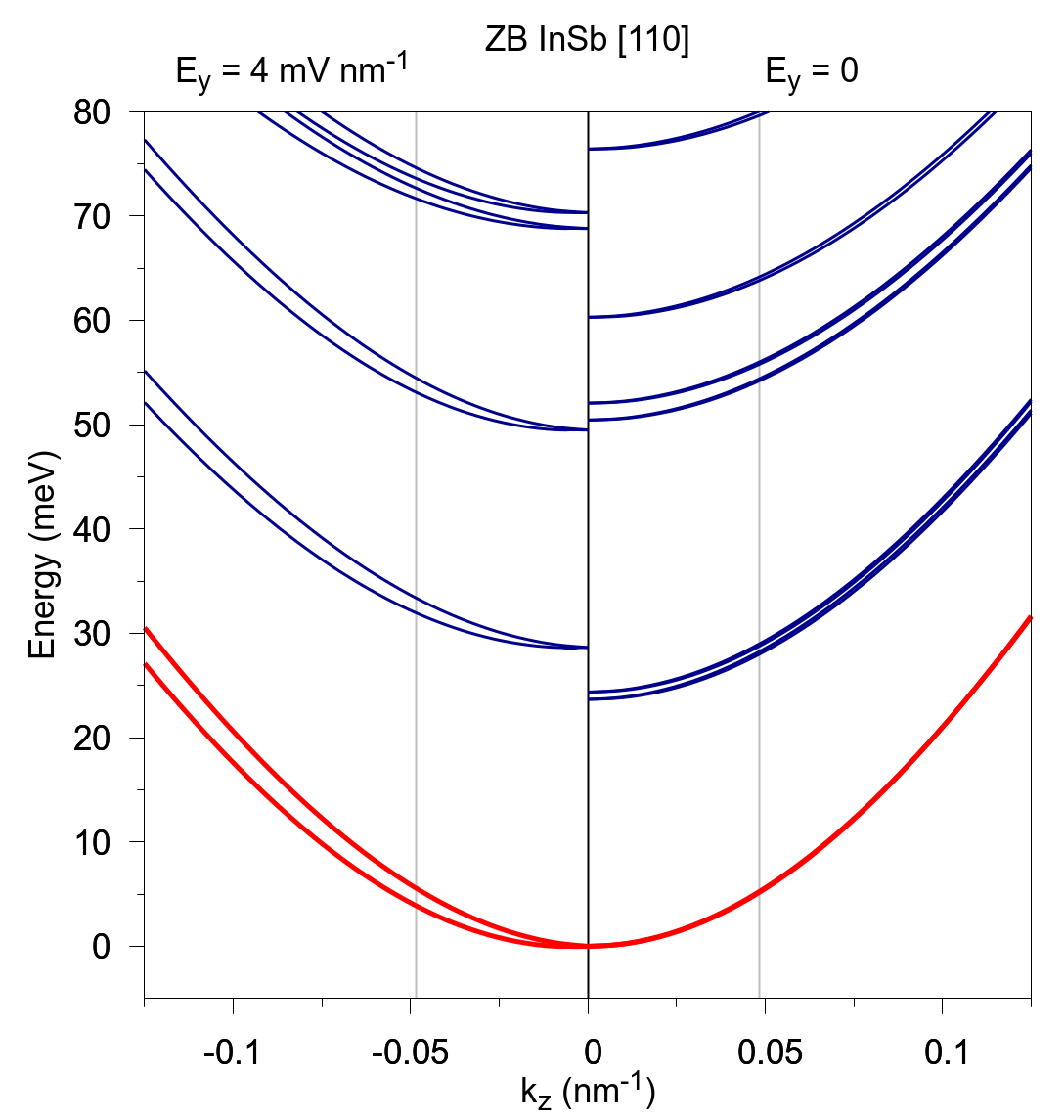}
\caption{Calculated electronic subband dispersion for a $L = 60$ nm ZB InSb
         hexagonal nanowire oriented along [110] direction. The leftmost subband dispersion
         (negative $k_{z}$)  corresponds to an electric field of $E_{y}=4 $mV/nm and
         the rightmost subband dispersion (positive $k_{z}$) to a zero applied electric
         field. The thin vertical lines correspond to the fitting range which was taken as
         $\approx 1\%$ of the Brillouin zone.}
\label{fig:bsInSb110Y}
\end{center}
\end{figure}

Figure \ref{fig:bsInSb110Y} shows the calculated electronic subband structure
 for a ZB InSb hexagonal nanowire along [110]. The  conduction subbands are
 shown in the absence and presence of a transverse electric field along
 the $y$ direction. In the absence of the electric field the lowest conduction
 subband has a small spin-splitting due to the hexagonal confinement. This
 is explained by directly quantizing the Dresselhaus field, Eq. \ref{eq:Dress110},
 in the $x$ and $y$ directions:
\begin{equation}
 \vec{\Omega}_{\textrm{110}}  =  \gamma^{ZB}\left[0,\kappa^2 k_{z},0\right],
\end{equation}
 where $\kappa^2$ is the expectation value of $\hat{k}_x^2 + \frac{1}{2}\hat{k}_y^2 $
 in the ground state: $\kappa^2 = \langle 0 |\hat{k}_x^2|0\rangle + \frac{1}{2} \langle 0 |\hat{k}_y^2|0\rangle $.

%-------------------------------------
%-------------------------------------

\begin{figure}[ht]
\begin{center}
\includegraphics[width=.5\textwidth]{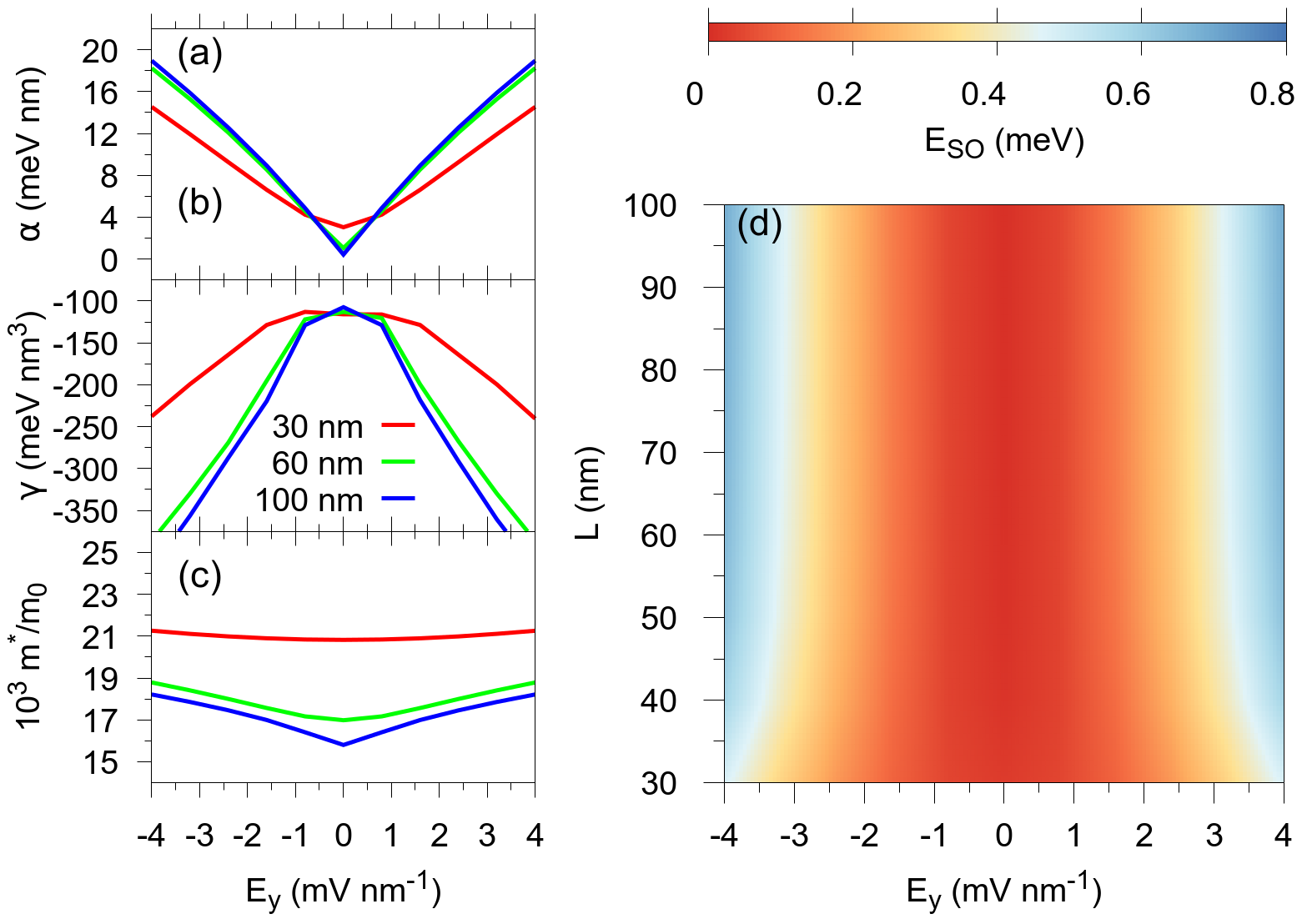}
\caption{Extracted (a) linear, $\alpha$,  and (b) cubic, $\gamma$, spin-orbit
         splitting coefficients, (c) effective masses and (d) spin-orbit
         coupling energy for different diameters $L$ as indicated, for InSb
         ZB nanowires oriented along [110].}
\label{fig:SOC_InSb_110_Y}
\end{center}
\end{figure}

As in the [001] case, the presence of the external electric field is the
 dominant factor in the spin-splitting also in nanowires along [110]. In
 fact, the linear and cubic spin-orbit parameters, effective masses, as
 well as the spin-orbit field, are in magnitude very similar to the [001]
 case, see Fig. \ref{fig:SOC_InSb_110_Y}, for the range of electric fields
 considered. However, due to non-vanishing Dresselhaus SOC for [110] direction,
 thinner nanowires have a non-zero spin-splitting with parameters $\alpha \approx 4$meV/nm
 and $\gamma \approx -100$meV/nm$^{3}$. The interplay between the Dresselhaus
 and Rashba SOC is additive for electric fields along $y$ direction while
 for electric field along $x$ direction the zero spin-splitting case is
 shifted to non-zero values of electric field.

%-------------------------------------
%-------------------------------------

%-------------------------------------------InSb 111-----------------------------------------------

\subsection {[111] growth direction}
\label{subsec:ZB111}

%-------------------------------------
\begin{figure}[ht]
%\begin{center}
\includegraphics[width=0.5\linewidth]{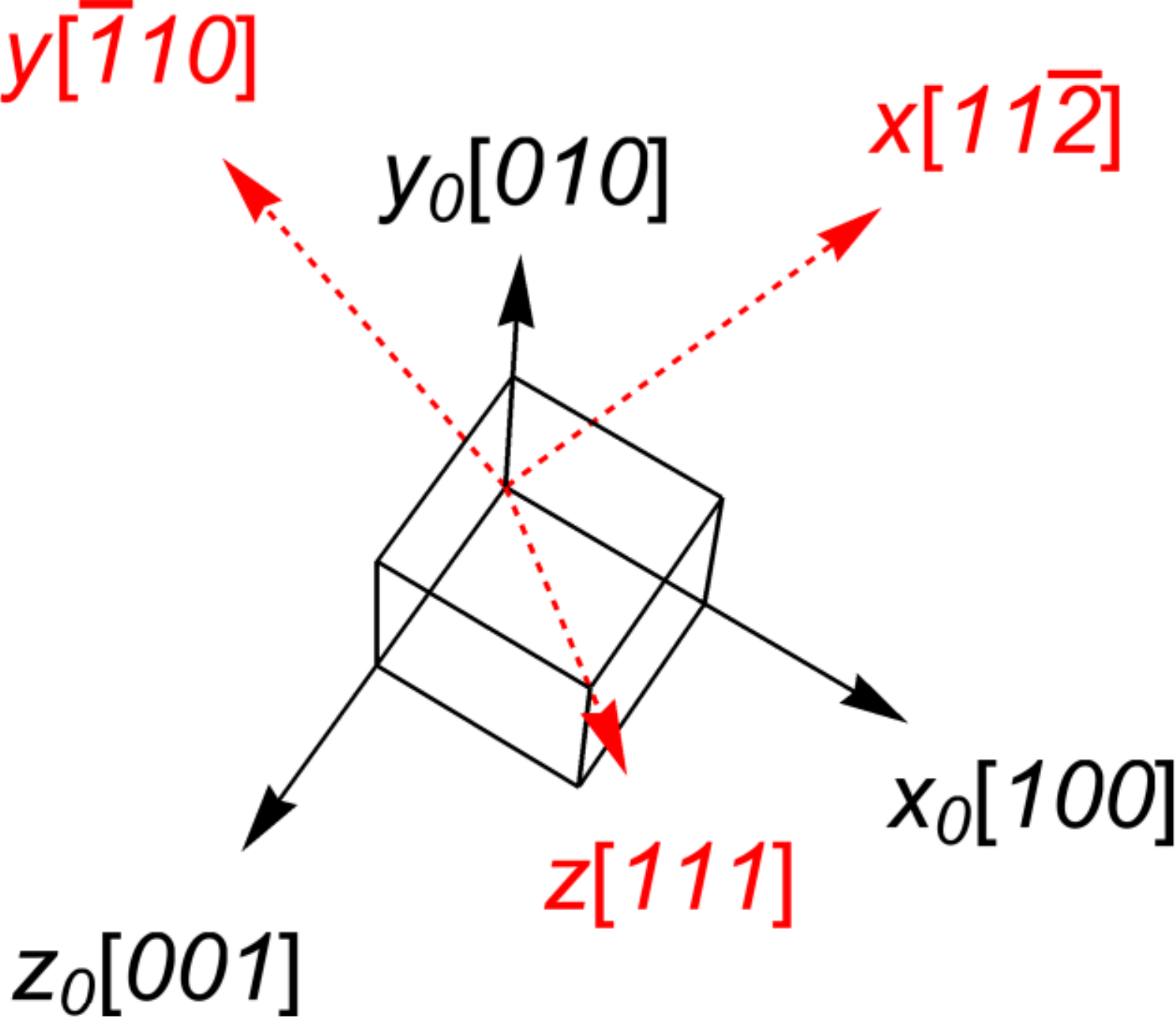}
\caption{ Scheme of the coordinate system with the growth direction along
         [111], and transverse plane spanned by indicated rotated
         $x$ and $y$ axes. }
         \label{fig:rotated[111]}
%\end{center}
\end{figure}
%-------------------------------------

\begin{figure}[ht]
%\begin{center}
\includegraphics[width=1\linewidth]{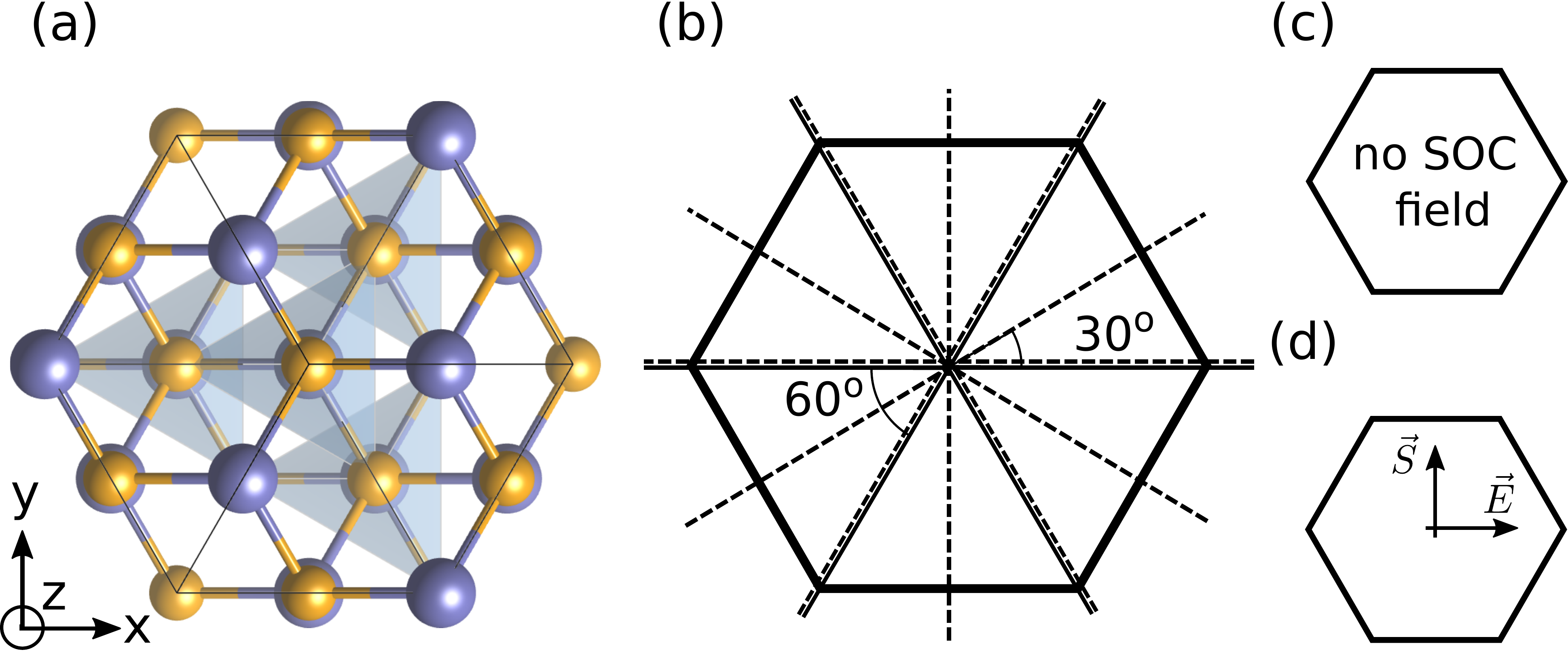}
\caption{Symmetry analysis of [111] oriented ZB nanowire. (a) Atomic arrangement
         along [111] orientation of a ZB structure with indicated $x$ and
         $y$ axes. (b) Mirror symmetry planes of the atomic structure (solid)
         and of the hexagonal confinement (dashed). (c) and (d) are the spin
         projections without and with the applied electric field, respectively.
         (a) was produced using the VESTA software.~\cite{Momma:db5098} }
\label{fig:ZB111}
%\end{center}
\end{figure}

Finally, we look at InSb nanowires oriented along [111]. The rotated coordinated
 axes are  $z = [111]$, $x = [11\overline{2}]$, and $y = [\overline{1}10]$,
 see Fig. \ref{fig:rotated[111]}.  The atomic structure profile is in
 Fig. \ref{fig:ZB111}(a). Here, the atoms arrange themselves with a trigonal
 symmetry, similar to the confinement profile. However, the atomic arrangement
 is less symmetric than the hexagonal confinement. The compatibility of the
 atomic structure along [111] and of the confinement is shown in Fig. \ref{fig:ZB111}(b).

The Dresselhaus spin-orbit field, Eq. \ref{eq:Dresselhaus_bulk}, for ZB
 structures with rotated coordinates as shown in Fig. \ref{fig:rotated[111]},
 transforms as,~\cite{Kammermeier2016}
\begin{widetext}
\begin{eqnarray}
\label{eq:Dress111}
\vec{\Omega}_{\textrm{111}}^{\textrm{ZB}} & = & \frac{\gamma^{\textrm{ZB}}}{\sqrt{6}}\left[-\frac{k_{y}\left(k_{x}^{2}+k_{y}^{2}+2\sqrt{2}k_{x}k_{z}-4k_{z}^{2}\right)}{\sqrt{2}},\frac{k_{y}^{2}\left(k_{x}+\sqrt{2}k_{z}\right)+k_{x}\left(k_{x}^{2}-\sqrt{2}k_{x}k_{z}-4k_{z}^{2}\right)}{\sqrt{2}},-k_{y}\left(k_{y}^{2}-3k_{x}^{2}\right)\right].
\end{eqnarray}
\end{widetext}
The coordinates of momenta $k_x$, $k_y$, and $k_z$, are with respect to
 the rotated axes with unit vector $\hat{k}_z$ pointing along [111].
 Unlike in previous examples, where we applied the electric field along $y$,
 here we direct it along $x$, to explicitly demonstrate that the orientation
 of the field, as well as of the wires, matters little once the fields are
 strong enough to raise the spin-orbit energies above 100 $\mu$eV or so.

%-------------------------------------
%-------------------------------------

When we quantize the spin-orbit field along $x$ and $y$, we get linear
 spin-orbit splitting for the free motion along $z$ proportional to
 $k_{z} \sigma_{y}$. The orientation of the spin caused by SOC
 in ZB nanowires without applied electric field is along $y$, as shown
 in Fig. \ref{fig:ZB111}(c). In the presence of the electric field along $x$
 the spin orients along $y$, see Fig. \ref{fig:ZB111}(d).

\begin{figure}[ht]
\begin{center}
\includegraphics[width=.4\textwidth]{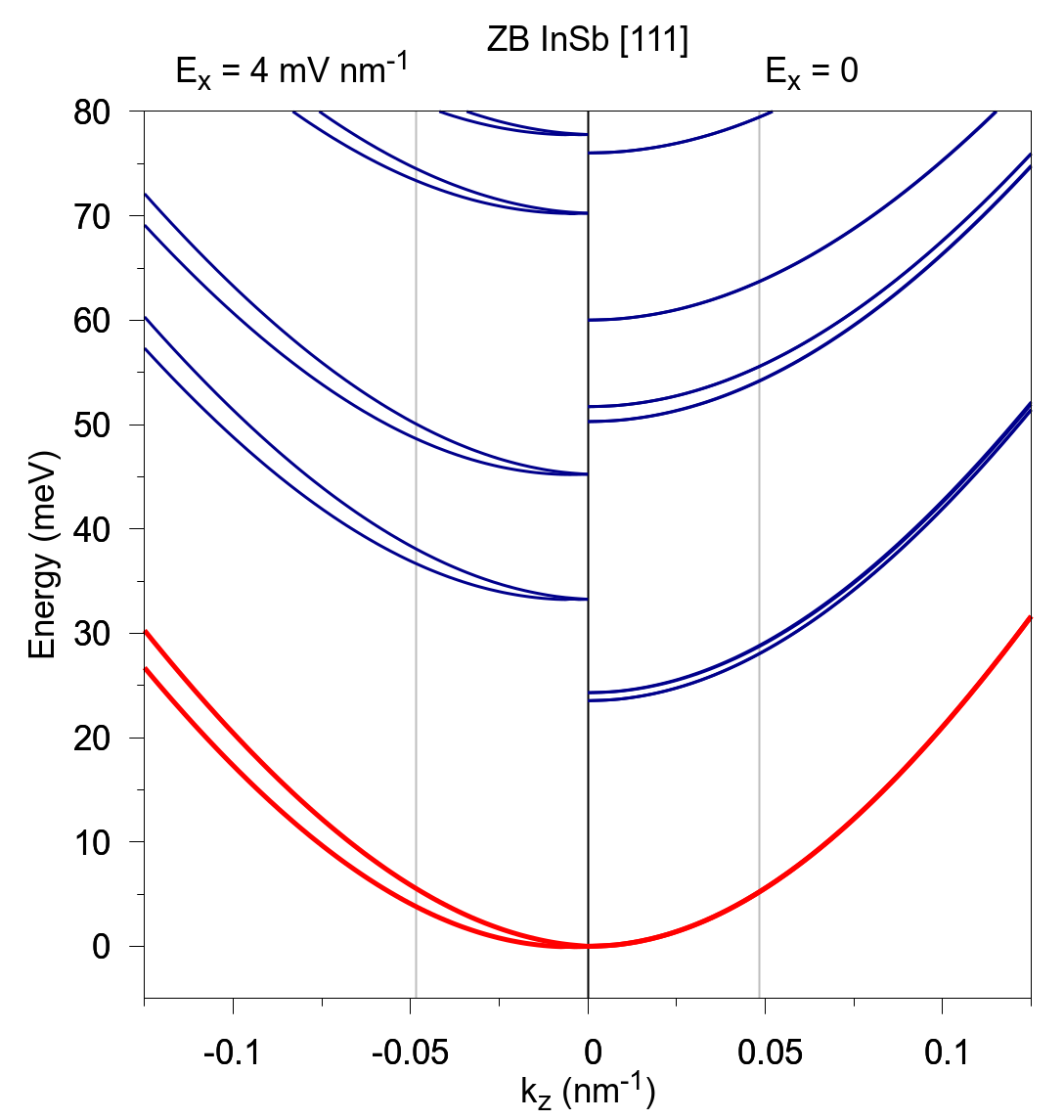}
\caption{Calculated electronic subband dispersion for a $L = 60$ nm ZB InSb
         hexagonal nanowire oriented along [111] direction. The leftmost subband dispersion
         (negative $k_{z}$)  corresponds to an electric field of $E_{x}=4 $mV/nm and
         the rightmost subband dispersion (positive $k_{z}$) to a zero applied electric
         field. The thin vertical lines correspond to the fitting range which was taken as
         $\approx 1\%$ of the Brillouin zone. }
\label{fig:bsInSb111X}
\end{center}
\end{figure}

In Fig. \ref{fig:bsInSb111X} we display the calculated electronic subband
 structure for a ZB InSb hexagonal nanowire along [111]. Again, the subbands
 are shown in the absence and presence of a transverse electric field along
 the $x$ direction. The zero spin-splitting is explained by quantizing the
 Dresselhaus field:
\begin{equation}
 \vec{\Omega}_{\textrm{111}} =  -\sqrt{\frac{1}{6}}\gamma^{\textrm{ZB}}\left[0,\kappa^2 k_{z},0\right],
\end{equation}
 where $\kappa^2$ is the expectation value of $\frac{1}{2}\left(\hat{k}_x^2 - \hat{k}_y^2\right)$
 in the ground state: $\kappa^2 = \frac{1}{2} \left(\langle 0 |\hat{k}_x^2|0\rangle - \langle 0 |\hat{k}_y^2|0\rangle\right)$.
 Because there is a $C_{3v}$ symmetry, $\kappa^2$ vanishes and the lowest
 conduction subband does not exhibits a spin-orbit field.

%-------------------------------------
%-------------------------------------

\begin{figure}[ht]
\begin{center}
\includegraphics[width=.5\textwidth]{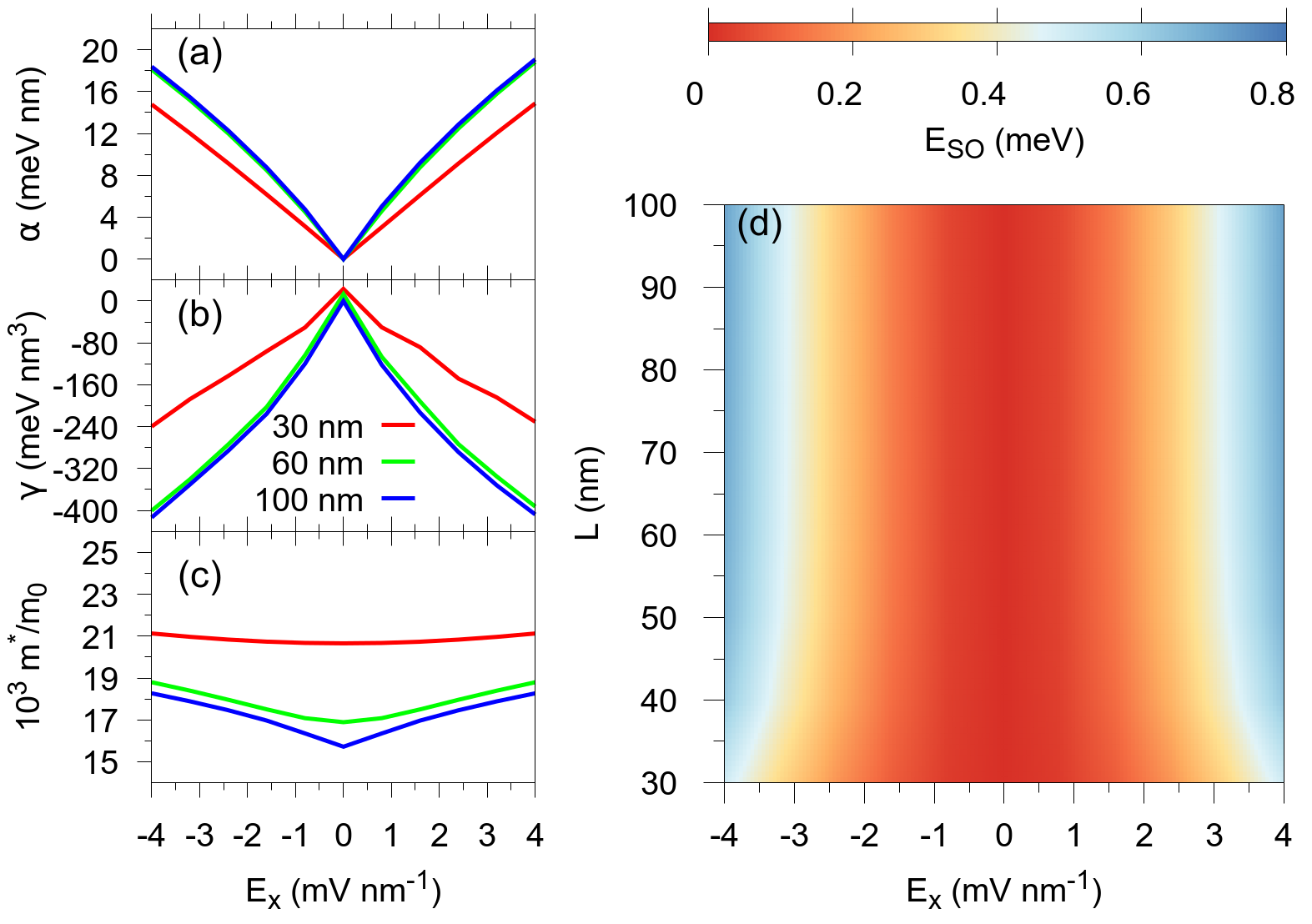}
\caption{Extracted (a) linear, $\alpha$,  and (b) cubic, $\gamma$, spin-orbit
         splitting coefficients, (c) effective masses and (d) spin-orbit
         coupling energy for different diameters $L$ as indicated, for InSb
         ZB nanowires oriented along [111]. Unlike in other cases, the electric field
         is now along $x$.}
\label{fig:SOC_InSb_111_X}
\end{center}
\end{figure}

What is the effect of the electric field oriented along $x$? Consulting
 Fig. \ref{fig:SOC_InSb_111_X} we see that the the overall behavior is very
 close to that seen in [001] and [011] wires with the field along $x$. This
 demonstrates that the growth direction, as well as the application of the
 electric field, are essentially irrelevant in determining the magnitude
 (but not direction!) of the spin-orbit fields. The magnitudes of the
 spin-orbit energy reach close to 1 meV for electric fields of 4 mV/nm.

%-------------------------------------
%-------------------------------------

%============WZ InAs==============================

\section{Wurtzite I\MakeLowercase{n}A\MakeLowercase{s} nanowires}
\label{sec:WZ}

For WZ crystals the $\hat{x}$ and $\hat{y}$ direction are geometrically distinct
 from $\hat{z}$ yielding different effective masses and, consequently, energy
 dispersions, described close to the $\Gamma$-point by the quadratic Hamiltonian,
\begin{equation}
        H_{0} = \frac{\hbar^{2}}{2\,m_{0}} \left[\frac{1}{m^*_{\perp}} \left( k^{2}_{x} + k^{2}_{y}\right) + \frac{1}{m^*_{\parallel}}\, k^{2}_{z} \right] .
\end{equation}
\\ For bulk WZ InAs the two values for the effective mass are: i) the
 perpendicular $m_{\perp}^{*} \approx 0.0416\,m_{0}$; ii) and the parallel
 $m_{\parallel}^{*} \approx 0.037\,m_{0}$ to the $c$-axis.~\cite{De2010,FariaJunior2016}

%-------------------------------------
\begin{figure}[ht]
%\begin{center}
\includegraphics[width=0.9\linewidth]{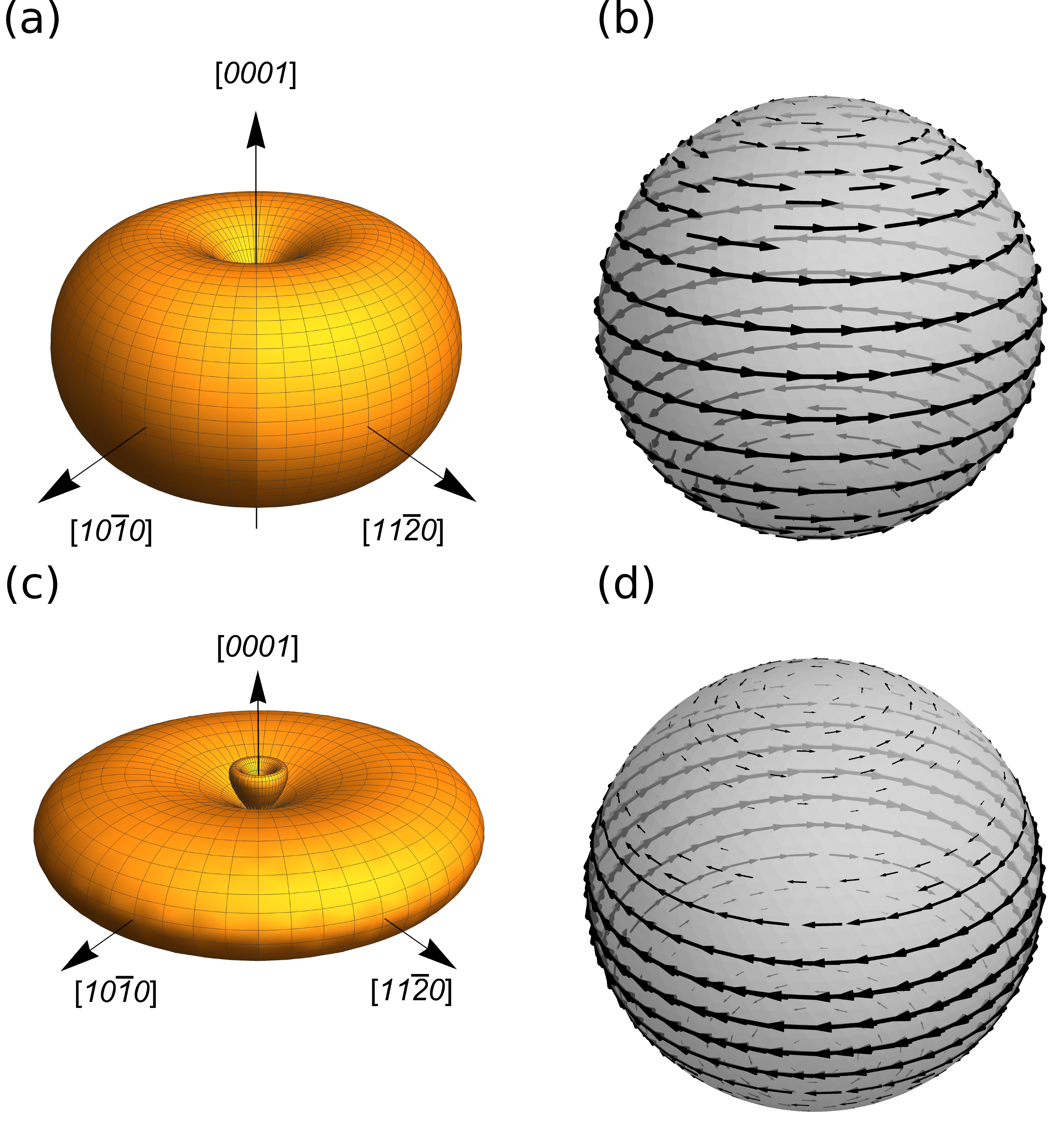}
\caption{Topology of the Dresselhaus spin-orbit coupling field. (a) and (c)
         Spherical  plot of the magnitude of the (Rashba) spin-orbit field
         of a WZ III-V crystal in the momentum space. The crystallographic axes
         are indicated. There are two cases where the field vanishes:
         i) the [0001] direction, shown in both (a) --- with $k = 1\,\AA$ --- and
         (c) --- with $k = 0.045\,\AA^{-1}$; ii) for a small range of momenta,
         shown in (c), there is also an additional surface over which the field vanishes
         (the hyperboloid $b k_z^2 = (k_x^2 + k_y^2)- \alpha/\gamma$).
         (b) and (d) Corresponding spin-orbit vector field  over a Fermi sphere. The
         vortices are along [0001] directions. The field has cylindrical symmetry. }
\label{fig:WZField}
%\end{center}
\end{figure}
%-------------------------------------

The functional form of the spin-orbit field of the conduction electrons in bulk WZ III-V
 semiconductor is~\cite{Fu2008}
\begin{equation}
\label{eq:WZBIA}
\vec{\Omega}_{\textrm{BIA}}  =  \left[\alpha^{\textrm{WZ}}+\gamma^{\textrm{WZ}}\left( b k_z^2 -k_x^2 -k_{y}^{2}\right)\right]\left(k_{y},-k_{x},0\right).
\end{equation}
The spin-orbit splitting vanishes for momenta along the hexagonal axis, [0001],
 that is for $k_x= k_y = 0$, as well as for the momenta in the hyperboloid,
 $b k_z^2 = (k_x^2 + k_y^2)- \alpha/\gamma$. A spherical plot of $\vec{\Omega}_{\rm BIA}$ is shown in
 Figs. \ref{fig:WZField}(a) and (c). Maximal spin-orbit splittings are along
 the directions $[10\overline{1}0]$ and $[11\overline{2}0]$. A recent DFT
 calculation\cite{Gmitra2016} found that for WZ InAs the bulk linear
 coefficient $\alpha^{WZ}_{\rm InAs} \approx 0.3$ eV$\cdot\,\textrm{\AA}$ while the cubic
 coefficient $\gamma^{WZ}_{\rm InAs} \approx 132.5$ eV$\cdot\,\textrm{\AA}^{3}$ and
 the anisotropy factor $b \approx -1.24$. We also project the vector field $\vec{\Omega}_{\rm BIA}$
 on a  Fermi sphere, in Figs. \ref{fig:WZField}(b) and (d). The field is
 solely in the basis plane of the hexagon, pointing perpendicular to the
 momentum.

In the 8-band $\vec{k} \cdot \vec{p}\,$ model, which is symmetric in the $xy$-plane,
 the Hamiltonian for $[10\overline{1}0]$ and $[11\overline{2}0]$
 directions is the same. Therefore, in the following we discuss separately
 the spin physics of hexagonal nanowires grown along [0001], and, as one
 case, together nanowires grown along $[10\overline{1}0]$ or $[11\overline{2}0]$
 directions.

%-----------------------------------InAs 0001-----------------------------------------------------------

\subsection {[0001] growth direction}
\label{susec:WZ0001}

%-------------------------------------
\begin{figure}[ht!]
%\begin{center}
\includegraphics[width=1\linewidth]{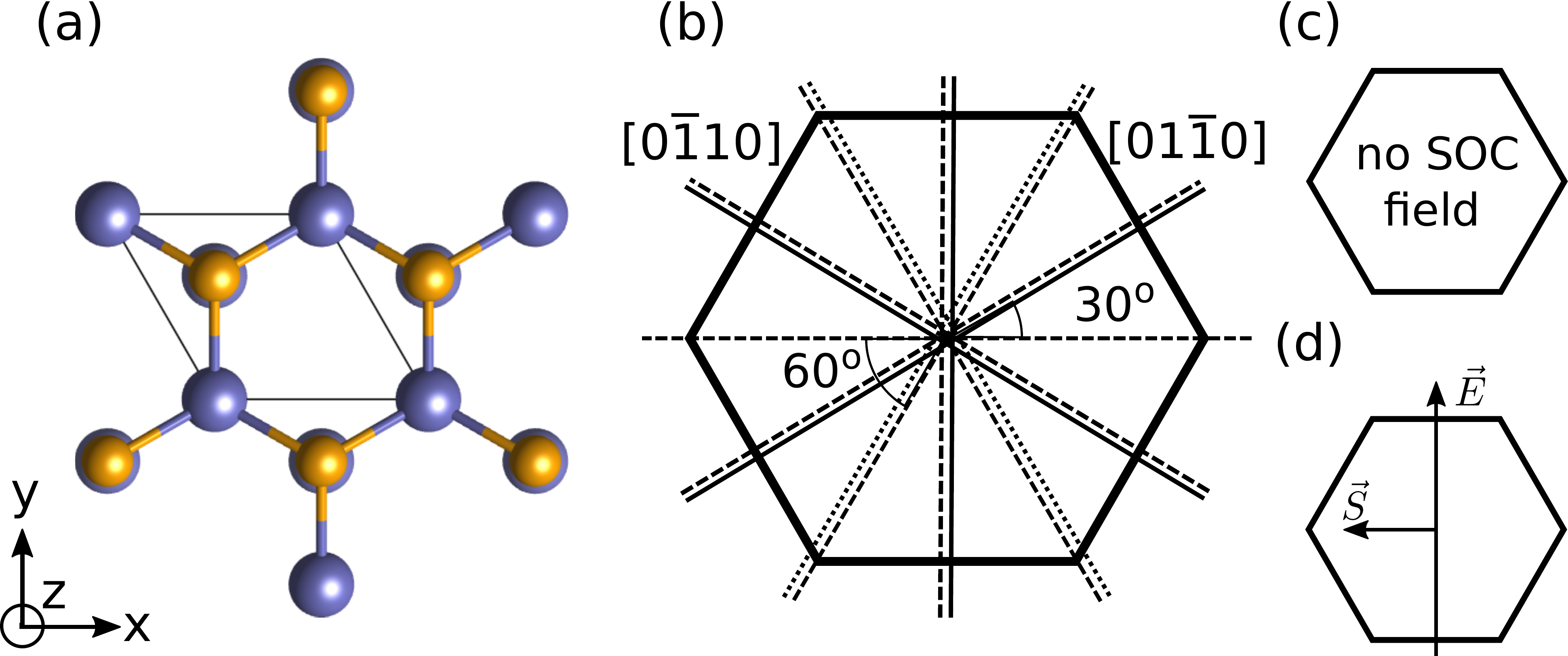}
\caption{Symmetry analysis of hexagonal WZ nanowire oriented along [0001] direction.
         (a) Atomic arrangement along [0001] orientation of a WZ structure
         with indicated $x$ and  $y$ axes. (b) Mirror symmetry planes of the
         atomic structure (solid) and of the hexagonal confinement (dashed).
         (c) and (d) are the spin  projections without and with the applied
         electric field, respectively.  In (b), the dotted lines represent the WZ
         glide planes, i. e., they need an extra translation of $\frac{c}{2}$
         along the $z$ direction.
         (a) was produced using the VESTA software.~\cite{Momma:db5098} }
\label{fig:WZ0001}
%\end{center}
\end{figure}
%-------------------------------------

The cross-section of the atomic structure of a WZ semiconductor
 along [0001] direction is shown in Fig. \ref{fig:WZ0001}(a). The atomic
 arrangement has an incomplete hexagonal symmetry that is not compatible
 with the chosen hexagonal confinement, resulting in an absence of some mirror
 symmetry planes in the nanowire structure, see  Fig. \ref{fig:WZ0001}(b).

As already mentioned, the spin-orbit field vanishes for momenta along [0001].
In the bulk WZ crystal, there are glide symmetry planes which require an
 extra $\frac{c}{2}$ translation along the $z$ axis. Since the nanowires considered
 in this section are grown
 along the $z$ direction, this glide symmetry plane also applies. Therefore, as indicated
 in Fig. \ref{fig:WZ0001}(c),
there is no spin-orbit field in the absence of electric field.
By applying a transverse electric field, say along $y$,
 the spin quantization axis will be $x$, see Fig. \ref{fig:WZ0001}(d).

\begin{figure}[ht]
\begin{center}
\includegraphics[width=.4\textwidth]{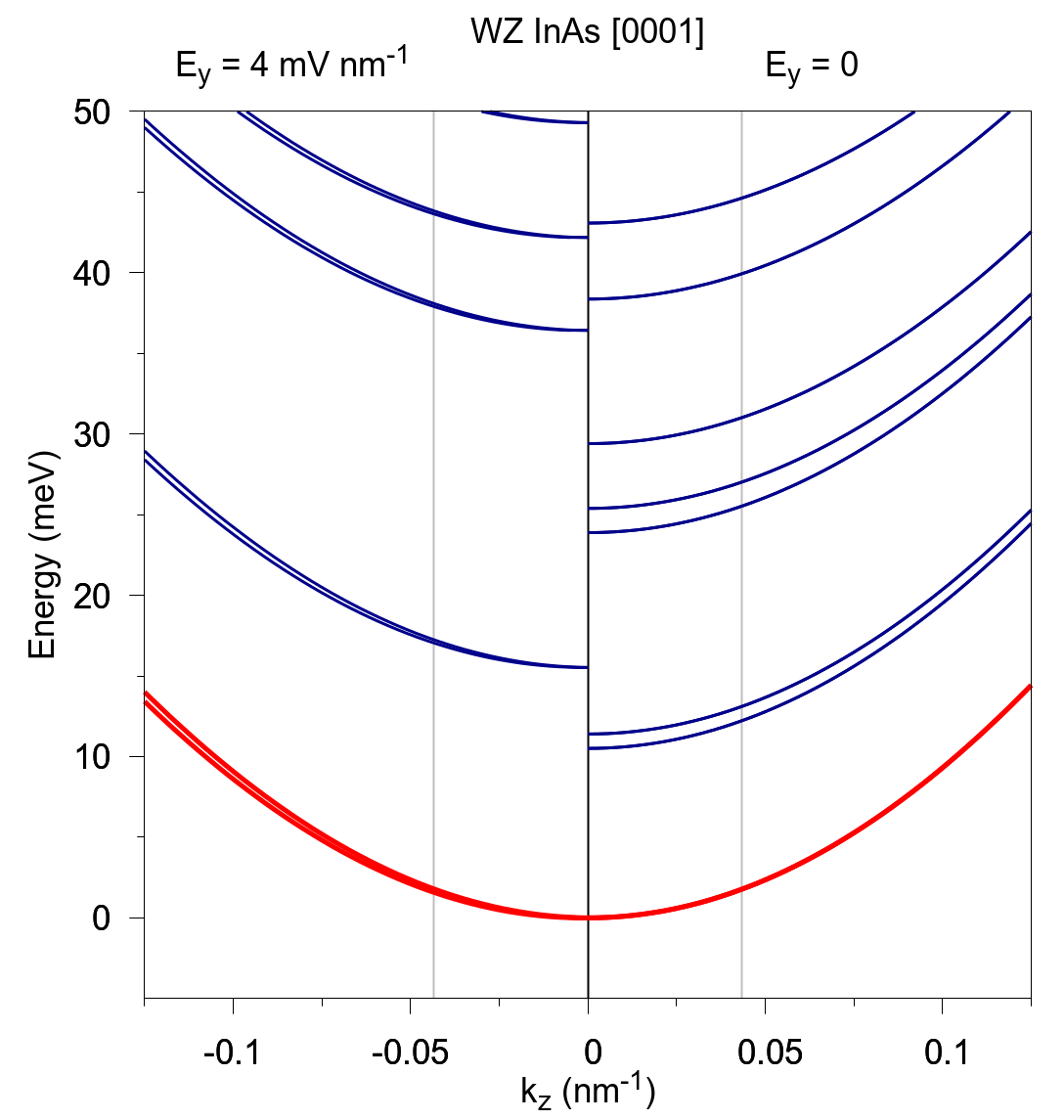}
\caption{Calculated electronic subband dispersion for a $L = 60$ nm WZ InAs
         hexagonal nanowire oriented along [0001] direction. The leftmost subband dispersion
         (negative $k_{z}$)  corresponds to an electric field of $E_{y}=4 $mV/nm and
         the rightmost subband dispersion (positive $k_{z}$) to a zero applied electric
         field. The thin vertical lines correspond to the fitting range which was taken as
         $\approx 1\%$ of the Brillouin zone.}
\label{fig:bsInAs0001Y}
\end{center}
\end{figure}

Figure \ref{fig:bsInAs0001Y} shows the calculated electronic subband structure
 for a WZ InAs hexagonal nanowire along [0001]. Conduction subband is
 shown in the absence and presence of a transverse electric field along the
 $y$ direction. In the absence of the electric field the lowest conduction
 subband is degenerated, while we see a small spin-splitting due to the
 applied electric field. This small spin-splitting indicates that the
 Rashba coefficient for WZ InAs is rather small.

%-------------------------------------
%-------------------------------------

\begin{figure}[ht]
\begin{center}
\includegraphics[width=.5\textwidth]{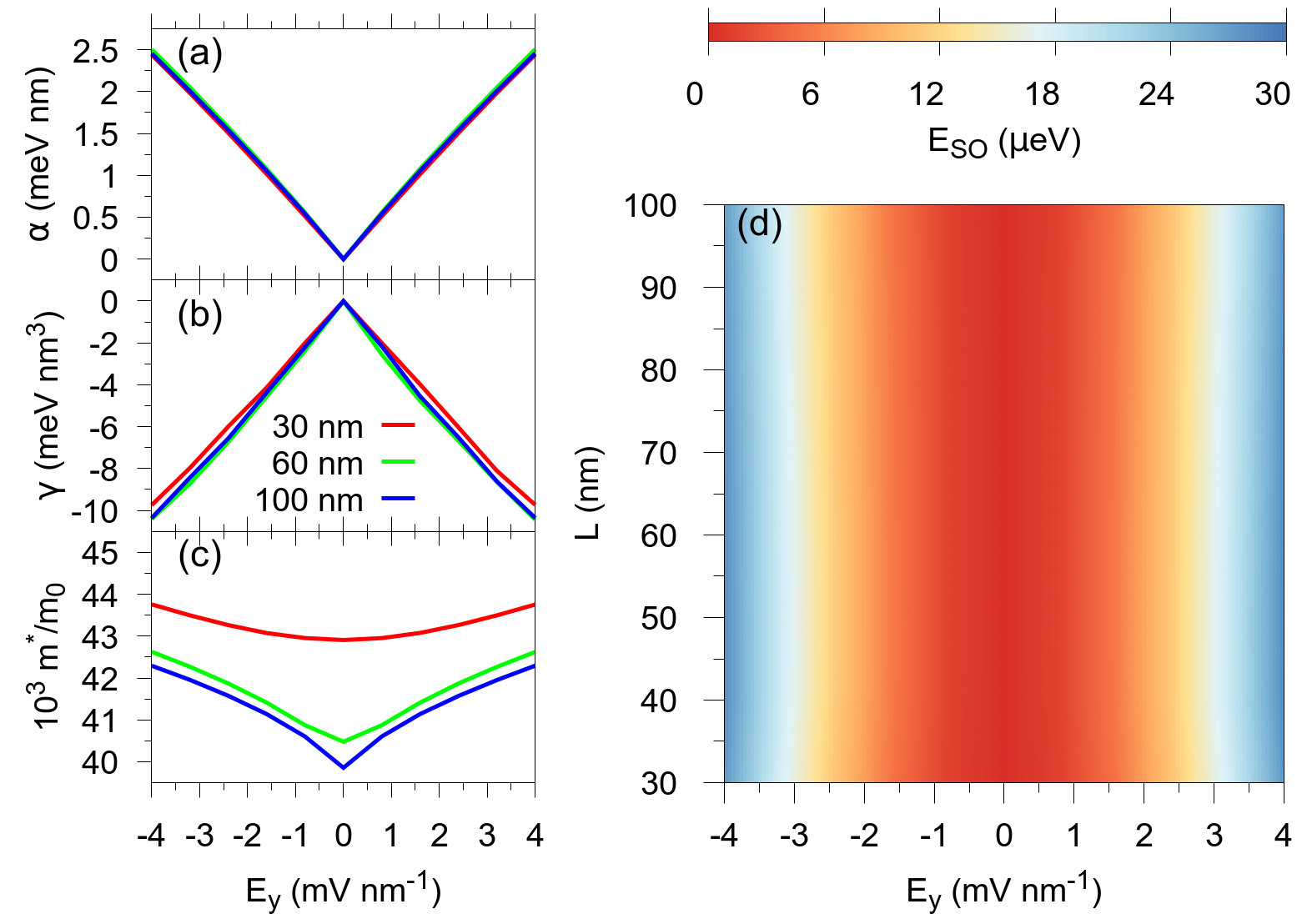}
\caption{Extracted (a) linear, $\alpha$,  and (b) cubic, $\gamma$, spin-orbit
         splitting coefficients, (c) effective masses and (d) spin-orbit
         coupling energy for different diameters $L$ as indicated, for InAs
         ZB nanowires oriented along [0001]. The electric field points along $y$. }
\label{fig:SOC_InAs_0001_Y}
\end{center}
\end{figure}

The extracted linear and cubic spin-orbit coefficients $\alpha$ and $\gamma$,
 as a function of $E_y$ are plotted in Fig. \ref{fig:SOC_InAs_0001_Y}(a)-(b).
 The linear coefficient is typically 1 meV$\cdot$nm for electric fields
 of a few mV/nm. Cubic coefficients are about 5 meV$\cdot$nm$^3$.
 These spin-orbit coefficients are more than an order of magnitude smaller
 than the ones we have encountered in InSb ZB nanowires. Why are spin-orbit
 effects in WZ nanowires grown along [0001] negligible? The reason stems
 in Eq. \ref{eq:WZBIA}. Quantizing the field along the confining $x$ and $y$ directions,
 even in the presence of the electric field, does not yield a term linear in $k_z$.
 Any such linear term present in the nanowire must come from higher order
 (and thus necessarily smaller) terms, not captured by Eq. \ref{eq:WZBIA}.

 In Fig. \ref{fig:SOC_InSb_001_Y}(c) we see that the confinement influences
 the effective mass of the lowest conduction subbands, although the influence
 is smaller than in the ZB case since WZ electrons have already
 a larger effective mass. The effective mass for nanowires with $L \agt 50$ nm
 is already within 10\% of the bulk electron mass. For thinner nanowires
 (30 nm) the effective mass reaches values 0.043 $m$.

%-------------------------------------
%-------------------------------------

Finally, in Fig. \ref{fig:SOC_InAs_0001_Y}(d) we provide the full map of
 the extracted spin-orbit strength $E_{\rm SO}$ as a function of both the
 electric field $E_y$ and the diameter of the nanowire $L$. For a given
 electric field in the considered range there is not much variation of the
 spin-orbit strength with respect to the nanowire diameter. The smallness
 of $ \alpha$ is reflected in the small spin-orbit energy. Indeed,
 the energy is only $E_{\rm SO} = 30\,\mu$eV for fields of 4 meV/nm.
 Nanowires based on WZ InAs, grown along [0001], are thus hardly
 suitable as a platform for studying topological superconducting proximity
 effects.

%------------------------------------------------InAs 1010-------------------------------------------

\subsection {$[10\overline{1}0]$ or $[11\overline{2}0]$ growth direction}
\label{susec:WZ1010}

\begin{figure}[ht]
%\begin{center}
\includegraphics[width=0.5\linewidth]{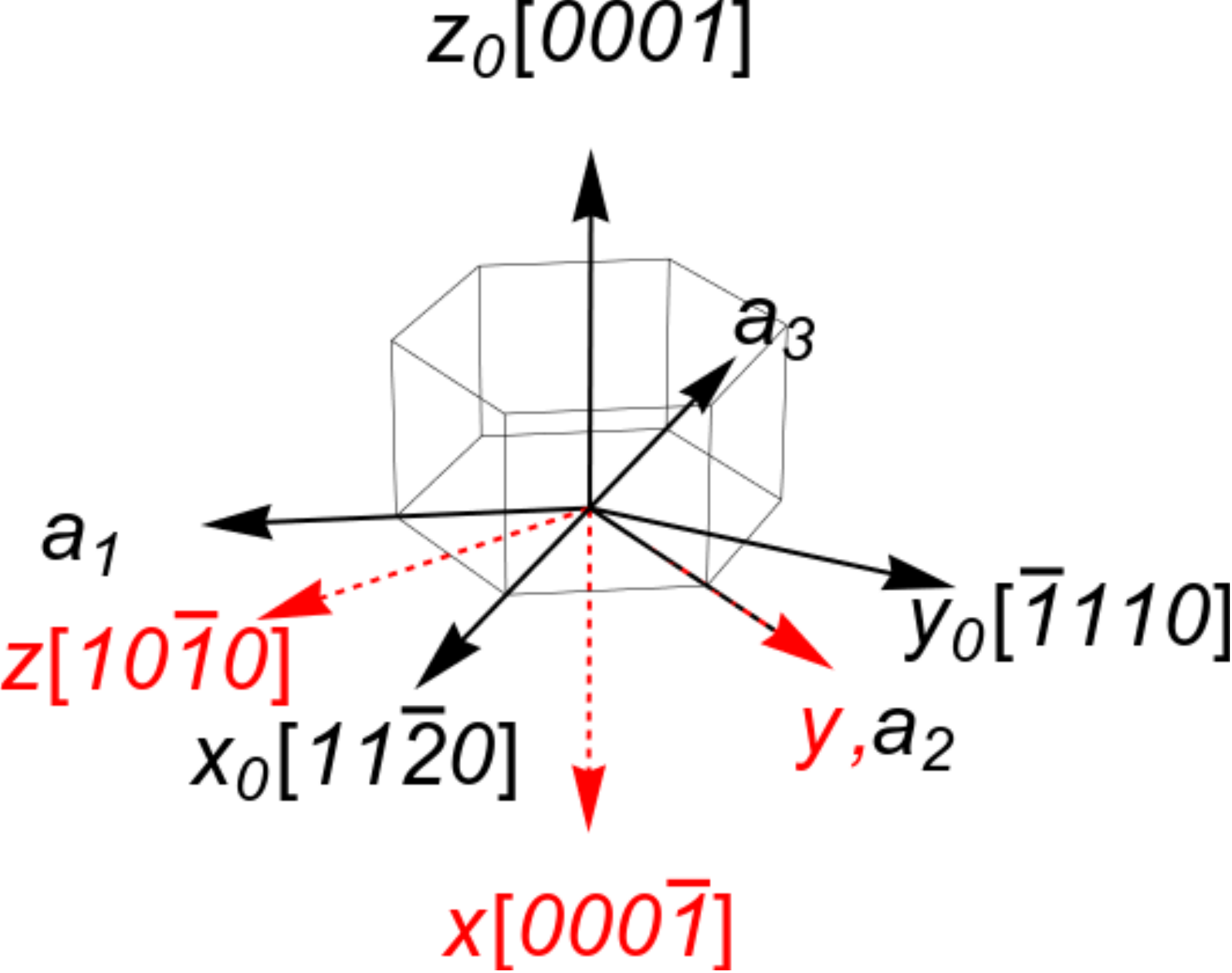}
\caption{ Scheme of the coordinate system with the growth direction along
          $[101\overline{1}0]$, and transverse plane spanned by indicated rotated
          $x$ and $y$ axes. }
\label{fig:rotated[1010]}
%\end{center}
\end{figure}

\begin{figure}[ht]
%\begin{center}
\includegraphics[width=1\linewidth]{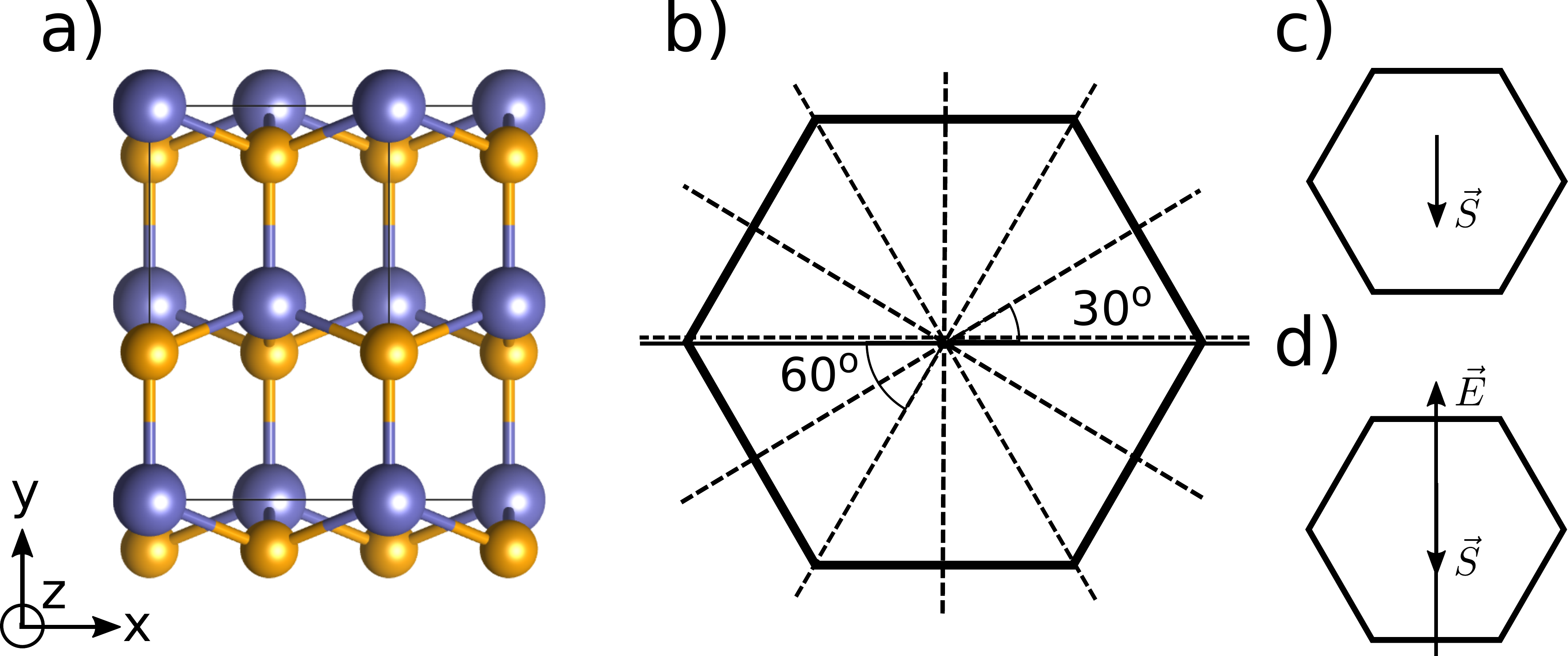}
\caption{Symmetry analysis of $[10\overline{1}0]$ oriented WZ nanowire. (a) Atomic
         arrangement along $[10\overline{1}0]$ orientation of a WZ structure
         with indicated $x$ and $y$ axes. (b) Mirror symmetry planes of the
         atomic structure (solid) and of the hexagonal confinement (dashed).
         (c) and (d) are the spin projections without and with the applied
         electric field, respectively.
         (a) was produced using the VESTA software.~\cite{Momma:db5098}}
\label{fig:WZ1010}
%\end{center}
\end{figure}

We have seen that spin-orbit effects in WZ nanowires grown along [0001] are
negligible. In contrast, spin-orbit energies are giant, in the absence of
electric field, for WZ nanowires grown
along [1010] or  $[11\overline{2}0]$ directions. Since WZ oriented along both $[10\overline{1}0]$ and $[11\overline{2}0]$
 are described by the same 8-band k.p Hamiltonians, here we choose to show
 only the coordinate system and atomic arrangements for the $[10\overline{1}0]$ case.
 Thus, the nanowire axis points along $z =  [10\overline{1}0]$.
 The new cartesian system is shown in Fig. \ref{fig:rotated[1010]}: axis
 $x = [000\overline{1}]$ and $y = \vec{a}_{2}$. The cross-section
 of the atomic structure of a WZ semiconductor along $[10\overline{1}0]$ direction
 is shown in Fig. \ref{fig:WZ1010}(a). The hexagonal confinement reduces
 the structural symmetry, retaining only one mirror plane, spanned by $y$
 and $z$ (making the system symmetric as $y \to - y$). The compatibility
 of the atomic structure along $[10\overline{1}0]$ and
 of the hexagonal confinement is shown in Fig. \ref{fig:WZ1010}(b).

%-------------------------------------
%-------------------------------------

The Dresselhaus spin-orbit field for $[10\overline{1}0]$ is,
%\begin{widetext}
\begin{eqnarray}
\label{eq:WZBIA1010}
\vec{\Omega}_{[10\overline{1}0]}^{\textrm{WZ}} & = & \left[\alpha^{\textrm{WZ}}+\gamma^{\textrm{WZ}}\left( b k_x^2 -k_y^2 -k_{z}^{2}\right)\right]\left[\begin{array}{c}
0\\
-k_{z}\\
k_{y}
\end{array}\right]%\left(0,-k_{z},k_{y}\right).
\end{eqnarray}
%\end{widetext}
The coordinates of momenta $k_x$, $k_y$, and $k_z$, are with respect to
 the rotated axes with unit vector $\hat{k}_z$ pointing along $[101\overline{1}0]$, $k_x$
 along [000$\overline{1}$], and $k_y$ along $ \vec{a}_{2}$.

%-------------------------------------
%-------------------------------------

When we quantize the spin-orbit field along $x$ and $y$, we get linear
 and cubic spin-orbit splitting for the free motion along $z$ proportional to
 $k_{z} \sigma_{y}$. The orientation of the spin caused by SOC
 in WZ nanowires without electric field is along $y$, as shown
 in Fig. \ref{fig:WZ1010}(c). By applying an electric field along $y$, the
 spin acquires a component along $x$. However, the Rashba coefficient,
 due to the applied electric field, for
 InAs WZ nanowires is rather small compared to intrinsic one, as seen
 in Fig. \ref{fig:WZ0001}(a), and the change in the spin orientation is
 negligible. Therefore the spin orientation, even with electric field
 is along $y$, for the range of electric field investigated,
 as depicted in Fig. \ref{fig:WZ1010}(d).

\begin{figure}[ht]
\begin{center}
\includegraphics[width=.4\textwidth]{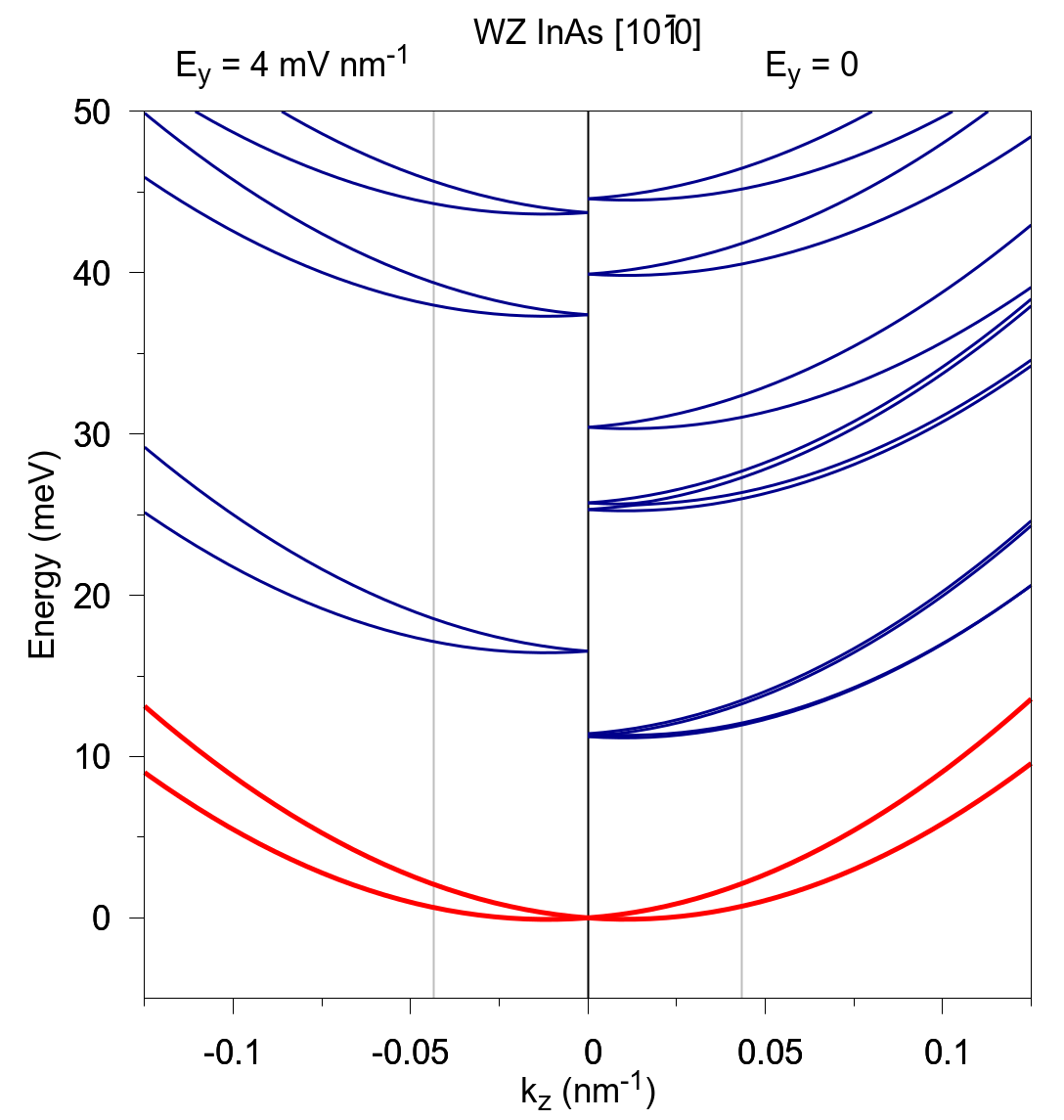}
\caption{Calculated electronic subband dispersion for a $L = 60$ nm WZ InAs
         hexagonal nanowire oriented along $[10\overline{1}0]$ direction. The leftmost subband dispersion
         (negative $k_{z}$)  corresponds to an electric field of $E_{y}=4 $mV/nm and
         the rightmost subband dispersion (positive $k_{z}$) to a zero applied electric
         field. The thin vertical lines correspond to the fitting range which was taken as
         $\approx 1\%$ of the Brillouin zone.}
\label{fig:bsInAs1010Y}
\end{center}
\end{figure}

Figure \ref{fig:bsInAs1010Y} shows the calculated electronic subband structure
 for a WZ InAs hexagonal nanowire along $[10\overline{1}0]$. The conduction band is
 shown in the absence and presence of a transverse electric field along the $y$
 direction. In the absence of the electric field the lowest conduction band already
 has a large spin-splitting. This is explained by
 directly quantizing the Dresselhaus field, Eq. (\ref{eq:WZBIA1010}). We get,
\begin{equation}
 \vec{\Omega}_{[10\overline{1}0]}  =  k_{z}\,\left[0,-\alpha^{\rm WZ} - \gamma^{\rm WZ} \left(\kappa^{2} - k_{z}^{2}\right),0\right],
\end{equation}
 where $\kappa^2$ is the expectation value of $b\,\hat{k}_y^2 + \hat{k}_x^2$
 in the ground state: $\kappa^2 = b\,\langle 0 |\hat{k}_x^{2}|0\rangle + \langle 0 |\hat{k}_y^2|0\rangle$
 which in general is not zero. There is always the linear term present, which is due
 to the bulk spin-orbit coupling $\alpha^{\rm WZ}$. This is the dominating spin-orbit
 contribution to the spin-orbit energy even in the presence of electric field (within
 the investigated ranges).

%-------------------------------------
%-------------------------------------

\begin{figure}[ht]
\begin{center}
\includegraphics[width=.5\textwidth]{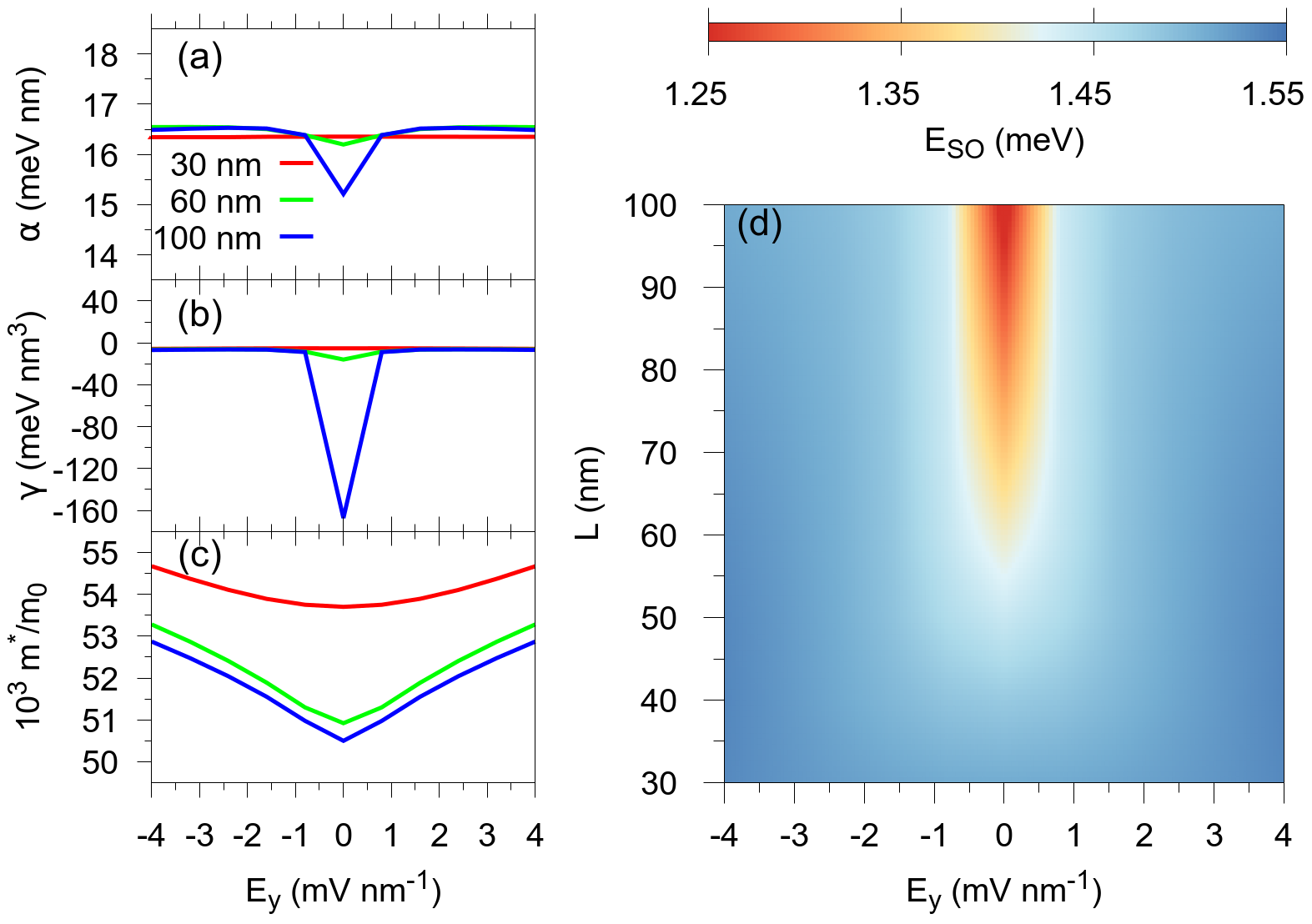}
\caption{Extracted (a) linear, $\alpha$,  and (b) cubic, $\gamma$, spin-orbit
         splitting coefficients, (c) effective masses and (d) spin-orbit
         coupling energy for different diameters $L$ as indicated, for InAs
         WZ nanowires oriented along $[10\overline{1}0]$.}
\label{fig:SOC_InAs_1010_Y}
\end{center}
\end{figure}

Indeed, the spin-splitting is not strongly enhanced in the presence of the electric
 field, as seen in the case of [110] ZB InSb nanowire, shown in Fig. \ref{fig:bsInSb110Y}.
 The extracted linear and cubic spin-orbit coefficients $\alpha$ and $\gamma$,
 as a function of $E_y$ are plotted in Fig. \ref{fig:SOC_InAs_1010_Y}(a)-(b).
 The linear coefficient is typically 15 meV$\cdot$nm for electric fields of a
 few mV/nm. In Fig. \ref{fig:SOC_InAs_1010_Y}(c) we see that the confinement influences
 the effective mass of the lowest conduction subbands. For thinner nanowires
 (30 nm) the effective mass reaches values 0.054 $m$ which is about
  10\% larger than the bulk effective mass.

%-------------------------------------
%-------------------------------------

In Fig. \ref{fig:SOC_InAs_1010_Y}(d) we give the full map of
 the extracted spin-orbit strength $E_{\rm SO}$ as a function of both the
 electric field $E_y$ and the diameter of the nanowire $L$.  Most
 important, the energy is in the range 1 - 2 meV; this magnitude is rather
 stable with respect to the nanowire diameter and the electric
 field.

%===============================================================================

\section{Discussion}
\label{sec:Discussion}

In ZB nanowires it is the confinement and electric field that dominate spin-orbit
splitting. Bulk effects are negligible, since they are only cubic in momentum.
Interface with vacuum leads to inter-facial spin-orbit coupling,~\cite{Rossler2002}
and electric field to more localized subband modes, inducing the Rashba effect.
~\cite{Bychkov1984a}
On the other hand, WZ crystals exhibit linear spin-orbit splitting already in the
bulk. Incidentally, what leads often to confusing terminology, this is also called Rashba
splitting,~\cite{LewYanVoon1996,Fu2008} as it was derived
by Rashba.~\cite{Rashba1959} In addition, in WZ confined systems and/or in the
presence of electric field, the spin-orbit splitting is proportional not only to the
electrostatic potential gradient, but depends on the potential itself.~\cite{Litvinov2003}
 The linear coefficient $\alpha$ is thus not necessarily a linear function of electric field.

When we induce a structural asymmetry via the electric field, we have at least
 two cases: i) the application of  $E=E_{0}\hat{x}$ induces a SIA spin-orbit
 coupling with spin polarization along the $y$ direction and; ii) the application
 of $E=E_{0}\hat{y}$ induces a SIA SOC with spin polarization
 along the $x$ direction. Both cases have a functional form for the spin-splitting
 which is linear in momentum. In case i) both BIA and SIA have
 spin polarization along the same direction (exception made for
 ZB [001] oriented nanowires), therefore  their contributions interfere
 with each other and we could get a subband dispersion which is
 asymmetric (or have an asymmetric spin-splitting) with respect to the sign
 of the applied electric field.~\cite{Andrada1992}
 In case ii) we do not have interferences between BIA and SIA, because
 they always point in distinct directions, and the spin-splitting parameters
 are always symmetric with respect to the applied electric field.

A distinction between our `hard wall' nanowires and electrically defined
 quantum wires is that, in the former, the confinement in the $xy$ plane have
 very similar strengths, therefore it couples the electron dynamics in
 all three dimensions, which is not the case  in the latter system where
 the confinement is much weaker than underlying quantum well confinement.~\cite{Andrada2003}
 This distinction means that the Rashba effect (structural asymmetry), describes very
 well the decoupled case (electrically defined quantum wire) but it should fail
 in general for the hard wall case. The failure is seen as a deviation
 from the linear dependence with the electric field of the associated spin-splitting
 parameter, see Figs. \ref{fig:SOC_InSb_001_Y}(a), \ref{fig:SOC_InSb_110_Y}(a),
 \ref{fig:SOC_InSb_111_X}(a), \ref{fig:SOC_InAs_0001_Y}(a) and \ref{fig:SOC_InAs_1010_Y}(a).
 Moreover, since in quantum wire systems the Rashba coefficient is given by the
 underlying asymmetry in the quantum well, it should remain invariant under
 changes in the electrical confinement. On the other hand, for hard wall
 confinement, the Rashba coefficient strictly depends on the geometric
 configuration of the system. Therefore, changes in the quantum confinement
 also change the Rashba coefficient.~\cite{Andrada2003,Zhang2006}

For ZB in Figs. \ref{fig:SOC_InSb_001_Y}(a)-(b), \ref{fig:SOC_InSb_110_Y}(a)-(b) and
 \ref{fig:SOC_InSb_111_X}(a)-(b), we see that for large confinements,
 $L = 30\,\textrm{nm}$, the spin-splitting coefficients (linear and cubic in momentum)
 present a linear dependence with the applied electric field.
 However as we increase the wire diameter to $L = 60\,\textrm{nm}$ we already
 see that this linear dependence holds only for small values of electric field.
 Moreover, comparing the spin-splitting parameters for diameters $L = 60\,\textrm{nm}$
 and $L = 100\,\textrm{nm}$ wee see that they almost do not change. Hence,
 we can say that the Rashba parameter has a dependence on the nanowire diameter:
 it is small for thin wires and grow up to a saturation value for large
 diameters. Also, the simplified Rashba model, when applied to nanowires, does not
 predict a cubic in momentum dependence for the spin-splitting parameters and the BIA
 term only show a cubic dependence for [110] oriented nanowires. However, since
 we are using the full multi band Hamiltonian and not the simplified Rashba model,
 we realistically capture all the features of the full model which includes: i)
 the dependence of the Rashba parameters on quantum confinement; ii) the
 deviation of linearity for large electric fields and iii) the presence
 of the cubic in momentum dependence of the spin-split parameters.
 For WZ in Figs. \ref{fig:SOC_InAs_0001_Y}(a)-(b) and \ref{fig:SOC_InAs_1010_Y}(a)-(b),
 the same applies, except that the Rashba coefficient does not vary with the
 nanowire diameter as discussed above.

\begin{figure}[ht]
   \begin{center}
      \includegraphics[width=.4\textwidth]{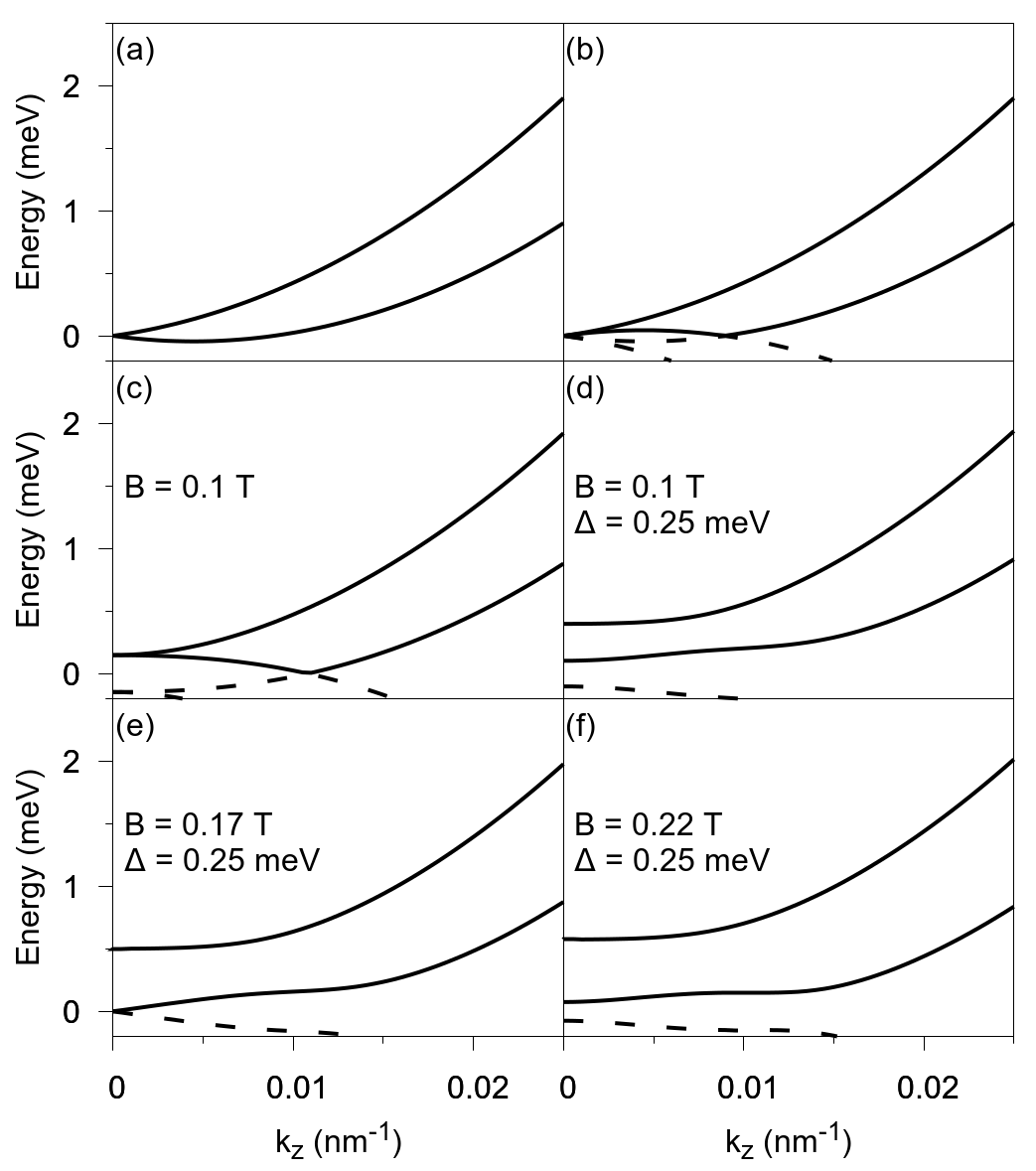}
   \end{center}
   \caption{\label{fig:BdGZBBZ} Zinb-blende InSb nanowire with $L = 100\,\textrm{nm}$
     with magnetic field applied along the nanowire axis (and perpendicular
     to the spin-orbit coupling) and superconductivity proximity effect.  Dashed
     lines represent negative energies (zero is set by the chemical potential $\mu$).
     (a) Lowest subband spectrum for $\mu=B=\Delta=0$.{}
     (b) Quasiparticle excitation spectrum for $\mu=B=\Delta=0$. (c) Excitation spectrum for $B=0.1\,\textrm{T},\,\Delta=\mu=0${}
     where Zeeman splitting opens a gap at $k_{z}=0$. (d) $B=0.1\,\textrm{T},\,\Delta=0.25\,\textrm{meV},\,\mu=0$
     with a superconducting gap for $k_{z}\neq0$ and a Zeeman gap near
     $k_{z}=0$. (e) $B=0.17\,\textrm{T},\,\Delta=0.25\,\textrm{meV},\,\mu=0${}
     meaning $V_{z} = \Delta$ where the gap at $k_{z}=0$ is closed meaning
     a phase transition. (f) $B=0.22\,\textrm{T},\,\Delta=0.25\,\textrm{meV},\,\mu=0$
     here the gap reopens confirming the phase transition. }
\end{figure}

We also briefly discuss the relevance of our results for superconducting proximity
effects. In Fig. \ref{fig:BdGZBBZ} we plot the spectrum of  ZB InSb nanowires
in the superconducting proximity regime (non-zero superconducting gap $\Delta$)
in the presence of a magnetic field causing Zeeman (but no orbital) splitting.
The spectrum is obtained by solving the BdG equation~\cite{Sau2010:PRL,Oreg2010:PRL},
\begin{widetext}
\begin{equation}
   \label{eq:kpBdG}
   H_{\textrm{BdG}}=\left\{ \left[\left(\frac{\hbar^{2}}{2\,m_{0}}\right)\left(\frac{1}{m^{*}}\right)k^{2}-\mu\right]\sigma_{0}+2\,\alpha\,k\sigma_{i}\right\} \tau_{z}-\frac{g^{*}\mu_{b}}{2}\vec{B}\cdot\vec{\sigma}+\Delta\sigma_{0}\tau_{x},
\end{equation}
\end{widetext}
where  $\vec{\sigma}$ is a vector
 containing the Pauli spin matrices (plus the identity, $\sigma_{0}$)
 acting on the spin degree of freedom and $\vec{\tau}$ is a vector
 also containing the Pauli matrices but acting on the particle-hole
 space. The wave function is in the Nambu spinor basis, i. e., it
 contains both particle and antiparticle wave functions and is written
 as $\Psi\left(\vec{r}\right)=\left[u_{\uparrow}\left(\vec{r}\right),u_{\downarrow}\left(\vec{r}\right),v_{\downarrow}\left(\vec{r}\right),-v_{\uparrow}\left(\vec{r}\right)\right]^{T}$.
 Here the Rashba term can be on $x$ or $y$ direction (depending on the
 direction of the applied electric field), and the magnetic field that is
 perpendicular to it. For the system to undergo the topological
 phase transition it has to be gapped before  we couple it to the superconductor,
 then with a change in the parameters it has to close the gap and reopen
 again. At $k_{z}=0$, the gap is  defined by $E\left(0\right) = |V_{Z} - \sqrt{\Delta^{2} + \mu^{2}}|$.
 The trivial phase is defined when $V_{Z} < \sqrt{\Delta^{2} + \mu^{2}}$,
 the phase transition (closing of the gap) when $V_{Z} = \sqrt{\Delta^{2} + \mu^{2}}$
 and the topological phase is defined when $V_{Z} > \sqrt{\Delta^{2} + \mu^{2}}$.~\cite{Sau2010:PRL,Alicea2010:PRB,Alicea2012:RPP,Elliott2015}

Since our $\vec{k} \cdot \vec{p}\,$ Hamiltonians describe the crystals with
 both bulk inversion asymmetry and structural inversion asymmetry, for instance
 when an external electric field is applied, its subbands are spin-split
 away from $\vec{k}=0$. Especially for the conduction subbands, they
 have a `Dirac-like' shape for very small momenta. In Ref.~\onlinecite{Alicea2010:PRB}
 the authors showed that the combination of this `Dirac-like' shape for the conduction
 subbands, the presence of a magnetic field, giving a Zeeman spin-split,
 and the proximity effects of a s-wave superconductor allows for a
 effective p-wave pairing in the lowest branch of the conduction subband.

%-------------------------------------
%-------------------------------------

The appearance of the topological superconducting
 phase, and therefore, the possibility of a zero-energy Majorana bound
 state follow from: i) the spin-orbit coupling spin polarize the subbands which
 in turn are split at $\vec{k}=0$ by the magnetic field; ii) with
 the Fermi level set in between the Zeeman gap, we get an effective
 spinless (or polarized spinful) metal; iii) the superconductor induces
 a p-wave pairing which is known to support Majorana fermions.~\cite{Fu2008:PRL}

Using realistic parameters fitted from our multiband $\vec{k} \cdot \vec{p}\,$ calculations
 we see that for $L=100\,\textrm{nm}$ ZB InSb nanowires, which are
 experimentally relevant,~\cite{Mourik2012:Science}
 the typical values which characterize the system are: $m^{*} \approx 0.017\,\textrm{m}_{0}$,
 $\alpha \approx 0.2\,\textrm{eV \AA}$, $g^{*}_{z} \approx -51$, and Zeeman
 splitting $V_{Z} \approx -1.48\times B_{0}\,\textrm{meV}$, with $B_{0}$ being the magnetic field magnitude. The proposed
 induced superconducting gap is $\Delta \approx 0.25\,\textrm{meV}$ and
 typical values for the magnetic field are $B_{0} \approx 0.1$.~\cite{Mourik2012:Science}
 In Fig. \ref{fig:BdGZBBZ}(d) we show that the gap is open for $B=0.1\,\textrm{T}$ and $\Delta=0.25\,\textrm{meV}$,
 and that matching the Zeeman energy to the pairing potential, the gap closes,
 see Fig. \ref{fig:BdGZBBZ}(e). Once the magnetic field further increases, the
 superconducting spectral gap reopens, see Fig. \ref{fig:BdGZBBZ}(f), demonstrating
 the possibility for topological phase transition.  However, experimentally this
 is still a challenging task due to imperfections in the growth process.~\cite{Drachmann2017,Lutchyn2017,gul2018ballistic,Zhang2018}

%===============================================================================

\section{Conclusions}
\label{sec:conc}

    We performed a systematic investigation of the spin-orbit
    interaction in hexagonal semiconductor nanowires under an applied transverse
    electric field. We used robust multiband $\vec{k} \cdot \vec{p}\,$
    Hamiltonians in the  envelope function approximation and plane waves
    expansion to extract relevant physical parameters describing the
    lowest energy conduction band with high fidelity.
    Specifically, we focused on ZB InSb and WZ InAs nanowires,
    extracting relevant spin-orbit parameters: linear $\alpha$, cubic $\gamma$, and
    spin-orbit energy $E_{SO}$.

    We found that in ZB InSb nanowires the spin-orbit splitting is strongly
    influenced by the quantum confinement. On the other hand, for
    WZ InAs nanowires there is already a large linear spin-orbit parameter
    $\alpha^{WZ}$,  which also dominates in the presence of confinement.
    Due to symmetry reasons, the spin-splitting remains largely unaffected
    in $[10\overline{1}0]$ or $[11\overline{2}0]$ oriented nanowires, while
    the splitting is absent for wires along [0001].

    In the presence of electric field, the spin-splitting gets strongly enhanced in
    ZB nanowires. The enhancement does not vary with the growth
    direction. The spin-orbit energies reach 0.8 meV for electric fields of 4 mV/nm.
    On the other hand, the electric field hardly influences the already large
    spin splitting of the WZ nanowires. For the [0001] direction, the spin-orbit
    energy remains small, reaching only 30 $\mu$eV in the field of 4 mV/nm.
    This growth orientation is least favorable for applications requiring large
    spin-orbit splitting.

    Finally, with our realistic set of parameters describing the first conduction
    band of the nanowires, we used the BdG formalism to describe the superconductivity
    induced effects and showed that system undergoes the topological phase
    transition. Our results could help guiding experimental efforts in demonstrating
    such superconducting topological effects.

%===============================================================================

\section*{Acknowledgements}

This work has been supported by CNPq (grant No. 149904/2013-4) ,
 CAPES (grant No. 88881.068174/2014-01), FAPESP (grant No. 2012/05618-0).
 T. C. thanks the LCCA for computational resources
 and D. R. Candido for useful insights. PEFJ acknowledges the financial
 support of the Alexander von Humboldt Foundation and Capes (grant No. 99999.000420/2016-06).
 J. F. and M. G. acknowledge support from DFG SFB 1277 (B07).

%===============================================================================

\section*{Appendix A: Plane wave expansion and numerical details}

%-------------------------------------
\begin{figure}[ht]
\begin{center}
\includegraphics[width=.45\textwidth]{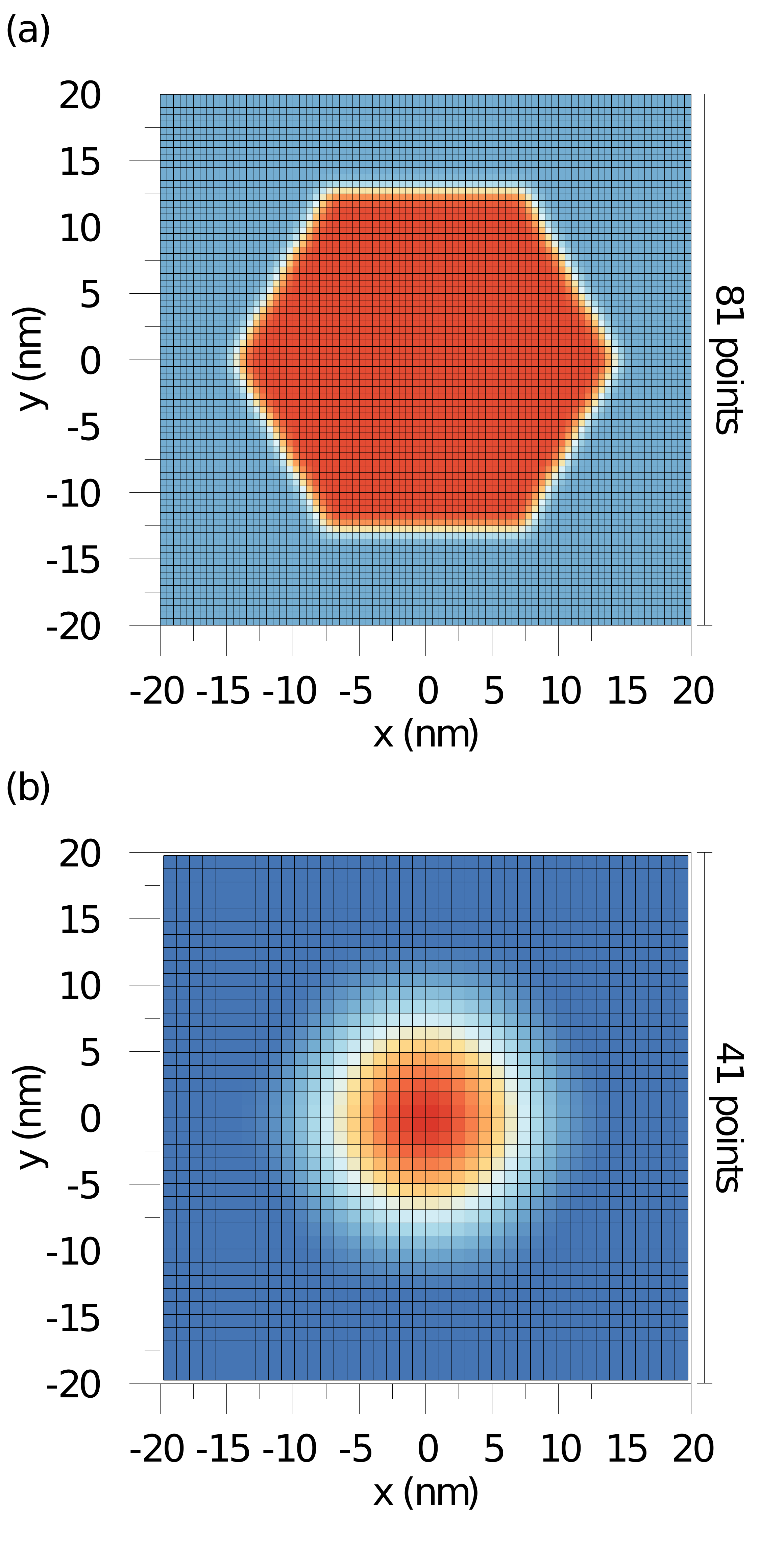}
\caption{Discretization grid for (a) potential profile and (b) wave function.
        Due to the plane wave expansion method the number of grid points
        is larger in the potential profile if compared with the number
        of grid points in the final wave function. }
\label{fig:PWE}
\end{center}
\end{figure}
%-------------------------------------

In the envelope function approximation description of the multiband $\vec{k} \cdot \vec{p}\,$
 we end up with a set of coupled differential equations given by~\cite{Bastard1981,Baraff1991,Burt1992,bastard}
 \begin{equation}
   \sum_{\alpha}^{A} \left[ K^{\alpha^{\prime}\alpha}(x,y) + V^{\alpha^{\prime}\alpha}(x,y)\right]f(x,y) = E\,f(x,y)
 \end{equation}
 \\where the summation over $A$ represents the multiband $\vec{k} \cdot \vec{p}\,$ model
 we are dealing with, in this paper it is either $A=8$ for the WZ
 $\vec{k} \cdot \vec{p}\,$ model or $A=14$ for the zinc-blend model;
 $K^{\alpha^{\prime}\alpha}(x,y)$ represents the spatial description of the
 kinetic terms---effective masses, interband and intraband couplings, $k$ dependent spin-orbit coupling terms, etc---and
 $V^{\alpha^{\prime}\alpha}(x,y)$ of the potential terms---quantum confinement profile,
 electric field, spin-orbit coupling terms, etc---and $f(x,y)$ is the envelope function.

Among the several ways that exists to solve such type of coupled differential
 equations, consider the plane wave expansion to the envelope functions
\begin{equation}
f(x,y)=\underset{K_{x},K_{y}}{\sum}e^{i\left(K_{x}x+K_{y}y\right)}\tilde{f}(K_{x},K_{y})\label{eq:fPWE}
\end{equation}

Carrying out the calculations with the above form of the envelope function, we
 can identify in the final equation of the Hamiltonian that the spatial
 dependent parameters and potentials (confinement and external electric field)
 can be written as~\cite{vukmirovc2008plane,ehrhardt2014multi,Budagosky2017}
\begin{equation}
\chi(x,y)=\underset{Q_{x},Q_{y}}{\sum}e^{i\left(Q_{x}x+Q_{y}y\right)}\tilde{\chi}(Q_{x},Q_{y})\label{eq:chiPWE}
\end{equation}
\\with the condition
\begin{equation}
Q_{\alpha}=K_{\alpha}-K_{\alpha}^{\prime},\;\alpha=x,y\label{eq:Q}
\end{equation}
\begin{equation}
\left\{ K_{\alpha},K_{\alpha}^{\prime}\right\} =j\frac{2\pi}{L_{\alpha}},\;j=0,\pm1,\pm2,\ldots\label{eq:KKprime}
\end{equation}

In a similar fashion that by performing the envelope function approximation
 we end up with a description of the spatial dependent functions and parameters
 in terms of derivatives ($k_{x(y)}\rightarrow-i\partial/\partial x(y)$), by performing
 the plane wave expansion can be summarized by the expansions given by Eqs. \ref{eq:fPWE} and \ref{eq:chiPWE}
 and the following substitutions to the k-vectors
\begin{eqnarray}
k_{\alpha} & \rightarrow & \frac{1}{2}\left(K_{\alpha}+K_{\alpha}^{\prime}\right)\nonumber \\
k_{\alpha}k_{\beta} & \rightarrow & \frac{1}{2}\left(K_{\alpha}K_{\beta}^{\prime}+K_{\beta}K_{\alpha}^{\prime}\right),\;\left\{ \alpha,\beta\right\} =x,y
\end{eqnarray}

From Eqs. \ref{eq:fPWE}-\ref{eq:KKprime} we notice that the number of
 coefficients of the parameters and potentials are bigger than the number
 of coefficients of the envelope function. For instance, considering 1
 plane wave for x and y directions we would have the set of $K_{x(y)}$
 and $K_{x(y)}^{\prime}$ vectors give by $\left\{ -1,0,1\right\} \times\frac{2\pi}{L_{x(y)}}$
 and consequently the set of $Q_{\alpha}$ vectors given by $\left\{ -2,-1,0,1,2\right\} \times\frac{2\pi}{L_{x(y)}}$,
 therefore leading to $3 \times 3$ coefficients for the wave functions and $5 \times 5$
 coefficients for the parameters and potentials. As a general rule,
 given a number of plane waves $N_{\textrm{pw}}$ for x and also y
 directions, the number of Fourier coefficients is $(2 \times N_{\textrm{pw}}+1)^{2}$
 for the wave functions and $(4 \times N_{\textrm{pw}}+1)^{2}$ for the parameters
 and potentials. The connection between the Fourier coefficients and the real
 space points is done by the Fourier transform routines.

In this paper we have used $20$ plane waves for x and y directions in
 a square grid for all simulations. This leads to $41 \times 41$ Fourier coefficients,
 or real space discretization values, for the wave functions and
 $81\times 81$ for the parameters and potentials. This value was sufficient
 to achieve energy convergence in our tests. In Fig. \ref{fig:PWE} we show
 the example of a WZ InAs nanowire along {[}0001{]} direction with $L=30$ nm.
 In Fig. \ref{fig:PWE}(a) we show the hexagonal confinement profile with
 each vertex of the square grid representing one of the $81 \times 81$
 discretization points. Similarly, in Fig. \ref{fig:PWE}(b) we show the
 $41 \times 41$ square grid discretization for the probability density at $k_{z}=0$ for
 the first conduction subband. The nanowire itself has $61$ discretization
 points along the diameter (distance between opposite vertices in hexagonal
 nanowires), with at least $10$ discretization points in the surrounding
 vacuum at each side along the line. In our simulations we always kept the
 ratio of points inside to points outside the nanowire constant.

Regarding the numerical calculations, we performed the diagonalization of
 the final Hamiltonian using the MAGMA~\cite{dghklty14} suite which implements
 the LAPACK routines in a multicore + GPU (graphical processing unit)
 computational environment. The numerical precision of the calculations
 is guaranteed up to single precision which translates to energies on the
 order of $10^{-6}$ eV, any value below this number was regarded as zero.

\section*{Appendix B}

%-------------------------------------
\begin{figure}[ht]
\begin{center}
\includegraphics[width=.45\textwidth]{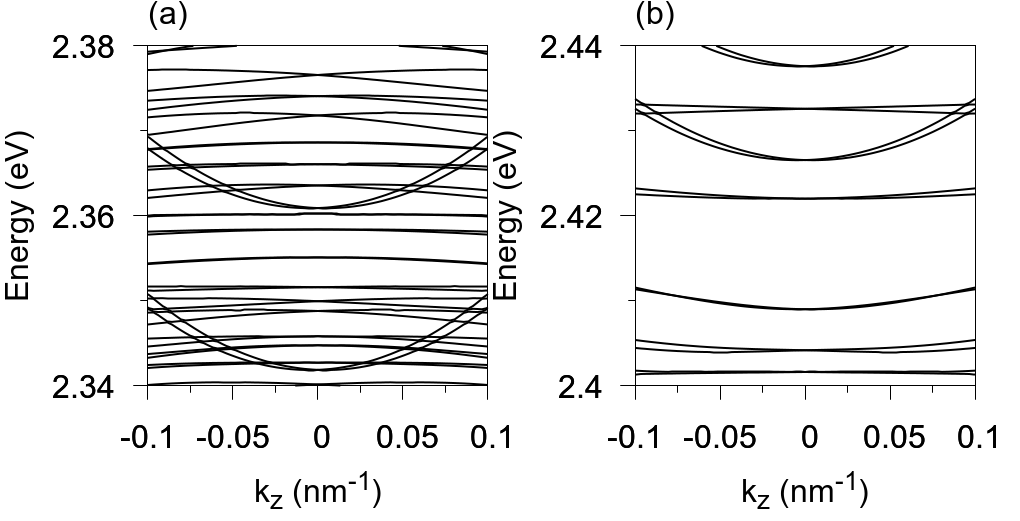}
\caption{Conduction and valence subbands crossing for a WZ InAs nanowire with
          $L = 60\,\rm nm$ due to high quantum confinement induced
          by electric field of $E = 16\,\textrm{mV/nm}$. In (a) we focused on
          showing the first two conduction subbands embedded in the valence
          subbands sea, whereas in (b) we show the first few valence subbands.}
\label{fig:bandcrossField}
\end{center}
\end{figure}
%-------------------------------------

In Fig. \ref{fig:bandcrossField} we show the band structure of a WZ InAs
 nanowire with $L = 60\,\rm nm$, with an applied electric field of
 $E = 16\,\textrm{mV/nm}$. The quantum confinement induced by the electric
 field is large enough to cause the conduction and valence
 subbands to cross. In this situation it is difficult to isolate the desired
 subband to apply the fitting method.

\section*{Appendix C}

%-------------------------------------
\begin{figure}[ht!]
\begin{center}
\includegraphics[width=.45\textwidth]{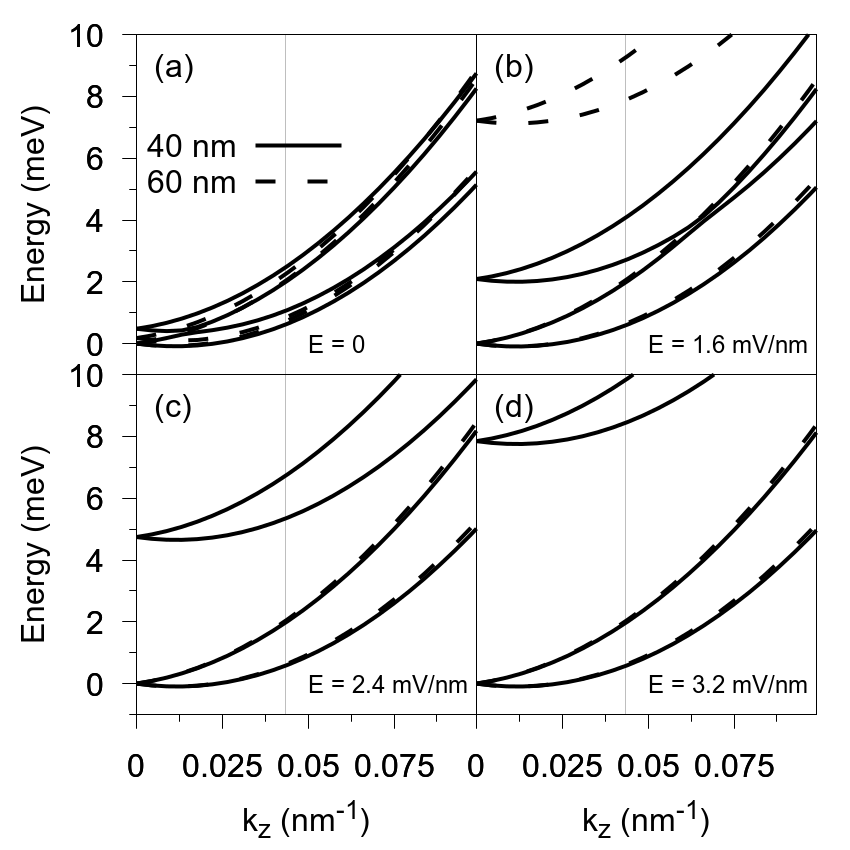}
\caption{(a) 2nd and 3rd conduction subband of a WZ InAs nanowire with 40 and 60 nm
         in diameter without applied electric field. (b) Same as (a) but with electric field
         of $E = 1.6\,\textrm{mV/nm}$. (c) Same as (a) but with electric field
         of $E = 2.4\,\textrm{mV/nm}$. (d) Same as (a) but with electric field
         of $E = 3.2\,\textrm{mV/nm}$. The vertical lines correspond  indicate
         1\%  of the Brillouin zone.}
\label{fig:bandcross}
\end{center}
\end{figure}
%-------------------------------------

What about higher conduction bands? In Fig. \ref{fig:bandcross} we show a
 band crossing evolution as a function
 of the applied electric field for two WZ InAs nanowires, of diameter 40 and 60 nm.
  Without an applied electric
 field the spin-split bands cross each other making it unrealistic to describe
 structure with a simple spin-half model. As we turn on and increase the magnitude
 of the electric field, the orbital quasi-degeneracy vanishes and the crossing
 point extends further away from the $\Gamma$-point. For this specific configuration,
 with an applied electric field of $E = 1.6\,\textrm{mV/nm}$ the band crossing
 occurs outside our the fitting range, therefore, making it possible, in principle,
 to apply the single-band model for spin-half electrons.
 However, due to quantum confinement effects, the crossing point
 shift is not the same for all nanowires crystal phases, neither applied electric
 field. Therefore, we  choose to not apply the fitting to higher excited
 conduction bands. A better approach would be to develop an effective Hamiltonian that
 takes into account all the desired bands in the desired range and fit the
 Hamiltonian itself rather than the energy dispersion. The disadvantage of such an
 approach is the loss of simplicity.

\bibliography{references}

\begin{thebibliography}{130}
\expandafter\ifx\csname natexlab\endcsname\relax\def\natexlab#1{#1}\fi
\expandafter\ifx\csname bibnamefont\endcsname\relax
  \def\bibnamefont#1{#1}\fi
\expandafter\ifx\csname bibfnamefont\endcsname\relax
  \def\bibfnamefont#1{#1}\fi
\expandafter\ifx\csname citenamefont\endcsname\relax
  \def\citenamefont#1{#1}\fi
\expandafter\ifx\csname url\endcsname\relax
  \def\url#1{\texttt{#1}}\fi
\expandafter\ifx\csname urlprefix\endcsname\relax\def\urlprefix{URL }\fi
\providecommand{\bibinfo}[2]{#2}
\providecommand{\eprint}[2][]{\url{#2}}

\bibitem[{\citenamefont{Zutic et~al.}(2004)\citenamefont{Zutic, Fabian, and
  Das~Sarma}}]{Zutic2004:RMP}
\bibinfo{author}{\bibfnamefont{I.}~\bibnamefont{Zutic}},
  \bibinfo{author}{\bibfnamefont{J.}~\bibnamefont{Fabian}}, \bibnamefont{and}
  \bibinfo{author}{\bibfnamefont{S.}~\bibnamefont{Das~Sarma}},
  \bibinfo{journal}{Reviews of Modern Physics} \textbf{\bibinfo{volume}{76}},
  \bibinfo{pages}{323} (\bibinfo{year}{2004}).

\bibitem[{\citenamefont{Fabian et~al.}(2007)\citenamefont{Fabian,
  Matos-Abiague, Ertler, Stano, and Zutic}}]{Fabian2007}
\bibinfo{author}{\bibfnamefont{J.}~\bibnamefont{Fabian}},
  \bibinfo{author}{\bibfnamefont{A.}~\bibnamefont{Matos-Abiague}},
  \bibinfo{author}{\bibfnamefont{C.}~\bibnamefont{Ertler}},
  \bibinfo{author}{\bibfnamefont{P.}~\bibnamefont{Stano}}, \bibnamefont{and}
  \bibinfo{author}{\bibfnamefont{I.}~\bibnamefont{Zutic}},
  \bibinfo{journal}{Acta Physica Slovaca} \textbf{\bibinfo{volume}{57}},
  \bibinfo{pages}{565} (\bibinfo{year}{2007}).

\bibitem[{\citenamefont{Ivchenko and Tarasenko}(2008)}]{Ivchenko2008}
\bibinfo{author}{\bibfnamefont{E.~L.} \bibnamefont{Ivchenko}} \bibnamefont{and}
  \bibinfo{author}{\bibfnamefont{S.~A.} \bibnamefont{Tarasenko}},
  \bibinfo{journal}{Semiconductor Science and Technology}
  \textbf{\bibinfo{volume}{23}}, \bibinfo{pages}{114007}
  (\bibinfo{year}{2008}).

\bibitem[{\citenamefont{Chen et~al.}(2014)\citenamefont{Chen, Wong, Chang,
  Dong, and Chen}}]{Chen2014}
\bibinfo{author}{\bibfnamefont{J.-Y.} \bibnamefont{Chen}},
  \bibinfo{author}{\bibfnamefont{T.-m.} \bibnamefont{Wong}},
  \bibinfo{author}{\bibfnamefont{C.-w.} \bibnamefont{Chang}},
  \bibinfo{author}{\bibfnamefont{C.-y.} \bibnamefont{Dong}}, \bibnamefont{and}
  \bibinfo{author}{\bibfnamefont{Y.-f.} \bibnamefont{Chen}},
  \bibinfo{journal}{Nature Nanotechnology} \textbf{\bibinfo{volume}{9}},
  \bibinfo{pages}{845} (\bibinfo{year}{2014}).

\bibitem[{\citenamefont{Zutic and {Faria Junior}}(2014)}]{Zutic2014}
\bibinfo{author}{\bibfnamefont{I.}~\bibnamefont{Zutic}} \bibnamefont{and}
  \bibinfo{author}{\bibfnamefont{P.~E.} \bibnamefont{{Faria Junior}}},
  \bibinfo{journal}{Nature Nanotechnology} \textbf{\bibinfo{volume}{9}},
  \bibinfo{pages}{750} (\bibinfo{year}{2014}).

\bibitem[{\citenamefont{Faria~Junior et~al.}(2015)\citenamefont{Faria~Junior,
  Xu, Lee, Gerhardt, Sipahi, and Zutic}}]{FariaJunior2015}
\bibinfo{author}{\bibfnamefont{P.~E.} \bibnamefont{Faria~Junior}},
  \bibinfo{author}{\bibfnamefont{G.}~\bibnamefont{Xu}},
  \bibinfo{author}{\bibfnamefont{J.}~\bibnamefont{Lee}},
  \bibinfo{author}{\bibfnamefont{N.~C.} \bibnamefont{Gerhardt}},
  \bibinfo{author}{\bibfnamefont{G.~M.} \bibnamefont{Sipahi}},
  \bibnamefont{and} \bibinfo{author}{\bibfnamefont{I.}~\bibnamefont{Zutic}},
  \bibinfo{journal}{Physical Review B} \textbf{\bibinfo{volume}{92}},
  \bibinfo{pages}{075311} (\bibinfo{year}{2015}).

\bibitem[{\citenamefont{{Faria Junior} et~al.}(2017)\citenamefont{{Faria
  Junior}, Xu, Chen, Sipahi, and Zutic}}]{FariaJunior2017}
\bibinfo{author}{\bibfnamefont{P.~E.} \bibnamefont{{Faria Junior}}},
  \bibinfo{author}{\bibfnamefont{G.}~\bibnamefont{Xu}},
  \bibinfo{author}{\bibfnamefont{Y.-F.} \bibnamefont{Chen}},
  \bibinfo{author}{\bibfnamefont{G.~M.} \bibnamefont{Sipahi}},
  \bibnamefont{and} \bibinfo{author}{\bibfnamefont{I.}~\bibnamefont{Zutic}},
  \bibinfo{journal}{Physical Review B} \textbf{\bibinfo{volume}{95}},
  \bibinfo{pages}{115301} (\bibinfo{year}{2017}).

\bibitem[{\citenamefont{Bernevig et~al.}(2006)\citenamefont{Bernevig, Hughes,
  and Zhang}}]{Bernevig2006}
\bibinfo{author}{\bibfnamefont{B.~A.} \bibnamefont{Bernevig}},
  \bibinfo{author}{\bibfnamefont{T.~L.} \bibnamefont{Hughes}},
  \bibnamefont{and} \bibinfo{author}{\bibfnamefont{S.-C.} \bibnamefont{Zhang}},
  \bibinfo{journal}{Science} \textbf{\bibinfo{volume}{314}},
  \bibinfo{pages}{1757} (\bibinfo{year}{2006}).

\bibitem[{\citenamefont{Oreg et~al.}(2010)\citenamefont{Oreg, Refael, and von
  Oppen}}]{Oreg2010:PRL}
\bibinfo{author}{\bibfnamefont{Y.}~\bibnamefont{Oreg}},
  \bibinfo{author}{\bibfnamefont{G.}~\bibnamefont{Refael}}, \bibnamefont{and}
  \bibinfo{author}{\bibfnamefont{F.}~\bibnamefont{von Oppen}},
  \bibinfo{journal}{Physical Review Letters} \textbf{\bibinfo{volume}{105}},
  \bibinfo{pages}{177002} (\bibinfo{year}{2010}).

\bibitem[{\citenamefont{Kloeffel et~al.}(2011)\citenamefont{Kloeffel, Trif, and
  Loss}}]{Kloeffel2011}
\bibinfo{author}{\bibfnamefont{C.}~\bibnamefont{Kloeffel}},
  \bibinfo{author}{\bibfnamefont{M.}~\bibnamefont{Trif}}, \bibnamefont{and}
  \bibinfo{author}{\bibfnamefont{D.}~\bibnamefont{Loss}},
  \bibinfo{journal}{Physical Review B} \textbf{\bibinfo{volume}{84}},
  \bibinfo{pages}{195314} (\bibinfo{year}{2011}).

\bibitem[{\citenamefont{Oreg et~al.}(2014)\citenamefont{Oreg, Sela, and
  Stern}}]{Oreg2014}
\bibinfo{author}{\bibfnamefont{Y.}~\bibnamefont{Oreg}},
  \bibinfo{author}{\bibfnamefont{E.}~\bibnamefont{Sela}}, \bibnamefont{and}
  \bibinfo{author}{\bibfnamefont{A.}~\bibnamefont{Stern}},
  \bibinfo{journal}{Physical Review B} \textbf{\bibinfo{volume}{89}},
  \bibinfo{pages}{115402} (\bibinfo{year}{2014}).

\bibitem[{\citenamefont{Schmidt and Pedder}(2016)}]{Schmidt2016}
\bibinfo{author}{\bibfnamefont{T.~L.} \bibnamefont{Schmidt}} \bibnamefont{and}
  \bibinfo{author}{\bibfnamefont{C.~J.} \bibnamefont{Pedder}},
  \bibinfo{journal}{Physical Review B} \textbf{\bibinfo{volume}{94}},
  \bibinfo{pages}{125420} (\bibinfo{year}{2016}).

\bibitem[{\citenamefont{Lutchyn et~al.}(2010)\citenamefont{Lutchyn, Sau, and
  Das~Sarma}}]{Lutchyn2010:PRL}
\bibinfo{author}{\bibfnamefont{R.~M.} \bibnamefont{Lutchyn}},
  \bibinfo{author}{\bibfnamefont{J.~D.} \bibnamefont{Sau}}, \bibnamefont{and}
  \bibinfo{author}{\bibfnamefont{S.}~\bibnamefont{Das~Sarma}},
  \bibinfo{journal}{Physical Review Letters} \textbf{\bibinfo{volume}{105}},
  \bibinfo{pages}{077001} (\bibinfo{year}{2010}).

\bibitem[{\citenamefont{Mourik et~al.}(2012)\citenamefont{Mourik, Zuo, Frolov,
  Plissard, Bakkers, and Kouwenhoven}}]{Mourik2012:Science}
\bibinfo{author}{\bibfnamefont{V.}~\bibnamefont{Mourik}},
  \bibinfo{author}{\bibfnamefont{K.}~\bibnamefont{Zuo}},
  \bibinfo{author}{\bibfnamefont{S.~M.} \bibnamefont{Frolov}},
  \bibinfo{author}{\bibfnamefont{S.~R.} \bibnamefont{Plissard}},
  \bibinfo{author}{\bibfnamefont{E.~P. A.~M.} \bibnamefont{Bakkers}},
  \bibnamefont{and} \bibinfo{author}{\bibfnamefont{L.~P.}
  \bibnamefont{Kouwenhoven}}, \bibinfo{journal}{Science}
  \textbf{\bibinfo{volume}{336}}, \bibinfo{pages}{1003} (\bibinfo{year}{2012}).

\bibitem[{\citenamefont{Deng et~al.}(2012)\citenamefont{Deng, Yu, Huang,
  Larsson, Caroff, and Xu}}]{Deng2012:NL}
\bibinfo{author}{\bibfnamefont{M.~T.} \bibnamefont{Deng}},
  \bibinfo{author}{\bibfnamefont{C.~L.} \bibnamefont{Yu}},
  \bibinfo{author}{\bibfnamefont{G.~Y.} \bibnamefont{Huang}},
  \bibinfo{author}{\bibfnamefont{M.}~\bibnamefont{Larsson}},
  \bibinfo{author}{\bibfnamefont{P.}~\bibnamefont{Caroff}}, \bibnamefont{and}
  \bibinfo{author}{\bibfnamefont{H.~Q.} \bibnamefont{Xu}},
  \bibinfo{journal}{Nano Letters} \textbf{\bibinfo{volume}{12}},
  \bibinfo{pages}{6414} (\bibinfo{year}{2012}).

\bibitem[{\citenamefont{Das et~al.}(2012)\citenamefont{Das, Ronen, Most, Oreg,
  Heiblum, and Shtrikman}}]{Das2012:NatPhys}
\bibinfo{author}{\bibfnamefont{A.}~\bibnamefont{Das}},
  \bibinfo{author}{\bibfnamefont{Y.}~\bibnamefont{Ronen}},
  \bibinfo{author}{\bibfnamefont{Y.}~\bibnamefont{Most}},
  \bibinfo{author}{\bibfnamefont{Y.}~\bibnamefont{Oreg}},
  \bibinfo{author}{\bibfnamefont{M.}~\bibnamefont{Heiblum}}, \bibnamefont{and}
  \bibinfo{author}{\bibfnamefont{H.}~\bibnamefont{Shtrikman}},
  \bibinfo{journal}{Nature Physics} \textbf{\bibinfo{volume}{8}},
  \bibinfo{pages}{887} (\bibinfo{year}{2012}).

\bibitem[{\citenamefont{Albrecht et~al.}(2016)\citenamefont{Albrecht,
  Higginbotham, Madsen, Kuemmeth, Jespersen, Nyg{\aa}rd, Krogstrup, and
  Marcus}}]{albrecht2016exponential}
\bibinfo{author}{\bibfnamefont{S.}~\bibnamefont{Albrecht}},
  \bibinfo{author}{\bibfnamefont{A.}~\bibnamefont{Higginbotham}},
  \bibinfo{author}{\bibfnamefont{M.}~\bibnamefont{Madsen}},
  \bibinfo{author}{\bibfnamefont{F.}~\bibnamefont{Kuemmeth}},
  \bibinfo{author}{\bibfnamefont{T.}~\bibnamefont{Jespersen}},
  \bibinfo{author}{\bibfnamefont{J.}~\bibnamefont{Nyg{\aa}rd}},
  \bibinfo{author}{\bibfnamefont{P.}~\bibnamefont{Krogstrup}},
  \bibnamefont{and} \bibinfo{author}{\bibfnamefont{C.}~\bibnamefont{Marcus}},
  \bibinfo{journal}{Nature} \textbf{\bibinfo{volume}{531}},
  \bibinfo{pages}{206} (\bibinfo{year}{2016}).

\bibitem[{\citenamefont{Deng et~al.}(2016)\citenamefont{Deng, Vaitiekenas,
  Hansen, Danon, Leijnse, Flensberg, Nyg{\r a}rd, Krogstrup, and
  Marcus}}]{Deng1557}
\bibinfo{author}{\bibfnamefont{M.~T.} \bibnamefont{Deng}},
  \bibinfo{author}{\bibfnamefont{S.}~\bibnamefont{Vaitiekenas}},
  \bibinfo{author}{\bibfnamefont{E.~B.} \bibnamefont{Hansen}},
  \bibinfo{author}{\bibfnamefont{J.}~\bibnamefont{Danon}},
  \bibinfo{author}{\bibfnamefont{M.}~\bibnamefont{Leijnse}},
  \bibinfo{author}{\bibfnamefont{K.}~\bibnamefont{Flensberg}},
  \bibinfo{author}{\bibfnamefont{J.}~\bibnamefont{Nyg{\r a}rd}},
  \bibinfo{author}{\bibfnamefont{P.}~\bibnamefont{Krogstrup}},
  \bibnamefont{and} \bibinfo{author}{\bibfnamefont{C.~M.}
  \bibnamefont{Marcus}}, \bibinfo{journal}{Science}
  \textbf{\bibinfo{volume}{354}}, \bibinfo{pages}{1557} (\bibinfo{year}{2016}).

\bibitem[{\citenamefont{Zhang et~al.}(2018)\citenamefont{Zhang, Liu,
  Gazibegovic, Xu, Logan, Wang, van Loo, Bommer, de~Moor, Car
  et~al.}}]{Zhang2018}
\bibinfo{author}{\bibfnamefont{H.}~\bibnamefont{Zhang}},
  \bibinfo{author}{\bibfnamefont{C.-X.} \bibnamefont{Liu}},
  \bibinfo{author}{\bibfnamefont{S.}~\bibnamefont{Gazibegovic}},
  \bibinfo{author}{\bibfnamefont{D.}~\bibnamefont{Xu}},
  \bibinfo{author}{\bibfnamefont{J.~A.} \bibnamefont{Logan}},
  \bibinfo{author}{\bibfnamefont{G.}~\bibnamefont{Wang}},
  \bibinfo{author}{\bibfnamefont{N.}~\bibnamefont{van Loo}},
  \bibinfo{author}{\bibfnamefont{J.~D.~S.} \bibnamefont{Bommer}},
  \bibinfo{author}{\bibfnamefont{M.~W.~A.} \bibnamefont{de~Moor}},
  \bibinfo{author}{\bibfnamefont{D.}~\bibnamefont{Car}}, \bibnamefont{et~al.},
  \bibinfo{journal}{Nature} \textbf{\bibinfo{volume}{556}}, \bibinfo{pages}{74}
  (\bibinfo{year}{2018}).

\bibitem[{\citenamefont{Dresselhaus}(1955)}]{Dresselhaus1955}
\bibinfo{author}{\bibfnamefont{G.}~\bibnamefont{Dresselhaus}},
  \bibinfo{journal}{Physical Review} \textbf{\bibinfo{volume}{100}},
  \bibinfo{pages}{580} (\bibinfo{year}{1955}).

\bibitem[{\citenamefont{Bychkov and Rashba}(1984)}]{Bychkov1984a}
\bibinfo{author}{\bibfnamefont{Y.~A.} \bibnamefont{Bychkov}} \bibnamefont{and}
  \bibinfo{author}{\bibfnamefont{E.}~\bibnamefont{Rashba}},
  \bibinfo{journal}{JETP Letters} \textbf{\bibinfo{volume}{39}},
  \bibinfo{pages}{78} (\bibinfo{year}{1984}).

\bibitem[{\citenamefont{Koralek et~al.}(2009)\citenamefont{Koralek, Weber,
  Orenstein, Bernevig, Zhang, Mack, and Awschalom}}]{Koralek2009}
\bibinfo{author}{\bibfnamefont{J.~D.} \bibnamefont{Koralek}},
  \bibinfo{author}{\bibfnamefont{C.}~\bibnamefont{Weber}},
  \bibinfo{author}{\bibfnamefont{J.}~\bibnamefont{Orenstein}},
  \bibinfo{author}{\bibfnamefont{B.}~\bibnamefont{Bernevig}},
  \bibinfo{author}{\bibfnamefont{S.-C.} \bibnamefont{Zhang}},
  \bibinfo{author}{\bibfnamefont{S.}~\bibnamefont{Mack}}, \bibnamefont{and}
  \bibinfo{author}{\bibfnamefont{D.}~\bibnamefont{Awschalom}},
  \bibinfo{journal}{Nature} \textbf{\bibinfo{volume}{458}},
  \bibinfo{pages}{610} (\bibinfo{year}{2009}).

\bibitem[{\citenamefont{Fu et~al.}(2016)\citenamefont{Fu, Penteado, Hachiya,
  Loss, and Egues}}]{Fu2016}
\bibinfo{author}{\bibfnamefont{J.}~\bibnamefont{Fu}},
  \bibinfo{author}{\bibfnamefont{P.~H.} \bibnamefont{Penteado}},
  \bibinfo{author}{\bibfnamefont{M.~O.} \bibnamefont{Hachiya}},
  \bibinfo{author}{\bibfnamefont{D.}~\bibnamefont{Loss}}, \bibnamefont{and}
  \bibinfo{author}{\bibfnamefont{J.~C.} \bibnamefont{Egues}},
  \bibinfo{journal}{Physical Review Letters} \textbf{\bibinfo{volume}{117}},
  \bibinfo{pages}{226401} (\bibinfo{year}{2016}).

\bibitem[{\citenamefont{Schliemann}(2017)}]{Schliemann2017:RMP}
\bibinfo{author}{\bibfnamefont{J.}~\bibnamefont{Schliemann}},
  \bibinfo{journal}{Reviews of Modern Physics} \textbf{\bibinfo{volume}{89}},
  \bibinfo{pages}{011001} (\bibinfo{year}{2017}).

\bibitem[{\citenamefont{Schliemann et~al.}(2003)\citenamefont{Schliemann,
  Egues, and Loss}}]{Schliemann2003}
\bibinfo{author}{\bibfnamefont{J.}~\bibnamefont{Schliemann}},
  \bibinfo{author}{\bibfnamefont{J.~C.} \bibnamefont{Egues}}, \bibnamefont{and}
  \bibinfo{author}{\bibfnamefont{D.}~\bibnamefont{Loss}},
  \bibinfo{journal}{Physical Review Letters} \textbf{\bibinfo{volume}{90}},
  \bibinfo{pages}{146801} (\bibinfo{year}{2003}).

\bibitem[{\citenamefont{Erlingsson et~al.}(2010)\citenamefont{Erlingsson,
  Egues, and Loss}}]{Erlingsson2010}
\bibinfo{author}{\bibfnamefont{S.~I.} \bibnamefont{Erlingsson}},
  \bibinfo{author}{\bibfnamefont{J.~C.} \bibnamefont{Egues}}, \bibnamefont{and}
  \bibinfo{author}{\bibfnamefont{D.}~\bibnamefont{Loss}},
  \bibinfo{journal}{Physical Review B} \textbf{\bibinfo{volume}{82}},
  \bibinfo{pages}{155456} (\bibinfo{year}{2010}).

\bibitem[{\citenamefont{Toloza~Sandoval
  et~al.}(2013)\citenamefont{Toloza~Sandoval, Ferreira~da Silva, de~Andrada~e
  Silva, and La~Rocca}}]{Sandoval2013}
\bibinfo{author}{\bibfnamefont{M.~A.} \bibnamefont{Toloza~Sandoval}},
  \bibinfo{author}{\bibfnamefont{A.}~\bibnamefont{Ferreira~da Silva}},
  \bibinfo{author}{\bibfnamefont{E.~A.} \bibnamefont{de~Andrada~e Silva}},
  \bibnamefont{and} \bibinfo{author}{\bibfnamefont{G.~C.}
  \bibnamefont{La~Rocca}}, \bibinfo{journal}{Physical Review B}
  \textbf{\bibinfo{volume}{87}}, \bibinfo{pages}{081304}
  (\bibinfo{year}{2013}).

\bibitem[{\citenamefont{Toloza~Sandoval
  et~al.}(2016)\citenamefont{Toloza~Sandoval, de~Andrada~e Silva, Ferreira~da
  Silva, and La~Rocca}}]{sandoval2016electron}
\bibinfo{author}{\bibfnamefont{M.~A.} \bibnamefont{Toloza~Sandoval}},
  \bibinfo{author}{\bibfnamefont{E.~A.} \bibnamefont{de~Andrada~e Silva}},
  \bibinfo{author}{\bibfnamefont{A.}~\bibnamefont{Ferreira~da Silva}},
  \bibnamefont{and} \bibinfo{author}{\bibfnamefont{G.~C.}
  \bibnamefont{La~Rocca}}, \bibinfo{journal}{Semiconductor Science and
  Technology} \textbf{\bibinfo{volume}{31}}, \bibinfo{pages}{115008}
  (\bibinfo{year}{2016}).

\bibitem[{\citenamefont{Furthmeier et~al.}(2016)\citenamefont{Furthmeier,
  Dirnberger, Gmitra, Bayer, Forsch, Hubmann, Sch{\"{u}}ller, Reiger, Fabian,
  Korn et~al.}}]{Furthmeier2016}
\bibinfo{author}{\bibfnamefont{S.}~\bibnamefont{Furthmeier}},
  \bibinfo{author}{\bibfnamefont{F.}~\bibnamefont{Dirnberger}},
  \bibinfo{author}{\bibfnamefont{M.}~\bibnamefont{Gmitra}},
  \bibinfo{author}{\bibfnamefont{A.}~\bibnamefont{Bayer}},
  \bibinfo{author}{\bibfnamefont{M.}~\bibnamefont{Forsch}},
  \bibinfo{author}{\bibfnamefont{J.}~\bibnamefont{Hubmann}},
  \bibinfo{author}{\bibfnamefont{C.}~\bibnamefont{Sch{\"{u}}ller}},
  \bibinfo{author}{\bibfnamefont{E.}~\bibnamefont{Reiger}},
  \bibinfo{author}{\bibfnamefont{J.}~\bibnamefont{Fabian}},
  \bibinfo{author}{\bibfnamefont{T.}~\bibnamefont{Korn}}, \bibnamefont{et~al.},
  \bibinfo{journal}{Nature Communications} \textbf{\bibinfo{volume}{7}},
  \bibinfo{pages}{12413} (\bibinfo{year}{2016}).

\bibitem[{\citenamefont{{Kammermeier} et~al.}(2018)\citenamefont{{Kammermeier},
  {Wenk}, {Dirnberger}, {Bougeard}, and {Schliemann}}}]{Kammermeier2018}
\bibinfo{author}{\bibfnamefont{M.}~\bibnamefont{{Kammermeier}}},
  \bibinfo{author}{\bibfnamefont{P.}~\bibnamefont{{Wenk}}},
  \bibinfo{author}{\bibfnamefont{F.}~\bibnamefont{{Dirnberger}}},
  \bibinfo{author}{\bibfnamefont{D.}~\bibnamefont{{Bougeard}}},
  \bibnamefont{and}
  \bibinfo{author}{\bibfnamefont{J.}~\bibnamefont{{Schliemann}}},
  \bibinfo{journal}{ArXiv e-prints}  (\bibinfo{year}{2018}),
  \eprint{1804.00148}.

\bibitem[{\citenamefont{Rainis and Loss}(2014)}]{Rainis2014}
\bibinfo{author}{\bibfnamefont{D.}~\bibnamefont{Rainis}} \bibnamefont{and}
  \bibinfo{author}{\bibfnamefont{D.}~\bibnamefont{Loss}},
  \bibinfo{journal}{Physical Review B} \textbf{\bibinfo{volume}{90}},
  \bibinfo{pages}{235415} (\bibinfo{year}{2014}).

\bibitem[{\citenamefont{Zawadzki and Pfeffer}(2004)}]{Zawadzki2004}
\bibinfo{author}{\bibfnamefont{W.}~\bibnamefont{Zawadzki}} \bibnamefont{and}
  \bibinfo{author}{\bibfnamefont{P.}~\bibnamefont{Pfeffer}},
  \bibinfo{journal}{Semiconductor Science and Technology}
  \textbf{\bibinfo{volume}{19}}, \bibinfo{pages}{R1} (\bibinfo{year}{2004}).

\bibitem[{\citenamefont{Thorgilsson et~al.}(2012)\citenamefont{Thorgilsson,
  Egues, Loss, and Erlingsson}}]{Thorgilsson2012}
\bibinfo{author}{\bibfnamefont{G.}~\bibnamefont{Thorgilsson}},
  \bibinfo{author}{\bibfnamefont{J.~C.} \bibnamefont{Egues}},
  \bibinfo{author}{\bibfnamefont{D.}~\bibnamefont{Loss}}, \bibnamefont{and}
  \bibinfo{author}{\bibfnamefont{S.~I.} \bibnamefont{Erlingsson}},
  \bibinfo{journal}{Physical Review B} \textbf{\bibinfo{volume}{85}},
  \bibinfo{pages}{045306} (\bibinfo{year}{2012}).

\bibitem[{\citenamefont{Nilsson et~al.}(2009)\citenamefont{Nilsson, Caroff,
  Thelander, Larsson, Wagner, Wernersson, Samuelson, and Xu}}]{Nilsson2009}
\bibinfo{author}{\bibfnamefont{H.~A.} \bibnamefont{Nilsson}},
  \bibinfo{author}{\bibfnamefont{P.}~\bibnamefont{Caroff}},
  \bibinfo{author}{\bibfnamefont{C.}~\bibnamefont{Thelander}},
  \bibinfo{author}{\bibfnamefont{M.}~\bibnamefont{Larsson}},
  \bibinfo{author}{\bibfnamefont{J.~B.} \bibnamefont{Wagner}},
  \bibinfo{author}{\bibfnamefont{L.-E.} \bibnamefont{Wernersson}},
  \bibinfo{author}{\bibfnamefont{L.}~\bibnamefont{Samuelson}},
  \bibnamefont{and} \bibinfo{author}{\bibfnamefont{H.~Q.} \bibnamefont{Xu}},
  \bibinfo{journal}{Nano Letters} \textbf{\bibinfo{volume}{9}},
  \bibinfo{pages}{3151} (\bibinfo{year}{2009}).

\bibitem[{\citenamefont{Nowak and Szafran}(2013)}]{Nowak2013}
\bibinfo{author}{\bibfnamefont{M.~P.} \bibnamefont{Nowak}} \bibnamefont{and}
  \bibinfo{author}{\bibfnamefont{B.}~\bibnamefont{Szafran}},
  \bibinfo{journal}{Physical Review B} \textbf{\bibinfo{volume}{87}},
  \bibinfo{pages}{205436} (\bibinfo{year}{2013}).

\bibitem[{\citenamefont{van Weperen et~al.}(2015)\citenamefont{van Weperen,
  Tarasinski, Eeltink, Pribiag, Plissard, Bakkers, Kouwenhoven, and
  Wimmer}}]{VanWeperen2015}
\bibinfo{author}{\bibfnamefont{I.}~\bibnamefont{van Weperen}},
  \bibinfo{author}{\bibfnamefont{B.}~\bibnamefont{Tarasinski}},
  \bibinfo{author}{\bibfnamefont{D.}~\bibnamefont{Eeltink}},
  \bibinfo{author}{\bibfnamefont{V.~S.} \bibnamefont{Pribiag}},
  \bibinfo{author}{\bibfnamefont{S.~R.} \bibnamefont{Plissard}},
  \bibinfo{author}{\bibfnamefont{E.~P. a.~M.} \bibnamefont{Bakkers}},
  \bibinfo{author}{\bibfnamefont{L.~P.} \bibnamefont{Kouwenhoven}},
  \bibnamefont{and} \bibinfo{author}{\bibfnamefont{M.}~\bibnamefont{Wimmer}},
  \bibinfo{journal}{Physical Review B} \textbf{\bibinfo{volume}{91}},
  \bibinfo{pages}{201413} (\bibinfo{year}{2015}).

\bibitem[{\citenamefont{Scher\"ubl et~al.}(2016)\citenamefont{Scher\"ubl,
  F\"ul\"op, Madsen, Nyg\aa{}rd, and Csonka}}]{Scherubl2016}
\bibinfo{author}{\bibfnamefont{Z.}~\bibnamefont{Scher\"ubl}},
  \bibinfo{author}{\bibfnamefont{G.}~\bibnamefont{F\"ul\"op}},
  \bibinfo{author}{\bibfnamefont{M.~H.} \bibnamefont{Madsen}},
  \bibinfo{author}{\bibfnamefont{J.}~\bibnamefont{Nyg\aa{}rd}},
  \bibnamefont{and} \bibinfo{author}{\bibfnamefont{S.}~\bibnamefont{Csonka}},
  \bibinfo{journal}{Physical Review B} \textbf{\bibinfo{volume}{94}},
  \bibinfo{pages}{035444} (\bibinfo{year}{2016}).

\bibitem[{\citenamefont{Bernardes et~al.}(2007)\citenamefont{Bernardes,
  Schliemann, Lee, Egues, and Loss}}]{Bernardes2007}
\bibinfo{author}{\bibfnamefont{E.}~\bibnamefont{Bernardes}},
  \bibinfo{author}{\bibfnamefont{J.}~\bibnamefont{Schliemann}},
  \bibinfo{author}{\bibfnamefont{M.}~\bibnamefont{Lee}},
  \bibinfo{author}{\bibfnamefont{J.~C.} \bibnamefont{Egues}}, \bibnamefont{and}
  \bibinfo{author}{\bibfnamefont{D.}~\bibnamefont{Loss}},
  \bibinfo{journal}{Physical Review Letters} \textbf{\bibinfo{volume}{99}},
  \bibinfo{pages}{076603} (\bibinfo{year}{2007}).

\bibitem[{\citenamefont{Calsaverini et~al.}(2008)\citenamefont{Calsaverini,
  Bernardes, Egues, and Loss}}]{Calsaverini2008}
\bibinfo{author}{\bibfnamefont{R.~S.} \bibnamefont{Calsaverini}},
  \bibinfo{author}{\bibfnamefont{E.}~\bibnamefont{Bernardes}},
  \bibinfo{author}{\bibfnamefont{J.~C.} \bibnamefont{Egues}}, \bibnamefont{and}
  \bibinfo{author}{\bibfnamefont{D.}~\bibnamefont{Loss}},
  \bibinfo{journal}{Physical Review B} \textbf{\bibinfo{volume}{78}},
  \bibinfo{pages}{155313} (\bibinfo{year}{2008}).

\bibitem[{\citenamefont{Jespersen et~al.}(2018)\citenamefont{Jespersen,
  Krogstrup, Lunde, Tanta, Kanne, Johnson, and Nyg\aa{}rd}}]{Jespersen2018}
\bibinfo{author}{\bibfnamefont{T.~S.} \bibnamefont{Jespersen}},
  \bibinfo{author}{\bibfnamefont{P.}~\bibnamefont{Krogstrup}},
  \bibinfo{author}{\bibfnamefont{A.~M.} \bibnamefont{Lunde}},
  \bibinfo{author}{\bibfnamefont{R.}~\bibnamefont{Tanta}},
  \bibinfo{author}{\bibfnamefont{T.}~\bibnamefont{Kanne}},
  \bibinfo{author}{\bibfnamefont{E.}~\bibnamefont{Johnson}}, \bibnamefont{and}
  \bibinfo{author}{\bibfnamefont{J.}~\bibnamefont{Nyg\aa{}rd}},
  \bibinfo{journal}{Physical Review B} \textbf{\bibinfo{volume}{97}},
  \bibinfo{pages}{041303} (\bibinfo{year}{2018}).

\bibitem[{\citenamefont{{W{\'o}jcik} et~al.}(2018)\citenamefont{{W{\'o}jcik},
  {Bertoni}, and {Goldoni}}}]{2018arXiv180109905W}
\bibinfo{author}{\bibfnamefont{P.}~\bibnamefont{{W{\'o}jcik}}},
  \bibinfo{author}{\bibfnamefont{A.}~\bibnamefont{{Bertoni}}},
  \bibnamefont{and}
  \bibinfo{author}{\bibfnamefont{G.}~\bibnamefont{{Goldoni}}},
  \bibinfo{journal}{ArXiv e-prints}  (\bibinfo{year}{2018}),
  \eprint{1801.09905}.

\bibitem[{\citenamefont{Governale and Zülicke}(2004)}]{Governale2004}
\bibinfo{author}{\bibfnamefont{M.}~\bibnamefont{Governale}} \bibnamefont{and}
  \bibinfo{author}{\bibfnamefont{U.}~\bibnamefont{Zülicke}},
  \bibinfo{journal}{Solid State Communications} \textbf{\bibinfo{volume}{131}},
  \bibinfo{pages}{581} (\bibinfo{year}{2004}).

\bibitem[{\citenamefont{Mireles and Kirczenow}(2001)}]{Mireles2001}
\bibinfo{author}{\bibfnamefont{F.}~\bibnamefont{Mireles}} \bibnamefont{and}
  \bibinfo{author}{\bibfnamefont{G.}~\bibnamefont{Kirczenow}},
  \bibinfo{journal}{Physical Review B} \textbf{\bibinfo{volume}{64}},
  \bibinfo{pages}{024426} (\bibinfo{year}{2001}).

\bibitem[{\citenamefont{Scheid et~al.}(2008)\citenamefont{Scheid, Kohda,
  Kunihashi, Richter, and Nitta}}]{Scheid2008}
\bibinfo{author}{\bibfnamefont{M.}~\bibnamefont{Scheid}},
  \bibinfo{author}{\bibfnamefont{M.}~\bibnamefont{Kohda}},
  \bibinfo{author}{\bibfnamefont{Y.}~\bibnamefont{Kunihashi}},
  \bibinfo{author}{\bibfnamefont{K.}~\bibnamefont{Richter}}, \bibnamefont{and}
  \bibinfo{author}{\bibfnamefont{J.}~\bibnamefont{Nitta}},
  \bibinfo{journal}{Physical Review Letters} \textbf{\bibinfo{volume}{101}},
  \bibinfo{pages}{266401} (\bibinfo{year}{2008}).

\bibitem[{\citenamefont{Kammermeier et~al.}(2016)\citenamefont{Kammermeier,
  Wenk, Schliemann, Heedt, and Sch\"apers}}]{Kammermeier2016}
\bibinfo{author}{\bibfnamefont{M.}~\bibnamefont{Kammermeier}},
  \bibinfo{author}{\bibfnamefont{P.}~\bibnamefont{Wenk}},
  \bibinfo{author}{\bibfnamefont{J.}~\bibnamefont{Schliemann}},
  \bibinfo{author}{\bibfnamefont{S.}~\bibnamefont{Heedt}}, \bibnamefont{and}
  \bibinfo{author}{\bibfnamefont{T.}~\bibnamefont{Sch\"apers}},
  \bibinfo{journal}{Physical Review B} \textbf{\bibinfo{volume}{93}},
  \bibinfo{pages}{205306} (\bibinfo{year}{2016}).

\bibitem[{\citenamefont{Sau et~al.}(2010)\citenamefont{Sau, Lutchyn, Tewari,
  and Das~Sarma}}]{Sau2010:PRL}
\bibinfo{author}{\bibfnamefont{J.~D.} \bibnamefont{Sau}},
  \bibinfo{author}{\bibfnamefont{R.~M.} \bibnamefont{Lutchyn}},
  \bibinfo{author}{\bibfnamefont{S.}~\bibnamefont{Tewari}}, \bibnamefont{and}
  \bibinfo{author}{\bibfnamefont{S.}~\bibnamefont{Das~Sarma}},
  \bibinfo{journal}{Physical Review Letters} \textbf{\bibinfo{volume}{104}},
  \bibinfo{pages}{040502} (\bibinfo{year}{2010}).

\bibitem[{\citenamefont{Niquet et~al.}(2006)\citenamefont{Niquet, Lherbier,
  Quang, Fern\'andez-Serra, Blase, and Delerue}}]{Niquet2006}
\bibinfo{author}{\bibfnamefont{Y.~M.} \bibnamefont{Niquet}},
  \bibinfo{author}{\bibfnamefont{A.}~\bibnamefont{Lherbier}},
  \bibinfo{author}{\bibfnamefont{N.~H.} \bibnamefont{Quang}},
  \bibinfo{author}{\bibfnamefont{M.~V.} \bibnamefont{Fern\'andez-Serra}},
  \bibinfo{author}{\bibfnamefont{X.}~\bibnamefont{Blase}}, \bibnamefont{and}
  \bibinfo{author}{\bibfnamefont{C.}~\bibnamefont{Delerue}},
  \bibinfo{journal}{Physical Review B} \textbf{\bibinfo{volume}{73}},
  \bibinfo{pages}{165319} (\bibinfo{year}{2006}).

\bibitem[{\citenamefont{Liao et~al.}(2015)\citenamefont{Liao, Luo, Yang, Chen,
  and Xu}}]{Liao2015}
\bibinfo{author}{\bibfnamefont{G.}~\bibnamefont{Liao}},
  \bibinfo{author}{\bibfnamefont{N.}~\bibnamefont{Luo}},
  \bibinfo{author}{\bibfnamefont{Z.}~\bibnamefont{Yang}},
  \bibinfo{author}{\bibfnamefont{K.}~\bibnamefont{Chen}}, \bibnamefont{and}
  \bibinfo{author}{\bibfnamefont{H.~Q.} \bibnamefont{Xu}},
  \bibinfo{journal}{Journal of Applied Physics} \textbf{\bibinfo{volume}{118}},
  \bibinfo{eid}{094308} (\bibinfo{year}{2015}).

\bibitem[{\citenamefont{Luo et~al.}(2016)\citenamefont{Luo, Liao, and
  Xu}}]{Luo2016}
\bibinfo{author}{\bibfnamefont{N.}~\bibnamefont{Luo}},
  \bibinfo{author}{\bibfnamefont{G.}~\bibnamefont{Liao}}, \bibnamefont{and}
  \bibinfo{author}{\bibfnamefont{H.~Q.} \bibnamefont{Xu}},
  \bibinfo{journal}{AIP Advances} \textbf{\bibinfo{volume}{6}},
  \bibinfo{pages}{125109} (\bibinfo{year}{2016}).

\bibitem[{\citenamefont{Soluyanov et~al.}(2016)\citenamefont{Soluyanov, Gresch,
  Troyer, Lutchyn, Bauer, and Nayak}}]{Soluyanov2016}
\bibinfo{author}{\bibfnamefont{A.~A.} \bibnamefont{Soluyanov}},
  \bibinfo{author}{\bibfnamefont{D.}~\bibnamefont{Gresch}},
  \bibinfo{author}{\bibfnamefont{M.}~\bibnamefont{Troyer}},
  \bibinfo{author}{\bibfnamefont{R.~M.} \bibnamefont{Lutchyn}},
  \bibinfo{author}{\bibfnamefont{B.}~\bibnamefont{Bauer}}, \bibnamefont{and}
  \bibinfo{author}{\bibfnamefont{C.}~\bibnamefont{Nayak}},
  \bibinfo{journal}{Physical Review B} \textbf{\bibinfo{volume}{93}},
  \bibinfo{pages}{115317} (\bibinfo{year}{2016}).

\bibitem[{\citenamefont{Kammhuber et~al.}(2016)\citenamefont{Kammhuber,
  Cassidy, Zhang, G\"ul, Pei, de~Moor, Nijholt, Watanabe, Taniguchi, Car
  et~al.}}]{Kammhuber2016}
\bibinfo{author}{\bibfnamefont{J.}~\bibnamefont{Kammhuber}},
  \bibinfo{author}{\bibfnamefont{M.~C.} \bibnamefont{Cassidy}},
  \bibinfo{author}{\bibfnamefont{H.}~\bibnamefont{Zhang}},
  \bibinfo{author}{\bibfnamefont{O.}~\bibnamefont{G\"ul}},
  \bibinfo{author}{\bibfnamefont{F.}~\bibnamefont{Pei}},
  \bibinfo{author}{\bibfnamefont{M.~W.~A.} \bibnamefont{de~Moor}},
  \bibinfo{author}{\bibfnamefont{B.}~\bibnamefont{Nijholt}},
  \bibinfo{author}{\bibfnamefont{K.}~\bibnamefont{Watanabe}},
  \bibinfo{author}{\bibfnamefont{T.}~\bibnamefont{Taniguchi}},
  \bibinfo{author}{\bibfnamefont{D.}~\bibnamefont{Car}}, \bibnamefont{et~al.},
  \bibinfo{journal}{Nano Letters} \textbf{\bibinfo{volume}{16}},
  \bibinfo{pages}{3482} (\bibinfo{year}{2016}).

\bibitem[{\citenamefont{Marcellina et~al.}(2017)\citenamefont{Marcellina,
  Hamilton, Winkler, and Culcer}}]{Marcelina2017}
\bibinfo{author}{\bibfnamefont{E.}~\bibnamefont{Marcellina}},
  \bibinfo{author}{\bibfnamefont{A.~R.} \bibnamefont{Hamilton}},
  \bibinfo{author}{\bibfnamefont{R.}~\bibnamefont{Winkler}}, \bibnamefont{and}
  \bibinfo{author}{\bibfnamefont{D.}~\bibnamefont{Culcer}},
  \bibinfo{journal}{Physical Review B} \textbf{\bibinfo{volume}{95}},
  \bibinfo{pages}{075305} (\bibinfo{year}{2017}).

\bibitem[{\citenamefont{Winkler et~al.}(2017)\citenamefont{Winkler, Varjas,
  Skolasinski, Soluyanov, Troyer, and Wimmer}}]{Winkler2017}
\bibinfo{author}{\bibfnamefont{G.~W.} \bibnamefont{Winkler}},
  \bibinfo{author}{\bibfnamefont{D.}~\bibnamefont{Varjas}},
  \bibinfo{author}{\bibfnamefont{R.}~\bibnamefont{Skolasinski}},
  \bibinfo{author}{\bibfnamefont{A.~A.} \bibnamefont{Soluyanov}},
  \bibinfo{author}{\bibfnamefont{M.}~\bibnamefont{Troyer}}, \bibnamefont{and}
  \bibinfo{author}{\bibfnamefont{M.}~\bibnamefont{Wimmer}},
  \bibinfo{journal}{Physical Review Letters} \textbf{\bibinfo{volume}{119}},
  \bibinfo{pages}{037701} (\bibinfo{year}{2017}).

\bibitem[{\citenamefont{Winkler}(2003)}]{winkler2003spin}
\bibinfo{author}{\bibfnamefont{R.}~\bibnamefont{Winkler}},
  \emph{\bibinfo{title}{Spin-orbit Coupling Effects in Two-Dimensional Electron
  and Hole Systems}}, no. \bibinfo{number}{191} in \bibinfo{series}{Physics and
  Astronomy Online Library} (\bibinfo{publisher}{Springer},
  \bibinfo{year}{2003}).

\bibitem[{\citenamefont{Pfeffer and Zawadzki}(1996)}]{Pfeffer1996}
\bibinfo{author}{\bibfnamefont{P.}~\bibnamefont{Pfeffer}} \bibnamefont{and}
  \bibinfo{author}{\bibfnamefont{W.}~\bibnamefont{Zawadzki}},
  \bibinfo{journal}{Physical Review B} \textbf{\bibinfo{volume}{53}},
  \bibinfo{pages}{12813} (\bibinfo{year}{1996}).

\bibitem[{\citenamefont{Faria~Junior et~al.}(2016)\citenamefont{Faria~Junior,
  Campos, Bastos, Gmitra, Fabian, and Sipahi}}]{FariaJunior2016}
\bibinfo{author}{\bibfnamefont{P.~E.} \bibnamefont{Faria~Junior}},
  \bibinfo{author}{\bibfnamefont{T.}~\bibnamefont{Campos}},
  \bibinfo{author}{\bibfnamefont{C.~M.~O.} \bibnamefont{Bastos}},
  \bibinfo{author}{\bibfnamefont{M.}~\bibnamefont{Gmitra}},
  \bibinfo{author}{\bibfnamefont{J.}~\bibnamefont{Fabian}}, \bibnamefont{and}
  \bibinfo{author}{\bibfnamefont{G.~M.} \bibnamefont{Sipahi}},
  \bibinfo{journal}{Physical Review B} \textbf{\bibinfo{volume}{93}},
  \bibinfo{pages}{235204} (\bibinfo{year}{2016}).

\bibitem[{\citenamefont{Chuang and Chang}(1996)}]{Chuang1996}
\bibinfo{author}{\bibfnamefont{S.~L.} \bibnamefont{Chuang}} \bibnamefont{and}
  \bibinfo{author}{\bibfnamefont{C.~S.} \bibnamefont{Chang}},
  \bibinfo{journal}{Physical Review B} \textbf{\bibinfo{volume}{54}},
  \bibinfo{pages}{2491} (\bibinfo{year}{1996}).

\bibitem[{\citenamefont{Beresford}(2004)}]{Beresford2004}
\bibinfo{author}{\bibfnamefont{R.}~\bibnamefont{Beresford}},
  \bibinfo{journal}{Journal of Applied Physics} \textbf{\bibinfo{volume}{95}},
  \bibinfo{pages}{6216} (\bibinfo{year}{2004}).

\bibitem[{\citenamefont{Rinke et~al.}(2008)\citenamefont{Rinke, Winkelnkemper,
  Qteish, Bimberg, Neugebauer, and Scheffler}}]{Rinke2008}
\bibinfo{author}{\bibfnamefont{P.}~\bibnamefont{Rinke}},
  \bibinfo{author}{\bibfnamefont{M.}~\bibnamefont{Winkelnkemper}},
  \bibinfo{author}{\bibfnamefont{A.}~\bibnamefont{Qteish}},
  \bibinfo{author}{\bibfnamefont{D.}~\bibnamefont{Bimberg}},
  \bibinfo{author}{\bibfnamefont{J.}~\bibnamefont{Neugebauer}},
  \bibnamefont{and}
  \bibinfo{author}{\bibfnamefont{M.}~\bibnamefont{Scheffler}},
  \bibinfo{journal}{Physical Review B} \textbf{\bibinfo{volume}{77}},
  \bibinfo{pages}{075202} (\bibinfo{year}{2008}).

\bibitem[{\citenamefont{Fu and Wu}(2008)}]{Fu2008}
\bibinfo{author}{\bibfnamefont{J.~Y.} \bibnamefont{Fu}} \bibnamefont{and}
  \bibinfo{author}{\bibfnamefont{M.~W.} \bibnamefont{Wu}},
  \bibinfo{journal}{Journal of Applied Physics} \textbf{\bibinfo{volume}{104}},
  \bibinfo{pages}{093712} (\bibinfo{year}{2008}).

\bibitem[{\citenamefont{Hansen et~al.}(2005)\citenamefont{Hansen, Bj\"ork,
  Fasth, Thelander, and Samuelson}}]{Hansen2005}
\bibinfo{author}{\bibfnamefont{A.~E.} \bibnamefont{Hansen}},
  \bibinfo{author}{\bibfnamefont{M.~T.} \bibnamefont{Bj\"ork}},
  \bibinfo{author}{\bibfnamefont{I.~C.} \bibnamefont{Fasth}},
  \bibinfo{author}{\bibfnamefont{C.}~\bibnamefont{Thelander}},
  \bibnamefont{and}
  \bibinfo{author}{\bibfnamefont{L.}~\bibnamefont{Samuelson}},
  \bibinfo{journal}{Physical Review B} \textbf{\bibinfo{volume}{71}},
  \bibinfo{pages}{205328} (\bibinfo{year}{2005}).

\bibitem[{\citenamefont{Dhara et~al.}(2009)\citenamefont{Dhara, Solanki, Singh,
  Narayanan, Chaudhari, Gokhale, Bhattacharya, and Deshmukh}}]{Dhara2009}
\bibinfo{author}{\bibfnamefont{S.}~\bibnamefont{Dhara}},
  \bibinfo{author}{\bibfnamefont{H.~S.} \bibnamefont{Solanki}},
  \bibinfo{author}{\bibfnamefont{V.}~\bibnamefont{Singh}},
  \bibinfo{author}{\bibfnamefont{A.}~\bibnamefont{Narayanan}},
  \bibinfo{author}{\bibfnamefont{P.}~\bibnamefont{Chaudhari}},
  \bibinfo{author}{\bibfnamefont{M.}~\bibnamefont{Gokhale}},
  \bibinfo{author}{\bibfnamefont{A.}~\bibnamefont{Bhattacharya}},
  \bibnamefont{and} \bibinfo{author}{\bibfnamefont{M.~M.}
  \bibnamefont{Deshmukh}}, \bibinfo{journal}{Physical Review B}
  \textbf{\bibinfo{volume}{79}}, \bibinfo{pages}{121311}
  (\bibinfo{year}{2009}).

\bibitem[{\citenamefont{Roulleau et~al.}(2010)\citenamefont{Roulleau, Choi,
  Riedi, Heinzel, Shorubalko, Ihn, and Ensslin}}]{Roulleau2010}
\bibinfo{author}{\bibfnamefont{P.}~\bibnamefont{Roulleau}},
  \bibinfo{author}{\bibfnamefont{T.}~\bibnamefont{Choi}},
  \bibinfo{author}{\bibfnamefont{S.}~\bibnamefont{Riedi}},
  \bibinfo{author}{\bibfnamefont{T.}~\bibnamefont{Heinzel}},
  \bibinfo{author}{\bibfnamefont{I.}~\bibnamefont{Shorubalko}},
  \bibinfo{author}{\bibfnamefont{T.}~\bibnamefont{Ihn}}, \bibnamefont{and}
  \bibinfo{author}{\bibfnamefont{K.}~\bibnamefont{Ensslin}},
  \bibinfo{journal}{Physical Review B} \textbf{\bibinfo{volume}{81}},
  \bibinfo{pages}{155449} (\bibinfo{year}{2010}).

\bibitem[{\citenamefont{Est\'evez~Hern\'andez
  et~al.}(2010)\citenamefont{Est\'evez~Hern\'andez, Akabori, Sladek, Volk,
  Alagha, Hardtdegen, Pala, Demarina, Gr\"utzmacher, and
  Sch\"apers}}]{PhysRevB.82.235303}
\bibinfo{author}{\bibfnamefont{S.}~\bibnamefont{Est\'evez~Hern\'andez}},
  \bibinfo{author}{\bibfnamefont{M.}~\bibnamefont{Akabori}},
  \bibinfo{author}{\bibfnamefont{K.}~\bibnamefont{Sladek}},
  \bibinfo{author}{\bibfnamefont{C.}~\bibnamefont{Volk}},
  \bibinfo{author}{\bibfnamefont{S.}~\bibnamefont{Alagha}},
  \bibinfo{author}{\bibfnamefont{H.}~\bibnamefont{Hardtdegen}},
  \bibinfo{author}{\bibfnamefont{M.~G.} \bibnamefont{Pala}},
  \bibinfo{author}{\bibfnamefont{N.}~\bibnamefont{Demarina}},
  \bibinfo{author}{\bibfnamefont{D.}~\bibnamefont{Gr\"utzmacher}},
  \bibnamefont{and}
  \bibinfo{author}{\bibfnamefont{T.}~\bibnamefont{Sch\"apers}},
  \bibinfo{journal}{Physical Review B} \textbf{\bibinfo{volume}{82}},
  \bibinfo{pages}{235303} (\bibinfo{year}{2010}).

\bibitem[{\citenamefont{Liang and Gao}(2012)}]{Liang2012}
\bibinfo{author}{\bibfnamefont{D.}~\bibnamefont{Liang}} \bibnamefont{and}
  \bibinfo{author}{\bibfnamefont{X.~P.} \bibnamefont{Gao}},
  \bibinfo{journal}{Nano Letters} \textbf{\bibinfo{volume}{12}},
  \bibinfo{pages}{3263} (\bibinfo{year}{2012}).

\bibitem[{\citenamefont{Luo et~al.}(2017)\citenamefont{Luo, Li, and
  Zunger}}]{Luo2017}
\bibinfo{author}{\bibfnamefont{J.-W.} \bibnamefont{Luo}},
  \bibinfo{author}{\bibfnamefont{S.-S.} \bibnamefont{Li}}, \bibnamefont{and}
  \bibinfo{author}{\bibfnamefont{A.}~\bibnamefont{Zunger}},
  \bibinfo{journal}{Physical Review Letters} \textbf{\bibinfo{volume}{119}},
  \bibinfo{pages}{126401} (\bibinfo{year}{2017}).

\bibitem[{\citenamefont{Zhang et~al.}(2015{\natexlab{a}})\citenamefont{Zhang,
  Tang, Jin, Duan, He, Rong, He, Zhang, Qin, Dai et~al.}}]{Zhang2015a}
\bibinfo{author}{\bibfnamefont{S.}~\bibnamefont{Zhang}},
  \bibinfo{author}{\bibfnamefont{N.}~\bibnamefont{Tang}},
  \bibinfo{author}{\bibfnamefont{W.}~\bibnamefont{Jin}},
  \bibinfo{author}{\bibfnamefont{J.}~\bibnamefont{Duan}},
  \bibinfo{author}{\bibfnamefont{X.}~\bibnamefont{He}},
  \bibinfo{author}{\bibfnamefont{X.}~\bibnamefont{Rong}},
  \bibinfo{author}{\bibfnamefont{C.}~\bibnamefont{He}},
  \bibinfo{author}{\bibfnamefont{L.}~\bibnamefont{Zhang}},
  \bibinfo{author}{\bibfnamefont{X.}~\bibnamefont{Qin}},
  \bibinfo{author}{\bibfnamefont{L.}~\bibnamefont{Dai}}, \bibnamefont{et~al.},
  \bibinfo{journal}{Nano Letters} \textbf{\bibinfo{volume}{15}},
  \bibinfo{pages}{1152} (\bibinfo{year}{2015}{\natexlab{a}}).

\bibitem[{\citenamefont{Dick et~al.}(2010)\citenamefont{Dick, Caroff,
  Bolinsson, Messing, Johansson, Deppert, Wallenberg, and
  Samuelson}}]{Dick2010}
\bibinfo{author}{\bibfnamefont{K.~A.} \bibnamefont{Dick}},
  \bibinfo{author}{\bibfnamefont{P.}~\bibnamefont{Caroff}},
  \bibinfo{author}{\bibfnamefont{J.}~\bibnamefont{Bolinsson}},
  \bibinfo{author}{\bibfnamefont{M.~E.} \bibnamefont{Messing}},
  \bibinfo{author}{\bibfnamefont{J.}~\bibnamefont{Johansson}},
  \bibinfo{author}{\bibfnamefont{K.}~\bibnamefont{Deppert}},
  \bibinfo{author}{\bibfnamefont{L.~R.} \bibnamefont{Wallenberg}},
  \bibnamefont{and}
  \bibinfo{author}{\bibfnamefont{L.}~\bibnamefont{Samuelson}},
  \bibinfo{journal}{Semiconductor Science and Technology}
  \textbf{\bibinfo{volume}{25}}, \bibinfo{pages}{024009}
  (\bibinfo{year}{2010}).

\bibitem[{\citenamefont{Krogstrup et~al.}(2010)\citenamefont{Krogstrup,
  Popovitz-Biro, Johnson, Madsen, Nyg{\aa}rd, and Shtrikman}}]{krogstrup2010}
\bibinfo{author}{\bibfnamefont{P.}~\bibnamefont{Krogstrup}},
  \bibinfo{author}{\bibfnamefont{R.}~\bibnamefont{Popovitz-Biro}},
  \bibinfo{author}{\bibfnamefont{E.}~\bibnamefont{Johnson}},
  \bibinfo{author}{\bibfnamefont{M.~H.} \bibnamefont{Madsen}},
  \bibinfo{author}{\bibfnamefont{J.}~\bibnamefont{Nyg{\aa}rd}},
  \bibnamefont{and}
  \bibinfo{author}{\bibfnamefont{H.}~\bibnamefont{Shtrikman}},
  \bibinfo{journal}{Nano Letters} \textbf{\bibinfo{volume}{10}},
  \bibinfo{pages}{4475} (\bibinfo{year}{2010}).

\bibitem[{\citenamefont{Hjort et~al.}(2014)\citenamefont{Hjort, Lehmann,
  Knutsson, Zakharov, Du, Sakong, Timm, Nylund, Lundgren, Kratzer
  et~al.}}]{Hjort2014}
\bibinfo{author}{\bibfnamefont{M.}~\bibnamefont{Hjort}},
  \bibinfo{author}{\bibfnamefont{S.}~\bibnamefont{Lehmann}},
  \bibinfo{author}{\bibfnamefont{J.}~\bibnamefont{Knutsson}},
  \bibinfo{author}{\bibfnamefont{A.~A.} \bibnamefont{Zakharov}},
  \bibinfo{author}{\bibfnamefont{Y.~A.} \bibnamefont{Du}},
  \bibinfo{author}{\bibfnamefont{S.}~\bibnamefont{Sakong}},
  \bibinfo{author}{\bibfnamefont{R.}~\bibnamefont{Timm}},
  \bibinfo{author}{\bibfnamefont{G.}~\bibnamefont{Nylund}},
  \bibinfo{author}{\bibfnamefont{E.}~\bibnamefont{Lundgren}},
  \bibinfo{author}{\bibfnamefont{P.}~\bibnamefont{Kratzer}},
  \bibnamefont{et~al.}, \bibinfo{journal}{ACS Nano}
  \textbf{\bibinfo{volume}{8}}, \bibinfo{pages}{12346} (\bibinfo{year}{2014}).

\bibitem[{\citenamefont{Panse et~al.}(2011)\citenamefont{Panse, Kriegner, and
  Bechstedt}}]{Panse2011}
\bibinfo{author}{\bibfnamefont{C.}~\bibnamefont{Panse}},
  \bibinfo{author}{\bibfnamefont{D.}~\bibnamefont{Kriegner}}, \bibnamefont{and}
  \bibinfo{author}{\bibfnamefont{F.}~\bibnamefont{Bechstedt}},
  \bibinfo{journal}{Physical Review B} \textbf{\bibinfo{volume}{84}},
  \bibinfo{pages}{075217} (\bibinfo{year}{2011}).

\bibitem[{\citenamefont{Wang et~al.}(2008)\citenamefont{Wang, Cai, and
  Zhang}}]{Wang20081}
\bibinfo{author}{\bibfnamefont{N.}~\bibnamefont{Wang}},
  \bibinfo{author}{\bibfnamefont{Y.}~\bibnamefont{Cai}}, \bibnamefont{and}
  \bibinfo{author}{\bibfnamefont{R.}~\bibnamefont{Zhang}},
  \bibinfo{journal}{Materials Science and Engineering: R: Reports}
  \textbf{\bibinfo{volume}{60}}, \bibinfo{pages}{1 } (\bibinfo{year}{2008}).

\bibitem[{\citenamefont{Caroff et~al.}(2009)\citenamefont{Caroff, Dick,
  Johansson, Messing, Deppert, and Samuelson}}]{Caroff2009:Nano}
\bibinfo{author}{\bibfnamefont{P.}~\bibnamefont{Caroff}},
  \bibinfo{author}{\bibfnamefont{K.~A.} \bibnamefont{Dick}},
  \bibinfo{author}{\bibfnamefont{J.}~\bibnamefont{Johansson}},
  \bibinfo{author}{\bibfnamefont{M.~E.} \bibnamefont{Messing}},
  \bibinfo{author}{\bibfnamefont{K.}~\bibnamefont{Deppert}}, \bibnamefont{and}
  \bibinfo{author}{\bibfnamefont{L.}~\bibnamefont{Samuelson}},
  \bibinfo{journal}{Nature Nanotechnology} \textbf{\bibinfo{volume}{4}},
  \bibinfo{pages}{50} (\bibinfo{year}{2009}).

\bibitem[{\citenamefont{Fortuna and Li}(2010)}]{0268-1242-25-2-024005}
\bibinfo{author}{\bibfnamefont{S.~A.} \bibnamefont{Fortuna}} \bibnamefont{and}
  \bibinfo{author}{\bibfnamefont{X.}~\bibnamefont{Li}},
  \bibinfo{journal}{Semiconductor Science and Technology}
  \textbf{\bibinfo{volume}{25}}, \bibinfo{pages}{024005}
  (\bibinfo{year}{2010}).

\bibitem[{\citenamefont{Krishnamachari
  et~al.}(2004)\citenamefont{Krishnamachari, Borgstrom, Ohlsson, Panev,
  Samuelson, Seifert, Larsson, and Wallenberg}}]{Krishnamachari2004}
\bibinfo{author}{\bibfnamefont{U.}~\bibnamefont{Krishnamachari}},
  \bibinfo{author}{\bibfnamefont{M.}~\bibnamefont{Borgstrom}},
  \bibinfo{author}{\bibfnamefont{B.~J.} \bibnamefont{Ohlsson}},
  \bibinfo{author}{\bibfnamefont{N.}~\bibnamefont{Panev}},
  \bibinfo{author}{\bibfnamefont{L.}~\bibnamefont{Samuelson}},
  \bibinfo{author}{\bibfnamefont{W.}~\bibnamefont{Seifert}},
  \bibinfo{author}{\bibfnamefont{M.~W.} \bibnamefont{Larsson}},
  \bibnamefont{and} \bibinfo{author}{\bibfnamefont{L.~R.}
  \bibnamefont{Wallenberg}}, \bibinfo{journal}{Applied Physics Letters}
  \textbf{\bibinfo{volume}{85}}, \bibinfo{pages}{2077} (\bibinfo{year}{2004}).

\bibitem[{\citenamefont{Zhang et~al.}(2015{\natexlab{b}})\citenamefont{Zhang,
  Zheng, Lu, Chen, Lu, and Zou}}]{Zhang2015}
\bibinfo{author}{\bibfnamefont{Z.}~\bibnamefont{Zhang}},
  \bibinfo{author}{\bibfnamefont{K.}~\bibnamefont{Zheng}},
  \bibinfo{author}{\bibfnamefont{Z.-Y.} \bibnamefont{Lu}},
  \bibinfo{author}{\bibfnamefont{P.-P.} \bibnamefont{Chen}},
  \bibinfo{author}{\bibfnamefont{W.}~\bibnamefont{Lu}}, \bibnamefont{and}
  \bibinfo{author}{\bibfnamefont{J.}~\bibnamefont{Zou}}, \bibinfo{journal}{Nano
  Letters} \textbf{\bibinfo{volume}{15}}, \bibinfo{pages}{876}
  (\bibinfo{year}{2015}{\natexlab{b}}).

\bibitem[{\citenamefont{Xu et~al.}(2012)\citenamefont{Xu, Wang, Guo, Liao, Gao,
  Tan, Jagadish, and Zou}}]{Xu2012}
\bibinfo{author}{\bibfnamefont{H.}~\bibnamefont{Xu}},
  \bibinfo{author}{\bibfnamefont{Y.}~\bibnamefont{Wang}},
  \bibinfo{author}{\bibfnamefont{Y.}~\bibnamefont{Guo}},
  \bibinfo{author}{\bibfnamefont{Z.}~\bibnamefont{Liao}},
  \bibinfo{author}{\bibfnamefont{Q.}~\bibnamefont{Gao}},
  \bibinfo{author}{\bibfnamefont{H.~H.} \bibnamefont{Tan}},
  \bibinfo{author}{\bibfnamefont{C.}~\bibnamefont{Jagadish}}, \bibnamefont{and}
  \bibinfo{author}{\bibfnamefont{J.}~\bibnamefont{Zou}}, \bibinfo{journal}{Nano
  Letters} \textbf{\bibinfo{volume}{12}}, \bibinfo{pages}{5744}
  (\bibinfo{year}{2012}).

\bibitem[{\citenamefont{Zhang et~al.}(2014)\citenamefont{Zhang, Lu, Xu, Chen,
  Lu, and Zou}}]{Zhang2014}
\bibinfo{author}{\bibfnamefont{Z.}~\bibnamefont{Zhang}},
  \bibinfo{author}{\bibfnamefont{Z.}~\bibnamefont{Lu}},
  \bibinfo{author}{\bibfnamefont{H.}~\bibnamefont{Xu}},
  \bibinfo{author}{\bibfnamefont{P.}~\bibnamefont{Chen}},
  \bibinfo{author}{\bibfnamefont{W.}~\bibnamefont{Lu}}, \bibnamefont{and}
  \bibinfo{author}{\bibfnamefont{J.}~\bibnamefont{Zou}}, \bibinfo{journal}{Nano
  Research} \textbf{\bibinfo{volume}{7}}, \bibinfo{pages}{1640}
  (\bibinfo{year}{2014}).

\bibitem[{\citenamefont{Yan et~al.}(2015)\citenamefont{Yan, Zhang, Li, Wu, and
  Ren}}]{Xin2015}
\bibinfo{author}{\bibfnamefont{X.}~\bibnamefont{Yan}},
  \bibinfo{author}{\bibfnamefont{X.}~\bibnamefont{Zhang}},
  \bibinfo{author}{\bibfnamefont{J.}~\bibnamefont{Li}},
  \bibinfo{author}{\bibfnamefont{Y.}~\bibnamefont{Wu}}, \bibnamefont{and}
  \bibinfo{author}{\bibfnamefont{X.}~\bibnamefont{Ren}},
  \bibinfo{journal}{Applied Physics Letters} \textbf{\bibinfo{volume}{107}},
  \bibinfo{pages}{023101} (\bibinfo{year}{2015}).

\bibitem[{\citenamefont{Hallberg et~al.}(2016)\citenamefont{Hallberg, Lehmann,
  Messing, and Dick}}]{Hallberg2016}
\bibinfo{author}{\bibfnamefont{R.~T.} \bibnamefont{Hallberg}},
  \bibinfo{author}{\bibfnamefont{S.}~\bibnamefont{Lehmann}},
  \bibinfo{author}{\bibfnamefont{M.~E.} \bibnamefont{Messing}},
  \bibnamefont{and} \bibinfo{author}{\bibfnamefont{K.~A.} \bibnamefont{Dick}},
  \bibinfo{journal}{Journal of Materials Research}
  \textbf{\bibinfo{volume}{31}}, \bibinfo{pages}{175} (\bibinfo{year}{2016}).

\bibitem[{\citenamefont{Krogstrup et~al.}(2015)\citenamefont{Krogstrup, Ziino,
  Chang, Albrecht, Madsen, Johnson, Nyg{\aa}rd, Marcus, and
  Jespersen}}]{krogstrup2015}
\bibinfo{author}{\bibfnamefont{P.}~\bibnamefont{Krogstrup}},
  \bibinfo{author}{\bibfnamefont{N.}~\bibnamefont{Ziino}},
  \bibinfo{author}{\bibfnamefont{W.}~\bibnamefont{Chang}},
  \bibinfo{author}{\bibfnamefont{S.}~\bibnamefont{Albrecht}},
  \bibinfo{author}{\bibfnamefont{M.}~\bibnamefont{Madsen}},
  \bibinfo{author}{\bibfnamefont{E.}~\bibnamefont{Johnson}},
  \bibinfo{author}{\bibfnamefont{J.}~\bibnamefont{Nyg{\aa}rd}},
  \bibinfo{author}{\bibfnamefont{C.}~\bibnamefont{Marcus}}, \bibnamefont{and}
  \bibinfo{author}{\bibfnamefont{T.}~\bibnamefont{Jespersen}},
  \bibinfo{journal}{Nature materials} \textbf{\bibinfo{volume}{14}},
  \bibinfo{pages}{400} (\bibinfo{year}{2015}).

\bibitem[{\citenamefont{Persson and Xu}(2006)}]{Persson2006}
\bibinfo{author}{\bibfnamefont{M.~P.} \bibnamefont{Persson}} \bibnamefont{and}
  \bibinfo{author}{\bibfnamefont{H.~Q.} \bibnamefont{Xu}},
  \bibinfo{journal}{Physical Review B} \textbf{\bibinfo{volume}{73}},
  \bibinfo{pages}{125346} (\bibinfo{year}{2006}).

\bibitem[{\citenamefont{Redli\ifmmode~\acute{n}\else \'{n}\fi{}ski and
  Peeters}(2008)}]{Peeters2008}
\bibinfo{author}{\bibfnamefont{P.}~\bibnamefont{Redli\ifmmode~\acute{n}\else
  \'{n}\fi{}ski}} \bibnamefont{and} \bibinfo{author}{\bibfnamefont{F.~M.}
  \bibnamefont{Peeters}}, \bibinfo{journal}{Physical Review B}
  \textbf{\bibinfo{volume}{77}}, \bibinfo{pages}{075329}
  (\bibinfo{year}{2008}).

\bibitem[{\citenamefont{{Antipov} et~al.}(2018)\citenamefont{{Antipov},
  {Bargerbos}, {Winkler}, {Bauer}, {Rossi}, and
  {Lutchyn}}}]{2018arXiv180102616A}
\bibinfo{author}{\bibfnamefont{A.~E.} \bibnamefont{{Antipov}}},
  \bibinfo{author}{\bibfnamefont{A.}~\bibnamefont{{Bargerbos}}},
  \bibinfo{author}{\bibfnamefont{G.~W.} \bibnamefont{{Winkler}}},
  \bibinfo{author}{\bibfnamefont{B.}~\bibnamefont{{Bauer}}},
  \bibinfo{author}{\bibfnamefont{E.}~\bibnamefont{{Rossi}}}, \bibnamefont{and}
  \bibinfo{author}{\bibfnamefont{R.~M.} \bibnamefont{{Lutchyn}}},
  \bibinfo{journal}{ArXiv e-prints}  (\bibinfo{year}{2018}),
  \eprint{1801.02616}.

\bibitem[{not({\natexlab{a}})}]{noteCS}
\bibinfo{note}{For ZB nanowires, previous theoretical
  studies~\cite{Persson2006,Peeters2008,Liao2015,Luo2016} have shown that
  changing the nanowire cross-section does not change the trend of the lowest
  conduction subband. The most significant changes are in the excited subbands
  and in their crossings (or anti-crossings) away from $\Gamma$-point. For WZ
  InAs there are no previous studies on the electronic band structure of
  nanowires that we can compare our results. Moreover, our calculations show
  similar trends to the experimental findings of Ref.~\onlinecite{Zhang2015a}
  for WZ nanowires along [0001] and $[11\overline{2}0]$ directions.}

\bibitem[{\citenamefont{Cardona et~al.}(1988)\citenamefont{Cardona,
  Christensen, and Fasol}}]{Cardona1988}
\bibinfo{author}{\bibfnamefont{M.}~\bibnamefont{Cardona}},
  \bibinfo{author}{\bibfnamefont{N.~E.} \bibnamefont{Christensen}},
  \bibnamefont{and} \bibinfo{author}{\bibfnamefont{G.}~\bibnamefont{Fasol}},
  \bibinfo{journal}{Physical Review B} \textbf{\bibinfo{volume}{38}},
  \bibinfo{pages}{1806} (\bibinfo{year}{1988}).

\bibitem[{\citenamefont{Pfeffer and Zawadzki}(1990)}]{Pfeffer1990}
\bibinfo{author}{\bibfnamefont{P.}~\bibnamefont{Pfeffer}} \bibnamefont{and}
  \bibinfo{author}{\bibfnamefont{W.}~\bibnamefont{Zawadzki}},
  \bibinfo{journal}{Physical Review B} \textbf{\bibinfo{volume}{41}},
  \bibinfo{pages}{1561} (\bibinfo{year}{1990}).

\bibitem[{not({\natexlab{b}})}]{note14kp}
\bibinfo{note}{One must be careful in dealing with the ZB 14-band k.p model due
  to its reduced built-in symmetry, as shown in
  Ref.~\onlinecite{ehrhardt2014multi}. Although this feature might lead to
  spurious spin-splittings in very thin nanowires, it does not affect the large
  diameter nanowires --- currently used in experimental setups --- we
  considered in this study.}

\bibitem[{del()}]{deltaminus}
\bibinfo{note}{We compared the conduction band bulk spin-splitting with
  internal, unpublished, ab initio calculations and with reported
  results.~\cite{Hsiu-Fen2012} To reproduce ab initio data we use a value of
  $\Delta^{-}=-0.37\,\textrm{eV}$, in contrast with Winkler's
  book~\cite{winkler2003spin} whose $\Delta^{-}=0.0$ and Jancu \textit{et
  al.}~\cite{Jancu2005} whose $\Delta^{-}=-0.27\,\textrm{eV}$.}

\bibitem[{\citenamefont{Luttinger and Kohn}(1955)}]{Luttinger1955}
\bibinfo{author}{\bibfnamefont{J.}~\bibnamefont{Luttinger}} \bibnamefont{and}
  \bibinfo{author}{\bibfnamefont{W.}~\bibnamefont{Kohn}},
  \bibinfo{journal}{Physical Review} \textbf{\bibinfo{volume}{97}},
  \bibinfo{pages}{869} (\bibinfo{year}{1955}).

\bibitem[{\citenamefont{Kane}(1966)}]{Kane1966}
\bibinfo{author}{\bibfnamefont{E.~O.} \bibnamefont{Kane}},
  \emph{\bibinfo{title}{{Physics of III-V compounds}}}
  (\bibinfo{publisher}{Academic Press}, \bibinfo{address}{New York},
  \bibinfo{year}{1966}), \bibinfo{note}{{v.} 1}.

\bibitem[{\citenamefont{Xia and Chang}(1993)}]{Xia1993}
\bibinfo{author}{\bibfnamefont{J.-B.} \bibnamefont{Xia}} \bibnamefont{and}
  \bibinfo{author}{\bibfnamefont{Y.-C.} \bibnamefont{Chang}},
  \bibinfo{journal}{Physical Review B} \textbf{\bibinfo{volume}{48}},
  \bibinfo{pages}{5179} (\bibinfo{year}{1993}).

\bibitem[{\citenamefont{Lew Yan~Voon and Willatzen}(2009)}]{voon2009k}
\bibinfo{author}{\bibfnamefont{L.~C.} \bibnamefont{Lew Yan~Voon}}
  \bibnamefont{and}
  \bibinfo{author}{\bibfnamefont{M.}~\bibnamefont{Willatzen}},
  \emph{\bibinfo{title}{The k p Method: Electronic Properties of
  Semiconductors}} (\bibinfo{publisher}{Springer}, \bibinfo{year}{2009}).

\bibitem[{\citenamefont{Miao et~al.}(2012)\citenamefont{Miao, Yan, VandeWalle,
  Lou, Li, and Chang}}]{Miao2012}
\bibinfo{author}{\bibfnamefont{M.~S.} \bibnamefont{Miao}},
  \bibinfo{author}{\bibfnamefont{Q.}~\bibnamefont{Yan}},
  \bibinfo{author}{\bibfnamefont{C.~G.} \bibnamefont{VandeWalle}},
  \bibinfo{author}{\bibfnamefont{W.~K.} \bibnamefont{Lou}},
  \bibinfo{author}{\bibfnamefont{L.~L.} \bibnamefont{Li}}, \bibnamefont{and}
  \bibinfo{author}{\bibfnamefont{K.}~\bibnamefont{Chang}},
  \bibinfo{journal}{Physical Review Letters} \textbf{\bibinfo{volume}{109}},
  \bibinfo{pages}{186803} (\bibinfo{year}{2012}).

\bibitem[{\citenamefont{{Faria Junior} et~al.}(2014)\citenamefont{{Faria
  Junior}, Campos, and Sipahi}}]{FariaJunior2014}
\bibinfo{author}{\bibfnamefont{P.~E.} \bibnamefont{{Faria Junior}}},
  \bibinfo{author}{\bibfnamefont{T.}~\bibnamefont{Campos}}, \bibnamefont{and}
  \bibinfo{author}{\bibfnamefont{G.~M.} \bibnamefont{Sipahi}},
  \bibinfo{journal}{Journal of Applied Physics} \textbf{\bibinfo{volume}{116}},
  \bibinfo{pages}{193501} (\bibinfo{year}{2014}).

\bibitem[{\citenamefont{Bastard}(1981)}]{Bastard1981}
\bibinfo{author}{\bibfnamefont{G.}~\bibnamefont{Bastard}},
  \bibinfo{journal}{Physical Review B} \textbf{\bibinfo{volume}{24}},
  \bibinfo{pages}{5693} (\bibinfo{year}{1981}).

\bibitem[{\citenamefont{Baraff and Gershoni}(1991)}]{Baraff1991}
\bibinfo{author}{\bibfnamefont{G.~A.} \bibnamefont{Baraff}} \bibnamefont{and}
  \bibinfo{author}{\bibfnamefont{D.}~\bibnamefont{Gershoni}},
  \bibinfo{journal}{Physical Review B} \textbf{\bibinfo{volume}{43}},
  \bibinfo{pages}{4011} (\bibinfo{year}{1991}).

\bibitem[{\citenamefont{Burt}(1992)}]{Burt1992}
\bibinfo{author}{\bibfnamefont{M.~G.} \bibnamefont{Burt}},
  \bibinfo{journal}{Journal of Physics: Condensed Matter}
  \textbf{\bibinfo{volume}{4}}, \bibinfo{pages}{6651} (\bibinfo{year}{1992}).

\bibitem[{\citenamefont{Bastard}(1988)}]{bastard}
\bibinfo{author}{\bibfnamefont{G.}~\bibnamefont{Bastard}},
  \emph{\bibinfo{title}{{Wave mechanics applied to semiconductor
  heterostructures}}} (\bibinfo{publisher}{Les {\'{E}}ditions de Physique},
  \bibinfo{year}{1988}).

\bibitem[{\citenamefont{Rodrigues et~al.}(2000)\citenamefont{Rodrigues,
  Scolfaro, Leite, and Sipahi}}]{Rodrigues2000}
\bibinfo{author}{\bibfnamefont{S.~C.~P.} \bibnamefont{Rodrigues}},
  \bibinfo{author}{\bibfnamefont{L.~M.~R.} \bibnamefont{Scolfaro}},
  \bibinfo{author}{\bibfnamefont{J.~R.} \bibnamefont{Leite}}, \bibnamefont{and}
  \bibinfo{author}{\bibfnamefont{G.~M.} \bibnamefont{Sipahi}},
  \bibinfo{journal}{Applied Physics Letters} \textbf{\bibinfo{volume}{76}},
  \bibinfo{pages}{1015} (\bibinfo{year}{2000}).

\bibitem[{\citenamefont{Mei}(2007)}]{Mei2007}
\bibinfo{author}{\bibfnamefont{T.}~\bibnamefont{Mei}},
  \bibinfo{journal}{Journal of Applied Physics} \textbf{\bibinfo{volume}{102}},
  \bibinfo{eid}{053708} (\bibinfo{year}{2007}).

\bibitem[{\citenamefont{Vukmirov{\'c} and
  Tomi{\'c}}(2008)}]{vukmirovc2008plane}
\bibinfo{author}{\bibfnamefont{N.}~\bibnamefont{Vukmirov{\'c}}}
  \bibnamefont{and}
  \bibinfo{author}{\bibfnamefont{S.}~\bibnamefont{Tomi{\'c}}},
  \bibinfo{journal}{Journal of Applied Physics} \textbf{\bibinfo{volume}{103}},
  \bibinfo{pages}{103718} (\bibinfo{year}{2008}).

\bibitem[{\citenamefont{Ehrhardt and Koprucki}(2014)}]{ehrhardt2014multi}
\bibinfo{author}{\bibfnamefont{M.}~\bibnamefont{Ehrhardt}} \bibnamefont{and}
  \bibinfo{author}{\bibfnamefont{T.}~\bibnamefont{Koprucki}},
  \emph{\bibinfo{title}{Multi-band Effective Mass Approximations: Advanced
  Mathematical Models and Numerical Techniques}}, vol.~\bibinfo{volume}{94}
  (\bibinfo{publisher}{Springer}, \bibinfo{year}{2014}).

\bibitem[{\citenamefont{Budagosky}(2017)}]{Budagosky2017}
\bibinfo{author}{\bibfnamefont{J.~A.} \bibnamefont{Budagosky}},
  \bibinfo{journal}{Physical Review B} \textbf{\bibinfo{volume}{96}},
  \bibinfo{pages}{115443} (\bibinfo{year}{2017}).

\bibitem[{\citenamefont{Jiang et~al.}(2014)\citenamefont{Jiang, Ma, Xu, and
  Song}}]{Jiang2014}
\bibinfo{author}{\bibfnamefont{Y.}~\bibnamefont{Jiang}},
  \bibinfo{author}{\bibfnamefont{X.}~\bibnamefont{Ma}},
  \bibinfo{author}{\bibfnamefont{Y.}~\bibnamefont{Xu}}, \bibnamefont{and}
  \bibinfo{author}{\bibfnamefont{G.}~\bibnamefont{Song}},
  \bibinfo{journal}{Journal of Applied Physics} \textbf{\bibinfo{volume}{116}},
  \bibinfo{pages}{173702} (\bibinfo{year}{2014}).

\bibitem[{\citenamefont{Ma et~al.}(2014)\citenamefont{Ma, Li, Zhang, Jiang, Xu,
  and Song}}]{Ma2014}
\bibinfo{author}{\bibfnamefont{X.}~\bibnamefont{Ma}},
  \bibinfo{author}{\bibfnamefont{K.}~\bibnamefont{Li}},
  \bibinfo{author}{\bibfnamefont{Z.}~\bibnamefont{Zhang}},
  \bibinfo{author}{\bibfnamefont{Y.}~\bibnamefont{Jiang}},
  \bibinfo{author}{\bibfnamefont{Y.}~\bibnamefont{Xu}}, \bibnamefont{and}
  \bibinfo{author}{\bibfnamefont{G.}~\bibnamefont{Song}},
  \bibinfo{journal}{Journal of Applied Physics} \textbf{\bibinfo{volume}{116}},
  \bibinfo{pages}{235702} (\bibinfo{year}{2014}).

\bibitem[{\citenamefont{Feng et~al.}(2012)\citenamefont{Feng, Zhu, Weitering,
  Stocks, Yao, and Xiao}}]{PhysRevB.85.195114}
\bibinfo{author}{\bibfnamefont{W.}~\bibnamefont{Feng}},
  \bibinfo{author}{\bibfnamefont{W.}~\bibnamefont{Zhu}},
  \bibinfo{author}{\bibfnamefont{H.~H.} \bibnamefont{Weitering}},
  \bibinfo{author}{\bibfnamefont{G.~M.} \bibnamefont{Stocks}},
  \bibinfo{author}{\bibfnamefont{Y.}~\bibnamefont{Yao}}, \bibnamefont{and}
  \bibinfo{author}{\bibfnamefont{D.}~\bibnamefont{Xiao}},
  \bibinfo{journal}{Physical Review B} \textbf{\bibinfo{volume}{85}},
  \bibinfo{pages}{195114} (\bibinfo{year}{2012}).

\bibitem[{\citenamefont{Liu et~al.}(2016)\citenamefont{Liu, Zhang, Abdalla, and
  Zunger}}]{ADFM:ADFM201505357}
\bibinfo{author}{\bibfnamefont{Q.}~\bibnamefont{Liu}},
  \bibinfo{author}{\bibfnamefont{X.}~\bibnamefont{Zhang}},
  \bibinfo{author}{\bibfnamefont{L.~B.} \bibnamefont{Abdalla}},
  \bibnamefont{and} \bibinfo{author}{\bibfnamefont{A.}~\bibnamefont{Zunger}},
  \bibinfo{journal}{Advanced Functional Materials}
  \textbf{\bibinfo{volume}{26}}, \bibinfo{pages}{3259} (\bibinfo{year}{2016}).

\bibitem[{\citenamefont{Gmitra and Fabian}(2016)}]{Gmitra2016}
\bibinfo{author}{\bibfnamefont{M.}~\bibnamefont{Gmitra}} \bibnamefont{and}
  \bibinfo{author}{\bibfnamefont{J.}~\bibnamefont{Fabian}},
  \bibinfo{journal}{Physical Review B} \textbf{\bibinfo{volume}{94}},
  \bibinfo{pages}{165202} (\bibinfo{year}{2016}).

\bibitem[{\citenamefont{Guillemin and
  Pollack}(1974)}]{guillemin1974differential}
\bibinfo{author}{\bibfnamefont{V.}~\bibnamefont{Guillemin}} \bibnamefont{and}
  \bibinfo{author}{\bibfnamefont{A.}~\bibnamefont{Pollack}},
  \emph{\bibinfo{title}{Differential Topology}}, Mathematics Series
  (\bibinfo{publisher}{Prentice-Hall}, \bibinfo{year}{1974}).

\bibitem[{\citenamefont{Momma and Izumi}(2011)}]{Momma:db5098}
\bibinfo{author}{\bibfnamefont{K.}~\bibnamefont{Momma}} \bibnamefont{and}
  \bibinfo{author}{\bibfnamefont{F.}~\bibnamefont{Izumi}},
  \bibinfo{journal}{Journal of Applied Crystallography}
  \textbf{\bibinfo{volume}{44}}, \bibinfo{pages}{1272} (\bibinfo{year}{2011}).

\bibitem[{\citenamefont{Luo et~al.}(2011)\citenamefont{Luo, Zhang, and
  Zunger}}]{Luo2011}
\bibinfo{author}{\bibfnamefont{J.-W.} \bibnamefont{Luo}},
  \bibinfo{author}{\bibfnamefont{L.}~\bibnamefont{Zhang}}, \bibnamefont{and}
  \bibinfo{author}{\bibfnamefont{A.}~\bibnamefont{Zunger}},
  \bibinfo{journal}{Physical Review B} \textbf{\bibinfo{volume}{84}},
  \bibinfo{pages}{121303} (\bibinfo{year}{2011}).

\bibitem[{\citenamefont{De and Pryor}(2010)}]{De2010}
\bibinfo{author}{\bibfnamefont{A.}~\bibnamefont{De}} \bibnamefont{and}
  \bibinfo{author}{\bibfnamefont{C.~E.} \bibnamefont{Pryor}},
  \bibinfo{journal}{Physical Review B} \textbf{\bibinfo{volume}{81}},
  \bibinfo{pages}{155210} (\bibinfo{year}{2010}).

\bibitem[{\citenamefont{{Rössler} and Kainz}(2002)}]{Rossler2002}
\bibinfo{author}{\bibfnamefont{U.}~\bibnamefont{{Rössler}}} \bibnamefont{and}
  \bibinfo{author}{\bibfnamefont{J.}~\bibnamefont{Kainz}},
  \bibinfo{journal}{Solid State Communications} \textbf{\bibinfo{volume}{121}},
  \bibinfo{pages}{313} (\bibinfo{year}{2002}).

\bibitem[{\citenamefont{Lew Yan~Voon et~al.}(1996)\citenamefont{Lew Yan~Voon,
  Willatzen, Cardona, and Christensen}}]{LewYanVoon1996}
\bibinfo{author}{\bibfnamefont{L.~C.} \bibnamefont{Lew Yan~Voon}},
  \bibinfo{author}{\bibfnamefont{M.}~\bibnamefont{Willatzen}},
  \bibinfo{author}{\bibfnamefont{M.}~\bibnamefont{Cardona}}, \bibnamefont{and}
  \bibinfo{author}{\bibfnamefont{N.~E.} \bibnamefont{Christensen}},
  \bibinfo{journal}{Physical Review B} \textbf{\bibinfo{volume}{53}},
  \bibinfo{pages}{10703} (\bibinfo{year}{1996}).

\bibitem[{\citenamefont{Rashba and Sheka}(1959)}]{Rashba1959}
\bibinfo{author}{\bibfnamefont{E.}~\bibnamefont{Rashba}} \bibnamefont{and}
  \bibinfo{author}{\bibfnamefont{V.}~\bibnamefont{Sheka}},
  \bibinfo{journal}{Fizika Tverdogo Tela} \textbf{\bibinfo{volume}{1}},
  \bibinfo{pages}{162} (\bibinfo{year}{1959}).

\bibitem[{\citenamefont{Litvinov}(2003)}]{Litvinov2003}
\bibinfo{author}{\bibfnamefont{V.~I.} \bibnamefont{Litvinov}},
  \bibinfo{journal}{Physical Review B} \textbf{\bibinfo{volume}{68}},
  \bibinfo{pages}{155314} (\bibinfo{year}{2003}).

\bibitem[{\citenamefont{de~Andrada~e Silva}(1992)}]{Andrada1992}
\bibinfo{author}{\bibfnamefont{E.~A.} \bibnamefont{de~Andrada~e Silva}},
  \bibinfo{journal}{Physical Review B} \textbf{\bibinfo{volume}{46}},
  \bibinfo{pages}{1921} (\bibinfo{year}{1992}).

\bibitem[{\citenamefont{de~Andrada~e Silva and La~Rocca}(2003)}]{Andrada2003}
\bibinfo{author}{\bibfnamefont{E.~A.} \bibnamefont{de~Andrada~e Silva}}
  \bibnamefont{and} \bibinfo{author}{\bibfnamefont{G.~C.}
  \bibnamefont{La~Rocca}}, \bibinfo{journal}{Physical Review B}
  \textbf{\bibinfo{volume}{67}}, \bibinfo{pages}{165318}
  (\bibinfo{year}{2003}).

\bibitem[{\citenamefont{Zhang and Xia}(2006)}]{Zhang2006}
\bibinfo{author}{\bibfnamefont{X.~W.} \bibnamefont{Zhang}} \bibnamefont{and}
  \bibinfo{author}{\bibfnamefont{J.~B.} \bibnamefont{Xia}},
  \bibinfo{journal}{Physical Review B} \textbf{\bibinfo{volume}{74}},
  \bibinfo{pages}{075304} (\bibinfo{year}{2006}).

\bibitem[{\citenamefont{Alicea}(2010)}]{Alicea2010:PRB}
\bibinfo{author}{\bibfnamefont{J.}~\bibnamefont{Alicea}},
  \bibinfo{journal}{Physical Review B} \textbf{\bibinfo{volume}{81}},
  \bibinfo{pages}{125318} (\bibinfo{year}{2010}).

\bibitem[{\citenamefont{Alicea}(2012)}]{Alicea2012:RPP}
\bibinfo{author}{\bibfnamefont{J.}~\bibnamefont{Alicea}},
  \bibinfo{journal}{Reports on Progress in Physics}
  \textbf{\bibinfo{volume}{75}}, \bibinfo{pages}{076501}
  (\bibinfo{year}{2012}).

\bibitem[{\citenamefont{Elliott and Franz}(2015)}]{Elliott2015}
\bibinfo{author}{\bibfnamefont{S.~R.} \bibnamefont{Elliott}} \bibnamefont{and}
  \bibinfo{author}{\bibfnamefont{M.}~\bibnamefont{Franz}},
  \bibinfo{journal}{Reviews of Modern Physics} \textbf{\bibinfo{volume}{87}},
  \bibinfo{pages}{137} (\bibinfo{year}{2015}).

\bibitem[{\citenamefont{Fu and Kane}(2008)}]{Fu2008:PRL}
\bibinfo{author}{\bibfnamefont{L.}~\bibnamefont{Fu}} \bibnamefont{and}
  \bibinfo{author}{\bibfnamefont{C.~L.} \bibnamefont{Kane}},
  \bibinfo{journal}{Physical Review Letters} \textbf{\bibinfo{volume}{100}},
  \bibinfo{pages}{096407} (\bibinfo{year}{2008}).

\bibitem[{\citenamefont{Drachmann et~al.}(2017)\citenamefont{Drachmann,
  Suominen, Kjaergaard, Shojaei, Palmstrøm, Marcus, and
  Nichele}}]{Drachmann2017}
\bibinfo{author}{\bibfnamefont{A.~C.~C.} \bibnamefont{Drachmann}},
  \bibinfo{author}{\bibfnamefont{H.~J.} \bibnamefont{Suominen}},
  \bibinfo{author}{\bibfnamefont{M.}~\bibnamefont{Kjaergaard}},
  \bibinfo{author}{\bibfnamefont{B.}~\bibnamefont{Shojaei}},
  \bibinfo{author}{\bibfnamefont{C.~J.} \bibnamefont{Palmstrøm}},
  \bibinfo{author}{\bibfnamefont{C.~M.} \bibnamefont{Marcus}},
  \bibnamefont{and} \bibinfo{author}{\bibfnamefont{F.}~\bibnamefont{Nichele}},
  \bibinfo{journal}{Nano Letters} \textbf{\bibinfo{volume}{17}},
  \bibinfo{pages}{1200} (\bibinfo{year}{2017}).

\bibitem[{\citenamefont{{Lutchyn} et~al.}(2017)\citenamefont{{Lutchyn},
  {Bakkers}, {Kouwenhoven}, {Krogstrup}, {Marcus}, and {Oreg}}}]{Lutchyn2017}
\bibinfo{author}{\bibfnamefont{R.~M.} \bibnamefont{{Lutchyn}}},
  \bibinfo{author}{\bibfnamefont{E.~P.~A.~M.} \bibnamefont{{Bakkers}}},
  \bibinfo{author}{\bibfnamefont{L.~P.} \bibnamefont{{Kouwenhoven}}},
  \bibinfo{author}{\bibfnamefont{P.}~\bibnamefont{{Krogstrup}}},
  \bibinfo{author}{\bibfnamefont{C.~M.} \bibnamefont{{Marcus}}},
  \bibnamefont{and} \bibinfo{author}{\bibfnamefont{Y.}~\bibnamefont{{Oreg}}},
  \bibinfo{journal}{ArXiv e-prints}  (\bibinfo{year}{2017}),
  \eprint{1707.04899}.

\bibitem[{\citenamefont{G{\"{u}}l et~al.}(2018)\citenamefont{G{\"{u}}l, Zhang,
  Bommer, de~Moor, Car, Plissard, Bakkers, Geresdi, Watanabe, Taniguchi
  et~al.}}]{gul2018ballistic}
\bibinfo{author}{\bibfnamefont{{\"{O}}.}~\bibnamefont{G{\"{u}}l}},
  \bibinfo{author}{\bibfnamefont{H.}~\bibnamefont{Zhang}},
  \bibinfo{author}{\bibfnamefont{J.~D.~S.} \bibnamefont{Bommer}},
  \bibinfo{author}{\bibfnamefont{M.~W.~A.} \bibnamefont{de~Moor}},
  \bibinfo{author}{\bibfnamefont{D.}~\bibnamefont{Car}},
  \bibinfo{author}{\bibfnamefont{S.~R.} \bibnamefont{Plissard}},
  \bibinfo{author}{\bibfnamefont{E.~P. A.~M.} \bibnamefont{Bakkers}},
  \bibinfo{author}{\bibfnamefont{A.}~\bibnamefont{Geresdi}},
  \bibinfo{author}{\bibfnamefont{K.}~\bibnamefont{Watanabe}},
  \bibinfo{author}{\bibfnamefont{T.}~\bibnamefont{Taniguchi}},
  \bibnamefont{et~al.}, \bibinfo{journal}{Nature Nanotechnology}
  \textbf{\bibinfo{volume}{13}}, \bibinfo{pages}{192} (\bibinfo{year}{2018}).

\bibitem[{\citenamefont{Dongarra et~al.}(2014)\citenamefont{Dongarra, Gates,
  Haidar, Kurzak, Luszczek, Tomov, and Yamazaki}}]{dghklty14}
\bibinfo{author}{\bibfnamefont{J.}~\bibnamefont{Dongarra}},
  \bibinfo{author}{\bibfnamefont{M.}~\bibnamefont{Gates}},
  \bibinfo{author}{\bibfnamefont{A.}~\bibnamefont{Haidar}},
  \bibinfo{author}{\bibfnamefont{J.}~\bibnamefont{Kurzak}},
  \bibinfo{author}{\bibfnamefont{P.}~\bibnamefont{Luszczek}},
  \bibinfo{author}{\bibfnamefont{S.}~\bibnamefont{Tomov}}, \bibnamefont{and}
  \bibinfo{author}{\bibfnamefont{I.}~\bibnamefont{Yamazaki}}, in
  \emph{\bibinfo{booktitle}{Numerical Computations with GPUs}}
  (\bibinfo{publisher}{Springer}, \bibinfo{year}{2014}), pp.
  \bibinfo{pages}{3--28}.

\bibitem[{\citenamefont{Kao et~al.}(2012)\citenamefont{Kao, Lo, Chiang, Chen,
  Wang, Hsu, Ren, Lee, Wu, and Gau}}]{Hsiu-Fen2012}
\bibinfo{author}{\bibfnamefont{H.-F.} \bibnamefont{Kao}},
  \bibinfo{author}{\bibfnamefont{I.}~\bibnamefont{Lo}},
  \bibinfo{author}{\bibfnamefont{J.-C.} \bibnamefont{Chiang}},
  \bibinfo{author}{\bibfnamefont{C.-N.} \bibnamefont{Chen}},
  \bibinfo{author}{\bibfnamefont{W.-T.} \bibnamefont{Wang}},
  \bibinfo{author}{\bibfnamefont{Y.-C.} \bibnamefont{Hsu}},
  \bibinfo{author}{\bibfnamefont{C.-Y.} \bibnamefont{Ren}},
  \bibinfo{author}{\bibfnamefont{M.-E.} \bibnamefont{Lee}},
  \bibinfo{author}{\bibfnamefont{C.-L.} \bibnamefont{Wu}}, \bibnamefont{and}
  \bibinfo{author}{\bibfnamefont{M.-H.} \bibnamefont{Gau}},
  \bibinfo{journal}{Journal of Physics: Condensed Matter}
  \textbf{\bibinfo{volume}{24}}, \bibinfo{pages}{415802}
  (\bibinfo{year}{2012}).

\bibitem[{\citenamefont{Jancu et~al.}(2005)\citenamefont{Jancu, Scholz,
  de~Andrada~e Silva, and {La Rocca}}}]{Jancu2005}
\bibinfo{author}{\bibfnamefont{J.-M.} \bibnamefont{Jancu}},
  \bibinfo{author}{\bibfnamefont{R.}~\bibnamefont{Scholz}},
  \bibinfo{author}{\bibfnamefont{E.~A.} \bibnamefont{de~Andrada~e Silva}},
  \bibnamefont{and} \bibinfo{author}{\bibfnamefont{G.~C.} \bibnamefont{{La
  Rocca}}}, \bibinfo{journal}{Physical Review B} \textbf{\bibinfo{volume}{72}},
  \bibinfo{pages}{193201} (\bibinfo{year}{2005}).

\end{thebibliography}

%===============================================================================

\end{document}